\documentclass[aps,eqsecnum,twocolumn,showpacs,preprintnumbers,nofootinbib,prd,superscriptaddress,groupedaddress,10pt]{revtex4-2}

\makeatletter
\def\l@subsubsection#1#2{}
\def\l@subsubsubsection#1#2{}
\makeatother

\usepackage{empheq}
\usepackage{graphicx,epsfig,epsf,fixmath}
\usepackage[dvipsnames]{xcolor}
\usepackage{epstopdf}
\usepackage{mathtools}
\usepackage{amssymb}
\usepackage{mathrsfs}
\usepackage{relsize}
\usepackage{graphics}
\usepackage{placeins} 
\usepackage{booktabs} 
\usepackage{siunitx} 
\DeclareSIUnit \parsec {pc} 
\usepackage{aas_macros}
\usepackage{bm}
\usepackage{dcolumn}
\usepackage{latexsym}
\usepackage{rotating}
\usepackage{longtable} 

\usepackage{enumerate}
\usepackage{tensor,multirow}
\usepackage{upgreek}
\usepackage{url}
\usepackage{float}
\usepackage[pdfencoding=auto, psdextra,linktocpage]{hyperref}
\definecolor{darkblue}{rgb}{0, 0.3, 0.6}
\hypersetup{
    colorlinks=true,
    linkcolor=darkblue,
    filecolor=darkblue,      
    urlcolor=darkblue,
    citecolor = darkblue,
    pdftitle={Going into a tailspin near the abyss: analytic solutions for spinning particles on near equatorial, plunging orbits in Kerr spacetime
    }
}
\usepackage{orcidlink}

\newcommand{\FE}{\text{FE}}

\newcommand{\gen}{\mathsf{gen}}
\newcommand{\per}{\mathsf{per}}
\newcommand{\crit}{\mathsf{crit}}
\newcommand{\homo}{\mathsf{hom}}
\newcommand{\isco}{\mathsf{isco}}
\newcommand{\pbo}{\mathsf{pbo}}
\newcommand{\spec}{\mathsf{spec}}
\renewcommand{\Im}{\mathrm{Im}\,} 

\newcommand{\dd}{\mathrm{d}}

\newcommand{\totder}[2]{\frac{\mathrm{d} #1}{\mathrm{d} #2}}

\newcommand{\Lzrd}{L_{z\text{rd}}}

\DeclareMathOperator\arctanh{tanh^{-1}}

\newcommand{\bea}{\begin{eqnarray}}
\newcommand{\eea}{\end{eqnarray}}
\newcommand{\be}{\begin{equation}}
\newcommand{\ee}{\end{equation}}
\newcommand{\ba}{\begin{align}}
\newcommand{\ea}{\end{align}}
\newcommand{\sgn}{\mathrm{sgn}}

\begin{document}
\title{
Going into a tailspin near the abyss: analytic solutions for spinning particles on near equatorial, plunging orbits in Kerr spacetime
}
\author{
Gabriel Andres Piovano$^1$ \orcidlink{0000-0003-1782-6813}
}
\email{gabrielandres.piovano@umons.ac.be}
\affiliation{
Physique de l’Univers, Champs et Gravitation, Universit\'{e} de Mons – UMONS,
Place du Parc 20, 7000 Mons, Belgium
}
\affiliation{
Université Libre de Bruxelles, BLU-ULB Brussels Laboratory of the Universe, International
Solvay Institutes, CP 231, B-1050 Brussels, Belgium
}
%

\begin{abstract} 
This work presents, for the first time, analytic solutions for the nearly equatorial, plunging motion of a spinning test-particle in Kerr spacetime. The equations of motion are solved at first-order in the small-body spin for all classes of plunging orbits with energy $E < 1$. The solutions incorporate the small precession of the orbital plane caused by the precession of the particle's spin. Additionally, we present the correction to the radius of the innermost bound circular orbit in closed-form, and introduce a novel, Keplerian-like parametrization for generic plunging orbits. Our solutions will be useful in the modelling of inspiral-merger-ringdown waveforms with self-force methods and black hole perturbation theory.
\end{abstract}
\maketitle

\section{Introduction}
Gravitational wave (GW) astronomy is a blooming research field. Since the first landmark detection of GW in 2015~\cite{LIGOScientific:2016aoc}, the LIGO-Virgo-Kagra (LVK) collaboration has detected almost 400 events~\cite{LIGOScientific:2026sit}, with 300  events reported through the end of the O4 run alone. Almost all of these events originated from astrophysical binaries of comparable masses, although a few asymmetric-mass systems have been observed.

In the milliHertz frequency band, the Laser Interferometer Space Antenna (LISA)~\cite{LISA:2017pwj,LISA:2024hlh} will detect new types of sources like the Extreme Mass Ratio Inspirals (or EMRIs). EMRI binaries are constituted by a stellar mass compact object (the secondary) orbiting a supermassive black hole (the primary), the latter with mass $M \sim 10^5 - 10^7 M_\odot$ \cite{LISA:2024hlh}. Intermediate mass ratio coalescences (IMRAC), with mass ratio $\epsilon \sim 10^{-2} - 10^{-4}$, represent another candidate sources observable by both LISA and next-generation ground based detectors such as the Einstein Telescope~\cite{ET:2019dnz,LISA:2024hlh}.

GW from asymmetric-mass sources like EMRIs and IMRACs are best modeled within the self-force (SF) framework, which naturally accounts for the disparate scales involved in the dynamics. SF schemes rely on perturbative expansions in $\epsilon$ to find approximate solutions of the Einstein field equations. We label with $n$SF the nth order SF term in the mass ratio. At zeroth order in $\epsilon$, the binary is treated as a point particle with mass $\mu$ free-falling in a Kerr spacetime generated by the primary. Corrections to the geodesic motion arise from radiation-reaction forces and finite-size effects of the smaller companion. In turn, the secondary induces metric perturbations on the background spacetime, leading to the generation of gravitational waves~\cite{Pound:2021qin,Barack:2018yvs}. On the other hand, accurate and efficient waveform models for IMRACs require hybrid approaches, which blend SF methods with other approximations~\cite{Honet:2025lmk,Rink:2024swg,Albertini:2022dmc,Leather:2025nhu,vandeMeent:2023ols}.

Recently, Ref.~\cite{Wardell:2021fyy} showed that SF waveform models incorporating 2SF corrections are in remarkable agreement with (numerical-relativity) NR simulations for mass ratios of order $\mathcal O(1/10)$, with hybrid SF-Post-Newtonian (PN) waveforms achieving even better performance almost up-to merger~\cite{Honet:2025dho}. An extensive program is underway to construct inspiral-merger-ringdown (IMR) waveform models for asymmetric-mass binaries based on SF theory, Post-Newtonian expansions and black hole perturbation theory~\cite{Honet:2025dho,Honet:2025lmk,Lhost:2024jmw,Roy:2025kra,Becker:2025zzw,DellaRocca:2025zbe}.

In the SF framework, the merger and ringdown phase can be modeled by considering a point-particle transitioning from the last stable orbit to the plunge phase. The leading-order contribution comes from the geodesic homoclininc motion and plunge, with radiation-reaction effects ensuring the smooth transition from the late inspiral into the plunge. After modeling the orbital motion, it is possible then to produce the merger part of the signal and gluing it with the inspiral part with a matched asymptotic expansion procedure~\cite{Kuchler:2025hwx,Honet:2025dho,Kuchler:2024esj,Lhost:2024jmw}.

In addition to SF effects, complete IMR waveform models must include post-geodesic corrections due to the spin and higher-order multipoles of the smaller object. Secondary spin corrections contribute to the inspiral waveform phase at first post-adiabatic order (or 1PA)~\cite{Mathews:2025nyb}, therefore being on par with conservative 1SF~\cite{vandeMeent:2017bcc,Nasipak:2025tby} and dissipative 2SF effects~\cite{Pound:2021qin,Wardell:2021fyy,Warburton:2021kwk}. In the transition-to-plunge phase, the field equations are expanded in powers of $\epsilon^{1/5}$ in a post-leading transition-to-plunge expansion (or PLT). Such a perturbative expansion must be carried out up to 7 PLT order to obtain a complete IMR waveform in the SF program. Linear corrections due to the secondary spin and 1SF terms enter at third PLT order, and at first post-geodesic order in the plunging orbits~\cite{Kuchler:2025hwx,Honet:2025dho}. 

This work focuses on the conservative effects due to the secondary spin, and neglects radiation reaction and other SF effects. Specifically, we consider bound plunges,
that is, orbits with energy $E < 1$ confined between two turning points, with one of them located inside the primary's horizon~\cite{Compere:2021bkk}.

Recently, great progress has been made in solving the equations of motion for a spinning test-body at linear-order in the particle's spin for periodic orbits~\cite{Drummond:2022neareq,Drummond:2022efc,Witzany:2023bmq,Witzany:2024ttz,Gonzo:2024zxo,Piovano:2025aro}, as well as in the computation of the spin corrections to gravitational fluxes and waveforms~\cite{Piovano:2020zin,Skoupy:2022adh,Skoupy:2024jsi,Drummond:2023wqc,Skoupy:2023lih,Skoupy:spherical-paper,Piovano:2024yks,Grant:2024ivt}.
By contrast, bound plunging orbits for a spinning particle have received little attention in the literature.  Ref.~\cite{Ciou:2025ygb} provides closed-form expressions for plunging orbits from the innermost stable circular orbit (ISCO) in Reissner–Nordstr{\"o}m spacetime. However, a complete analysis of bound (and unbound) plunges for a spinning particle is still missing, and there are no analytic solutions in Kerr spacetime. Geodesic plunges have been studied in much greater detail. Refs.~\cite{Levin:2008yp,Mummery:2023hlo,Li:2023bgn} derived closed-form solutions for plunges on the equatorial plane, while Dyson and van de Meent found analytic solutions for generic plunging orbits~\cite{Dyson:2023fws}, which were later extended to Kerr-Newman spacetime~\cite{Ko:2023igf}.

Skoup{\'y} and Witzany found in Ref.~\cite{Skoupy:analytic-solutions} a brilliant method to construct analytic solutions at linear order in the secondary spin for generic orbits, expressed in terms of ``virtual geodesics''. This formalism was later used to calculate the spin corrections to the GW fluxes~\cite{Skoupy:2026ewu}, while the shifts to the frequencies of~\cite{Skoupy:analytic-solutions} were compared with numerical~\cite{Drummond:2022efc,Drummond:2022neareq, Piovano:2024yks} and analytical results~\cite{Witzany:2023bmq,Witzany:2024ttz}. 
However, it is still unknown how to map their shifts to the orbits with the corrections derived with other approaches~\cite{Drummond:2022neareq,Drummond:2022efc,Witzany:2023bmq,Piovano:2024yks,Skoupy:2022adh,Bini:2017pee}, except for averaged quantities such as the frequencies and GW fluxes. Additionally, Ref.~\cite{Skoupy:analytic-solutions} numerically checked their analytic trajectories against numerical simulations only for periodic orbits, and it does not provide details on unbound, homoclinic, or plunging orbits, or on the shifts to the separatrix.

For the reasons mentioned above, this work focuses on nearly equatorial orbits in Kerr spacetime, for which the equations of motion are partially decoupled~\cite{Drummond:2022neareq,Tanaka:1996ht}. General solutions by quadrature for the nearly equatorial motion were given in Ref.~\cite{Piovano:2025aro}, together with the spin correction to the separatrix. Here, we provide an in-depth study of bound plunging orbits for a spinning particle, including the precession of the particle's spin and the ensuing precession of the orbital plane. We introduce a new, Keplerian-like parametrization for plunging orbits with complex radial roots, valid for spinning and spinless particles. Thanks to this convenient parametrization, we extend the Fixed Eccentricity (FE) spin-gauge, introduced in Ref.~\cite{Piovano:2025aro}, for spin-perturbed orbits with complex radial roots. We then applied the general solutions for planar and polar motion in Refs.~\cite{Piovano:2025aro,Piovano:2024yks}, respectively, to derive new analytic trajectories for bound plunges, which are completely separated in geodesic and linear-in-spin corrections. These closed-form expressions are described in terms of elliptic integrals and elliptic and elementary functions. Moreover, we showed, for the first time, that perturbative solutions for bound plunges of a spinning particle include pole singularities near the horizon. To circumvent this problem, we found novel analytic trajectories that are partially expanded in the secondary spin. These solutions are particularly convenient in waveform modeling applications. Finally, we derived, for the first time, the shift to the innermost bound circular orbit (IBCO) in closed-form.

The paper is structured as follows. Section~\ref{sec:spinning_motion} starts with a brief review of the motion of spinning test-particle in Kerr spacetime.  Then, it introduces the equations of motion for nearly equatorial orbits. Section~\ref{sec:equatorial_orbits} presents the analytic solutions for bound plunges on the equatorial plane, while Section~\ref{sec:polar_motion_spin_prec} gives the solution for the precession of the orbital plane and the spin-precession phase. We wrap up in Section~\ref{sec:discussion} with a summary of the results of this paper, and a discussion of possible follow-ups.

\subsection{Notation}
This work uses geometrized units $G=c=1$. Moreover, the mass $M$ of the primary is set to $M=1$. All spacetime indices are denoted by Greek letters. Partial and covariant derivatives are indicated with a comma and semicolon, respectively, i.e. $f_{\mu,\nu}= \partial_\nu f_\mu$ and  $f_{\mu;\nu}= \nabla_\nu f_\mu$. The metric signature is $(-,+,+,+)$, with the Riemann tensor defined as 
\begin{equation}
 \tensor{R}{^\delta_\sigma_\mu_\nu} \omega_\delta = 2\nabla_{[\mu} \nabla_{\nu]} \omega_\sigma
\end{equation}
where $\omega_\delta$ an arbitrary 1-form, while the square brackets denote antisymmetrization.
$\epsilon_{\alpha\beta\gamma\delta} = \sqrt{-g} e_{\alpha\beta\gamma\delta}$ is the antisymmetric Levi-Civita tensor and $e_{\alpha\beta\gamma \delta}$ the Levi-Civita symbol ($e_{0123} =1$).

\section{Equations of motion for a spinning particle}\label{sec:spinning_motion}
This Section introduces the equations of motion for a spinning test-particle in an ambient spacetime. In our case, we consider as background the Kerr geometry, which is described in Boyer-Lindquist coordinates by the line element
\begin{align}
   \dd s^2 &= -\left(1 - \frac{2 r}{\Sigma}  \right) \dd t^2 + \frac{\Sigma}{\Delta} \dd r^2 + \frac{\Sigma}{1-z^2} \dd z^2 + \nonumber \\ 
   &+\frac{1-z^2}{\Sigma}\left[2 a^2 r (1-z^2)+(a^2+r^2)\Sigma\right] \dd \phi^2 - \nonumber\\
   &- \frac{4 a r (1-z^2)}{\Sigma} \dd t \dd \phi \, ,
\end{align}
with  $z = \cos(\theta)$, $\Delta = r^2 -2 r + a^2$ and $\Sigma = r^2 + a^2 z^2$ while $a$ is the primary spin, which is aligned to the z-axis of a Cartesian coordinate system centered on the primary, namely $a \geq 0$.

The dynamics of a spinning test-body in a background spacetime obey the Mathisson-Papapetrou-Dixon equations \cite{Mathisson:2010, Papapetrou:1951pa, Dixon:1970I}
\begin{subequations}
\label{eq:MPDeq}
  \begin{empheq}{align}
     \frac{D p^\mu}{\dd \tau} &= -\frac{1}{2} R^{\mu}_{~\nu \rho \sigma} v^\nu S^{\rho \sigma}  \, ,  \label{eq:spin_curvature} \\ 
     \frac{D S^{\mu\nu}}{\dd \tau} &= 2 p^{[\mu}v^{\nu]}   \, , 
  \end{empheq}
\end{subequations}
where $D/\dd \tau = v^\mu \nabla_\mu$, $\tau$ is the proper time, $R^{\mu}{}_{\nu\kappa\lambda}$ is the Riemann tensor, $x^\mu(\tau)$ is the representative worldline of the body with tangent vector $v^\mu$. Moreover, $p^\mu$ and $S^{\mu \nu}$ are, respectively, the four-momentum and anti-symmetric spin tensor of the body. Eqs.~\eqref{eq:MPDeq} describe the motion a point-particle endowed with a monopole $\mu$ and a dipole $S$, that is
\begin{equation}
    \mu^2 = -p^\sigma p_\sigma \ , \qquad S=\frac{1}{2} S^{\mu \nu}S_{\mu\nu} \ .
\end{equation}
A spin vector can then be defined in terms of the linear momentum and spin-tensor as
\begin{equation}
    S^\mu = - \frac{1}{2} \epsilon^{\mu\nu\rho\sigma} u_\nu S_{\rho\sigma} \, . \label{eq:spinvectordef}
\end{equation}
where $u^\nu = p^\nu/\mu$. 
Eqs.~\eqref{eq:MPDeq} do not provide a unique solution for the worldlines of the spinning body, since it is possible to describe the motion of the physical object using different centroids~\cite{Semerak:1999qc,Kyrian:2007zz}. To solve this problem, it is necessary to supplement the MPD equations with an additional constraint, which is equivalent to choosing an observer attached to a centroid. Here we consider the convenient Tulczyjew-Dixon condition~\cite{tulczyjew1959motion,Dixon:1970I}:
\begin{equation}
S^{\mu\nu}p_\nu =0 \, . \label{eq:TDconstrain}
\end{equation}
The constraint~\eqref{eq:TDconstrain} implies a closed-form relation between the 4-velocity $v^\mu$ and 4-momentum $p^\mu$ (see Ref.~\cite{Semerak:1999qc}), which in general does not exist for a generic supplementary spin condition.
Moreover, the mass $\mu$ and spin-magnitude $S$ are constants of motion in the Tulczyjew-Dixon condition, while the specific spin vector $s^\rho =S^\rho/\mu$ can be expressed in terms of the specific spin tensor $s^{\rho\sigma} = S^{\rho \sigma}/\mu$ as
\begin{equation}
    s^{\mu\nu} = \frac{1}{\mu}\epsilon^{\mu\nu\alpha\beta}p_\alpha s_{\beta} \, . \label{eq:spintensordef}
\end{equation}

Eqs.~\eqref{eq:MPDeq} can be significantly simplified in the case of asymmetric-mass binaries. In fact,  the spin-magnitude $S$ has the same dimensions of an angular momentum in geometric units, and scales as the mass ratio $\epsilon$ 
\begin{equation}
\frac{S}{\mu M} = \epsilon \chi \,, \quad \chi \equiv S/ \mu^2 \ , \label{eq:spin_magnitude}
\end{equation} 
where the mass $M$ was restored for clarity, while $\chi$ is the dimensionless spin of the test body. The spin $\chi$ is order $\mathcal O(1/10)$ when the secondary is either a black hole or neutron star. Less compact objects may have larger spin, but do not reach the strong field regime near the primary for binaries observable by LISA. See footnote 2 of Ref.~\cite{Skoupy:spherical-paper} and Ref.~\cite{Hartl:2002ig} for more details. Additionally, the spin-curvature force, described by the right-hand side of Eq.~\eqref{eq:spin_curvature}, is conservative in nature. This observation, combined with Eq.~\eqref{eq:spin_magnitude}, implies that only spin effects at order $\mathcal O(\epsilon \chi)$ contribute to the GW phase at 1PA order during the inspiral~\cite{Hinderer:2008dm,Mathews:2025nyb}, and at first post-geodesic order during the plunge~\cite{Honet:2025dho,Kuchler:2025hwx}. Thus, it is sufficient to solve the MPD equations with $\mathcal O(\epsilon \chi)$ accuracy for EMRIs and a large class of IMRAC binaries observable by LISA and the Einstein Telescope. 

At linear order in the secondary spin, Eqs.~\eqref{eq:MPDeq} reduce to
\begin{subequations}
\label{eq:MPDeq_lin}
  \begin{empheq}{align}
    \frac{D_{\rm g} v^\mu_{\rm g}}{\dd \tau} &= 0   \, , \label{eq:2ndEoMgeo}  \\
    \frac{D_{\rm g} \delta v^\mu}{\dd \tau} &  = -\frac{1}{2} R^{\mu}_{~\nu \rho \sigma} v^\nu_{\rm g} s^{\rho \sigma} \, , \label{eq:2nd_order_EoM_linear_momentum} \\
    \frac{D_{\rm g} s^{\mu}}{\dd \tau} &= 0 \, ,  \label{eq:parallel_transport_spin_vector}  
  \end{empheq}
\end{subequations}
with $v^\mu_{\rm g}$ the geodesic 4-velocity and $\delta v^\mu$ the shifts due to the secondary spin, while $D_g/\dd \tau = v^\mu_{\rm g} \nabla_\mu$. Notice that the velocity $v^\mu$ and linear momentum $p^\mu$ are collinear at order $\mathcal O(\epsilon \chi)$, $u^\mu = v^\mu + \mathcal O(\epsilon^2 \chi^2)$~\cite{Semerak:1999qc}, while Eq.~\eqref{eq:parallel_transport_spin_vector} is equivalent to $D s^{\mu \nu} /\dd \tau = 0$ thanks to Eq.~\eqref{eq:spinvectordef}.

Both the spin-vector and spin tensor are parallel transported along a reference geodesic $x^\mu_{\rm g}$. A solution for Eq.~\eqref{eq:parallel_transport_spin_vector} in Kerr spacetime was found by Marck~\cite{marck1983solution} in terms of a  parallel-transported tetrad $e_0^\mu$, $e_1^\mu$, $e_2^\mu$, $e_3^\mu$. The spin vector can then be decomposed in terms of the Marck tetrad as
\begin{equation}
    s^\mu = s_\parallel e_3^\mu + s_\perp (e_1^\mu \cos\psi_{\rm p}(\lambda) + e_2^\mu \sin\psi_{\rm p}(\lambda)) \, , \label{eq:spin-vector}
\end{equation}
where $\psi_{\rm p}(\lambda)$ is the precession phase of the spin vector. We use the same convention of Ref.~\cite{vandeMeent:2019cam} for the Marck tetrad. $\psi_{\rm p}(\lambda)$ obeys a separable equation of motion~\cite{marck1983solution}, which can be solved analytically in terms of elliptic integrals or elementary functions, depending on the type of motion. Ref.~\cite{vandeMeent:2019cam} found analytical expressions for generic periodic orbits, while Ref.~\cite{Piovano:2025aro} derived closed-form expressions for nearly equatorial homoclinic motion.

In a Kerr background, both the system~\eqref{eq:MPDeq} and its linearized counterpart admit two integrals of motion, $E$ and $J_z$, thanks to the Killing vectors $\kappa_{(t)}^\mu \partial_\mu = \partial_t$ and $\kappa_{(\phi)}^\mu \partial_\mu = \partial_\phi$. These constants of motion are defined as
\begin{subequations}
\begin{align}
    E = - u_\mu \kappa_{(t)}^\mu + \frac{1}{2} \kappa^{(t)}_{\mu;\nu} s^{\mu\nu} \,, \label{eq:energy_MPD} \\
    J_z = u_\mu \kappa_{(\phi)}^\mu - \frac{1}{2} \kappa^{(\phi)}_{\mu;\nu} s^{\mu\nu} \label{eq:angular_momentum_MPD} \,,
\end{align}
\end{subequations}
which, at infinity, can be interpreted, respectively, as the orbital energy (normalized by $\mu$) and total orbital angular momentum parallel to the z-axis (rescaled by $\mu M$) of the spinning test-body. Additionally, the system~\eqref{eq:MPDeq_lin} admits two additional constants of motion
\begin{align}
    s_\parallel &= \frac{Y_{\mu\nu} u^\mu s^\nu}{\sqrt{K_{\mu\nu} u^\mu u^\nu}} \,, \label{eq:parallel_spin_MPD}\\
    K &= K_{\mu\nu} u^\mu u^\nu + 4 u^\mu s^{\rho\sigma} Y^\kappa{}_{\left[\mu\right.} Y_{\left.\sigma\right]\rho;\kappa} \,, \label{eq:carter_constant_MPD}
\end{align}
which are conserved at order $\mathcal O(\epsilon\chi)$. The quantities $K$ and $s_\parallel$ are the R\"{u}diger constants~\cite{Rudiger:1981,Rudiger:1983}, which can be interpreted, respectively, as a generalization of the Carter constant for a spinning test-body and the projection of the spin vector onto the orbital angular momentum. The R\"{u}diger constants are a direct consequence of the existence of a Killing-Yano tensor $Y_{\mu\nu}$, $Y_{\mu(\nu;\kappa)} = 0$, with related Killing tensor $K_{\mu\nu}$, for a Kerr metric
(see also~\cite{Compere:2023alp} for the derivation of R\"{u}diger constants valid at order $\mathcal O(\epsilon^2\chi^2)$).  

As observed in Ref.~\cite{Kubiznak:2011ay,Witzany:2019nml}, Eqs.~\eqref{eq:2nd_order_EoM_linear_momentum} are completely integrable due to the existence of the first integrals~\eqref{eq:energy_MPD}-\eqref{eq:carter_constant_MPD} (see Refs.~\cite{Compere:2023alp,Ramond:2024ozy,Ramond:2026fpi} for integrability of the MPD equations at quadratic order in spin). Thus, Eqs.~\eqref{eq:MPDeq_lin} can be recast as a system of first-order differential equations, as elegantly shown by Witzany using the Hamilton-Jacobi formalism~\cite{Witzany:2019nml}. However, the first-order equations of motion of Ref.~\cite{Witzany:2019nml} are not separable, except in certain cases. Eqs.~\eqref{eq:MPDeq_lin} are partially separable for nearly equatorial orbits~\cite{Tanaka:1996ht,Drummond:2022neareq} and quasi-spherical, periodic motion~\cite{Skoupy:spherical-paper}.

In the following, we analytically solved the first-order system of equations derived in Ref.~\cite{Witzany:2019nml} for nearly equatorial bound plunges.

\subsection{Geodesic motion on the equatorial plane} \label{sec:geodesic_equatorial_motion}
This Section presents the relevant features  of plunging geodesics on the equatorial plane of the Kerr black hole. In the case of generic plunges, we also introduce a novel parametrization for the orbits.

The geodesic equations of motion are
\begin{subequations} \label{eq:1st-order-EoMgeo_rad}
    \begin{empheq}[]{align}
        \totder{r_{\rm g}}{\lambda} & = \pm \sqrt{R_{\rm g}(r_{\rm g})}  \, ,\\
        \totder{t_{\rm g}}{\lambda} &= T_{\rm g}(r_{\rm g}) \, , \\
        \totder{\phi_{\rm g}}{\lambda} &=  \Phi_{\rm g}(r_{\rm g})  \, ,
    \end{empheq}
\end{subequations}
where $\lambda$ is the Mino-time~\cite{Mino:2003yg} defined as $ \dd \tau/ \dd \lambda = \Sigma$, while 
\begin{align}
    R_{\rm g}(r_{\rm g}) &= r_{\rm g} V_{\rm g}(r_{\rm g}) \label{eq:1st-order-EoMgeo}  \, , \\
    V_{\rm g}(r_{\rm g}) &=  (E^2_{\rm g} - 1)r^3_{\rm g} + 2r^2_{\rm g} - a^2(1 - E^2_{\rm g}) r_{\rm g}  \nonumber \\
    & - L^2_{z\rm g} r_{\rm g} + 2 \Lzrd^2 \label{eq:1st-order-EoMgeo-reduced}  \, , \\
    T_{\rm g}(r_{\rm g}) &= \frac{1}{\Delta}\Big( E_{\rm g} r^4_{\rm g}  + E_{\rm g} a^2 r^2_{\rm g}  -2 a \Lzrd r_{\rm g} \Big) \label{eq:1st-order-EoMgeoT}  \, ,\\
    \Phi_{\rm g}(r_{\rm g}) &= \frac{1}{\Delta}\Big( (a E_{\rm g} + \Lzrd)r^2_{\rm g} - 2 \Lzrd r_{\rm g}  \Big) \label{eq:1st-order-EoMgeoPhi} \, . 
\end{align}
The quantities $E_{\rm g}$ and $L_{z \rm g}$ are the geodesic constants of motion, with $\Lzrd = L_{z \rm g} - a E_{\rm g}$ the reduced z-component of the angular momentum. In particular, $\Lzrd>0$ ($\Lzrd<0$) corresponds to prograde (retrograde) orbits. We focus on bound plunging orbits\footnote{Refs.~\cite{Liu:2023tcy,Compere:2021bkk} use the term ``trapped" to denote orbits occurring between two turning points, with one of them inside the black hole horizons regardless of the orbital energy. We prefer the term bound plunges to distinguish from trapped orbits with energy $E_{\rm g} \geq 1$.} with energy $0 < E_{\rm g} < 1$, that is, orbits for which the particle starts at a radial turning point, cross the outer horizon $r_+$ and ultimately either reach another turning point inside the event horizons or the ring singularity~\cite{Liu:2023tcy,Compere:2021bkk,Dyson:2023fws,Mummery:2023hlo}.

Geodesic orbits are characterized by the roots of the radial function $R_{\rm g}(r_{\rm g})$, which is a quartic polynomial in the radius $r_{\rm g}$. At least two roots are real, $r_{\rm g} = 0$ and $r_{1 \rm g}$, the latter being the largest of the four roots. Several kinds of bound plunging motion exist, depending on the nature of the roots. In general, bound plunging orbits can occur whenever one or more of the radial roots are smaller than the inner horizon $r_+$~\cite{Compere:2021bkk}. Thus, bound plunges are always admissible on the equatorial plane. 

The roots of $R_{\rm g}(r_{\rm g})$ can be ordered as
\begin{equation}
    r_{1 \rm g} \geq r_{2 \rm g} \geq r_{3 \rm g} > 0
\end{equation}
when all four of them are real. Moreover, the third root $r_{3 \rm g}$ can be written as
\begin{equation}
    r_{3 \rm g} = \frac{2}{1-E^2_{\rm g}} - (r_{1 \rm g} + r_{2 \rm g}) \ .
\end{equation}
Instead of the energy and  angular momentum $(E_{\rm g}, L_{z \rm g})$ or the roots $(r_{1 \rm g}, r_{2 \rm g})$, the orbits can be parametrized using the Keplerian-like orbital elements $(p_{\rm g}, e_{\rm g})$
\begin{align}
    r_{1 \rm g} = \frac{p_{\rm g}}{1 - e_{\rm g}} \, \qquad  r_{2 \rm g} = \frac{p_{\rm g}}{1 + e_{\rm g}} \,
\end{align}
with $e_{\rm g}$ and $p_{\rm g}$ being the eccentricity and semi-latus rectum, respectively. The above parametrization is particularly convenient for periodic orbits, but it is also useful to model bound plunging orbits with real roots. We will show later on how to extend this parametrization in the case of complex roots.

Before discussing the different kinds of bound plunging orbits, it is worth mentioning four important radii. The largest of these is the innermost stable circular orbit (ISCO) $r_{\rm g,\text{isco}}$, which identifies the triple root for the radial potential. The ISCO radius $r_{\rm g,\text{isco}} = r_{\rm g,\text{isco}}(a)$, and the corresponding energy and angular momentum at the ISCO, depend only on the primary spin $a$.  All circular, equatorial orbits with radii smaller than $r_{\rm g,\text{isco}}$ are unstable. A second relevant radius is the innermost bound circular orbit, or IBCO, which corresponds to the double root of the radial potential for $E_{\rm g} = 1$. In such a case, $R_{\rm g}(r_{\rm g})$ simplifies to
\begin{equation}
     R_{\rm g}(r_{\rm g}) = r_{\rm g}(2 r^2_{\rm g} - L^2_{z \rm g} r_{\rm g} + 2 \Lzrd^2) \ , \; \text{ for }  E_{\rm g} = 1 
\end{equation}
Imposing the condition that the latter has a double root, we get $r_{\rm g,\text{ibco}} = 2 \mp a + 2 \sqrt{1 + \mp a}$, where the upper (lower) sign correspond to prograde (retrograde) orbits. Unstable circular orbits exist only for radii between $[r_{\rm g,\text{ibco}},r_{\rm g,\text{isco}})$. Massive particles located at $r_{\rm g} < r_{\rm g,\text{ibco}}$ cannot stay on circular orbits, and inevitably plunge. Notice that particles with $E_{\rm g} =1$ have zero-velocity at infinity.
Finally, the inner and outer event horizons are the last relevant radii for massive particles. As mentioned in~\cite{Mummery:2023hlo}, the following inequalities are always satisfied (for a non-extremal primary): $r_- < r_+ < r_{\rm g,\text{ibco}} < r_{\rm g,\text{isco}}$.

In the following, we provide more details on each type of bound plunge. Analytic solutions for these orbits are presented in Appendix~\ref{app:geo_analytic}.

\subsubsection{Critical plunges}
Critical plunges occur when all four roots of $R_{\rm g}(r_{\rm g})$ are real and $r_{2 \rm g} = r_{3 \rm g}$, which identifies the radius of an unstable, circular orbit located  between $[r_{\rm g,\text{ibco}},r_{\rm g,\text{isco}})$. The orbital elements $(p^*_{\rm g},e^*_{\rm g})$ for which $r_{2 \rm g} = r_{3 \rm g}$ defines the location of the separatrix. In general, a separatrix is the boundary that divides regions of the phase space with distinct qualitative behaviors. In our case, the separatrix corresponds to a homoclinic orbits~\cite{Levin:2008yp}, which demarcate the phase space of periodic motion and bound plunging orbits. Typically, the semilatus rectum $p^*_{\rm g} = p^*_{\rm g}(a,e^*_{\rm g})$ for the separatrix is considered as a function of $(a,e^*_{\rm g})$, with eccentricity $e^*_{\rm g}$ a free parameter. The location of the separatrix $p^*_{\rm g}$ can be found as root of a polynomial~\cite{Stein:2019buj}.

Critical plunging orbits have the same constants of motion of homoclinic orbits, but different initial conditions, with $r_{\rm g}(0) \in [0,r_{2 \rm g})$ for $\lambda =0$. Moreover, the Mino time $\Lambda^\crit$ required to reach the ring singularity $r_{\rm g} = 0$ from the double root $r_{2\rm g}$ is infinite
\begin{equation}
    \Lambda^\crit = \int^{r_{2\rm g}}_0 \frac{\dd r'}{\sqrt{R_{\rm g}(r')}} = \infty        
\end{equation}
Another characteristic of a critical plunge is the winding of the particle's motion around the turning point $r_{2 \rm g}$, since the azimuthal trajectory $\phi(r_{\rm g})$ (and the coordinate time trajectory $t_{\rm g}(r_{\rm g})$ as well) diverge logarithmically at the double root (see Appendices~\ref{app:integrals_critical_plunge},\ref{app:geo_analytic} and Fig.~\ref{fig:critical_plunge}).

\subsubsection{ISCO plunges}
A plunging orbit from the ISCO can be considered as a special case of a critical plunge with $(p^*_{\rm g},e^*_{\rm g}) = (r_{\rm g, \text{isco}},0)$. ISCO plunges have the same constants of motion of the ISCO, but their orbital radius evolve from $r_{\rm g, \text{isco}}$ to $r_{\rm g} = 0$.
Since the radial potential has a triple root, the coordinate time trajectory $t_{\rm g}(r_{\rm g})$ and azimuthal trajectory $\phi_{\rm g}(r_{\rm g})$ diverge as a pole near the ISCO (see Appendices~\ref{app:integrals_critical_plunge},\ref{app:geo_analytic}). As a result, a particle on an ISCO plunge spends much more time winding near the ISCO than near the double root in a critical plunge (compare Fig.~\ref{fig:isco} with Fig.~\ref{fig:critical_plunge}.

\subsubsection{Plunges related to bound orbits (PBOs)}
Plunges related to bound orbits (or PBOs in short) have the same constants of motion as periodic orbits but initial condition $r_{\rm g}(0) \leq r_{3 \rm g}$ for $ \lambda = 0$ (see \cite{Li:2023bgn, Dyson:2023fws}). Depending on the root structure of the radial potential, PBOs are described by either elliptic integrals and functions, or by elementary functions. The latter case occurs when the radial potential admits stable, circular orbits, that is $r_{1 \rm g} = r_{2 \rm g}$. PBOs with a double root reduce to the circular motion at the ISCO when $r_{1 \rm g} = r_{2 \rm g} \to r_{3 \rm g}$, while PBOs with all distinct roots reduce to the critical plunges for $r_{2 \rm g} \to r_{3 \rm g}$. Finally, there is a maximum value of the orbital angular momentum, $L^+_{z \rm g}$, for which PBOs are allowed~\cite{Compere:2021kjz}. Such a value corresponds to the angular momentum for which $r_{3 \rm g} = r_+$. When $L^+_{z \rm g} \leq L_{z \rm g}$, PBOs are forbidden, but periodic orbits are still possible.

\subsubsection{Generic plunges}
Generic plunging orbits are characterized by the presence of complex conjugate roots $r_{2 \rm g} = \overline{r_{3 \rm g}} = \rho_{r \rm g} + i \rho_{i \rm g}$, with $\rho_{r \rm g}$ and $\rho_{i \rm g}$ the real and imaginary parts, and the overbar denoting complex conjugation. We introduce here a novel parametrization, which relies on the same Keplerian-like orbital elements commonly used for periodic orbits. This parametrization is convenient for the derivation of the spin-corrections to the motion. 
First, we define the real root $r_{1 \rm g}$ and real part $\rho_{r \rm g}$ as
\begin{align}
    r_{1 \rm g} = \frac{p_{\rm g}}{1 - e_{\rm g}} \ , \; \qquad \rho_{r \rm g} = \frac{p_{\rm g}}{1+e_{\rm g}} \ ,
\end{align}
with $p_{\rm g} \leq p^*_{\rm g}$ and $e_{\rm g}\leq e^*_{\rm g}$. Moreover, we use Vieta's formulas to find relations between the roots $r_{1 \rm g}, r_{2 \rm g}, r_{3 \rm g}$ of the polynomial \eqref{eq:1st-order-EoMgeo-reduced} with its coefficients. By inverting these relations, it is then possible to write $E_{\rm g}, L_{z \rm g}$ and $\rho_{i \rm g}$ in terms of $r_{1 \rm g}$ and $\rho_{r \rm g}$, and so $(p_{\rm g}, e_{\rm g})$. The final result is
\begin{widetext}
\begin{align}
    E_{\rm g}(p_{\rm g}, {e_{\rm g}}) &= \sqrt{\frac{2(\rho_{\rm g} -1) + r_{1 \rm g}}{r_{1 \rm g}+ 2 \rho_{\rm g}}} \ , \label{eq:geo_energy_plunge} \\
    L_{z \rm g}(p_{\rm g}, {e_{\rm g}}) & = \pm 2\sqrt{\frac{(r_{1 \rm g}-2)r^2_{1 \rm g} \rho_{\rm g} \mp 2 \sqrt{a^2 r_{1 \rm g} \rho_{\rm g} \Delta(r_{1 \rm g})\big( 2(\rho_{\rm g} -1) + r_{1 \rm g}\big)} + a^2\big(r_{1 \rm g} \rho_{\rm g} + r_{1 \rm g} + 2 (\rho_{\rm g} -1)\big)}{(r_{1 \rm g} -2)^2(r_{1 \rm g} + 2 \rho_{\rm g})}}  \ ,  \label{eq:geo_angular_momentum_plunge} \\
    \rho_{i \rm g}(p_{\rm g}, {e_{\rm g}}) & = - \sqrt{a^2 - \rho_{\rm g}(\rho_{\rm g} -4) + 2\frac{4\rho_{\rm g} + a^2(\rho_{\rm g} + 1)}{r_{1 \rm g} - 2} + 4\frac{2 a^2 \rho_{\rm g}\mp \sqrt{a^2 r_{1 \rm g}\rho_{r \rm g}\Delta(r_{1 \rm g})\big( 2(\rho_{\rm g} -1) + r_{1 \rm g}\big)}}{(r_{1 \rm g} - 2)^2}} \ ,
\end{align}    
\end{widetext}
with the upper (lower) sign for prograde (retrograde) orbits, and $\Delta(r_{1 \rm g}) = r^2_{1 \rm g} - 2 r_{1 \rm g} + a^2$. 

In the limit $(p_{\rm g}, {e_{\rm g}}) \to (p^*_{\rm g}, {e^*_{\rm g}})$, the imaginary part $\rho_{i \rm g} \to 0$, while Eqs.~\eqref{eq:geo_energy_plunge}-\eqref{eq:geo_angular_momentum_plunge} reduce to the constants of motion for homoclinic orbits. 
However, the Jacobian matrix for Eqs.~\eqref{eq:geo_energy_plunge}-\eqref{eq:geo_angular_momentum_plunge} in the limit $(p_{\rm g}, {e_{\rm g}}) \to (p^*_{\rm g}, {e^*_{\rm g}})$ is different from the Jacobian matrix for the constants of motion of homoclinic orbits. In fact, the determinant of the Jacobian matrix of Eqs.~\eqref{eq:geo_energy_plunge}-\eqref{eq:geo_angular_momentum_plunge} with respect to $(p_{\rm g}, {e_{\rm g}})$ is not vanishing at the geodesic separatrix.

\begin{figure}[!bth]
    {\centering
        \includegraphics[width=.47\textwidth]{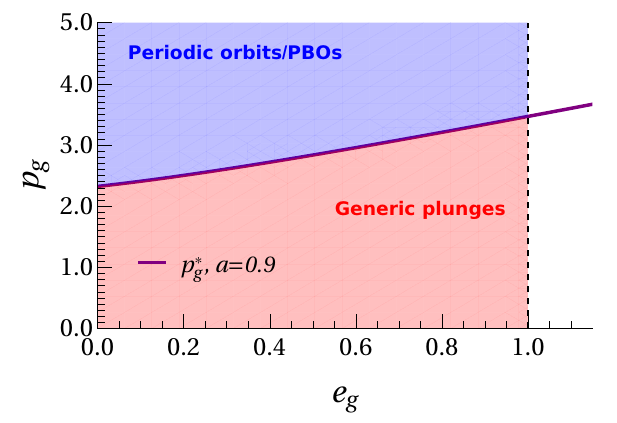}}
        \caption{Parameter space for periodic orbits and PBOs (blue region), which is defined between $0<e_{\rm g}<1, p_{\rm g} > p^*_{\rm g}$ and generic plunges (red region), which is defined for $0<e_{\rm g}<1, p_{\rm g} < p^*_{\rm g}$. The thick, violet line is the separatrix curve $p^*_{\rm g}(e^*_{\rm g})$, while the dashed line $e_{\rm g} =1$ divide the regions of bound and unbound motion. The primary spin is $a=0.9$. 
}       \label{fig:generic_plunge_region}
\end{figure}

\subsubsection{Special plunges}
Finally, a special type of plunge occurs when the condition $L_{z \rm g} = a E_{\rm g}$ is satisfied. In such a case, the radial potential $R_{\rm g}(r_{\rm g})$ simplifies to
\begin{equation}
    R_{\rm g}(r_{\rm g}) = \big[(E^2_{\rm g} -1)r^2_{\rm g} + 2 r_{\rm g} - a^2\big]r^2_{\rm g} \ ,
\end{equation}
whose turning points are the double root $r_{\rm g} = 0$, and
\begin{align}
    r_{1 \rm g} &= \frac{1 + \sqrt{1 - a^2(1 - E^2_{\rm g})}}{1 - E^2_{\rm g}} \ , \\
    r_{2 \rm g} &= \frac{1 - \sqrt{1 - a^2(1 - E^2_{\rm g})}}{1 - E^2_{\rm g}} \ .
\end{align}
The roots $(r_{1 \rm g}, r_{2 \rm g})$ are always real and $r_+ > r_- > r_{2 \rm g}$, since $E^2_{\rm g} < 1$.

\subsection{Equations of motion for nearly equatorial orbits}\label{sec:EoM_near_eq}
We solve the system~\eqref{eq:MPDeq_lin} perturbatively by assuming that the particle's orbits can be written in the form 
\begin{align}    
    x^\mu(\tau) &= x^\mu_{\rm g}(\tau) + \epsilon \chi \delta x^\mu(x^\mu_{\rm g}(\tau), s^\mu(\tau))   \label{eq:trajectory_expansion}  \ .
\end{align}
In particular, we consider orbits $\mathcal O(\epsilon \chi)$ close to the equatorial plane, so that the polar trajectory $z =z (\tau)$ takes the form
\begin{equation}
 z(\tau) = 0 + \epsilon \chi_\parallel \delta z_\parallel(\tau) + \epsilon \chi_\perp \delta z_\perp(\tau) \ , \label{eq:condition_almost_equatorial_orbits}
\end{equation}
with $\delta z_\parallel$ ($\delta z_\perp$) the correction due the projection of the spin vector parallel (orthogonal) to the primary spin. The nearly equatorial motion is then governed by two sets of equations. One  system describes the planar motion constrained on the equatorial plane; a second set describes the precession of the orbital plane induced by the misalignment of the secondary spin with respect to the orbital angular momentum.

\subsubsection{Motion on the equatorial plane}
The first-order differential equations describing the equatorial motion of a spinning particle are~\cite{Piovano:2025aro}
\begin{subequations}\label{eq:1st-order-EoM}
    \begin{empheq}[]{align}
        & \bigg(\totder{r}{\lambda} \bigg)^{\!\!2} = \frac{R_5(r)}{r} \label{eq:R_of_r_velocity}   \, ,\\
        &\totder{t}{\lambda} = T_{\rm g}(r) + \epsilon \chi_\parallel \delta T(r)  \label{eq:T_of_r_velocity} \, ,\\
        &\totder{\phi}{\lambda} = \Phi_{\rm g}(r) + \epsilon \chi_\parallel \delta \Phi(r)  \label{eq:Phi_of_r_velocity} \, , 
    \end{empheq}
\end{subequations} 
where
\begin{align}
    & \delta T(r) = \frac{2}{\Delta}( a E_{\rm g} r - \Lzrd)  -\frac{\Lzrd}{r} + \displaystyle \sum^2_{i=1} \frac{\partial T_{\rm g}}{\partial C_{i \rm g}} \delta C_i  \, ,\\
    & \delta \Phi(r) = \frac{a}{r \Delta}( a E_{\rm g} r  - \Lzrd) - E_{\rm g} + \displaystyle \sum^2_{i=1} \frac{\partial \Phi_{\rm g}}{\partial C_{i \rm g}} \delta C_i   \, , 
\end{align}
with $(C_{1 \rm g}, C_{2 \rm g}) = (E_{\rm g}, L_{z \rm g})$. We denote with  $\delta C_i$ the spin corrections to the constants of motion, with $(\delta C_1, \delta C_2) = (\delta E, \delta L_z)$.
The radial potential $R_5(r)$ is a quintic polynomial 
\begin{equation}
   R_5(r) = \displaystyle \sum^5_{i=0} c_i r^i   \, ,  
\end{equation}
whose coefficients $c_i, \, i = 0, \dots ,5$ are defined as
\begin{align}
    c_5 &= E^2_{\rm g} - 1 + \epsilon \chi_\parallel 2 E_{\rm g} \delta E \ , \\
    c_4 &= 2 \ , \\
    c_3 &= a^2(E^2_{\rm g} - 1) - L^2_{z \rm g} \nonumber \\
    & + \epsilon \chi_\parallel 2 \big(a^2 E_{\rm g} \delta E + L_{z \rm g} (E_{\rm g} - \delta L_z )\big)    \ , \\
    c_2 &= 2 \Lzrd^2 - \epsilon \chi_\parallel 2 \Lzrd \big(3 E_{\rm g} + 2 a \delta E - 2 \delta L_z\big) \ ,   \\
    c_1 &=0 \ ,  \\
    c_0 &=  \epsilon \chi_\parallel 2 a \Lzrd^2 \ .
\end{align}
Analytic solutions of the system~\eqref{eq:1st-order-EoM} are presented in Sec.~\ref{sec:equatorial_orbits}.

\subsubsection{Spin vector and precession phase}
The solution for the spin-vector is given by Eq.~\eqref{eq:spin-vector}, which is valid for any orbit. Thus, we only need to find a solution for the spin-precession phase $\psi_{\rm p}(\lambda)$, which satisfies the following evolution equation in the limit of almost equatorial orbits 
\begin{equation}
    \totder{\psi_p}{\lambda} = \Psi_r(r_{\rm g})  \, , \label{eq:spin-precession-angle}
\end{equation}
where 
\begin{equation}
    \Psi_r(r_{\rm g}) = |\Lzrd| \frac{(r^2_{\rm g}+a^2)E_{\rm g}-a L_{z \rm g}}{\Lzrd^2 + r^2_{\rm g}} + a \sgn(\Lzrd) \ ,
\end{equation}
Analytic expressions for $\psi_p(\lambda)$ along different plunging orbits are presented in Sec.~\ref{sec:polar_motion_spin_prec}. 

\subsubsection{Polar motion}
The polar motion evolves according to the following equation of motion~\cite{Piovano:2024yks} (see also Eq. (5.15) of Ref.~\cite{Drummond:2022neareq})
\begin{equation}
\frac{\dd^2 \delta z}{\dd \lambda^2} + \Upsilon^2_{z\rm g}\delta z = - 3 s_\perp\frac{\cos\psi_{\rm p}}{r^2}L_{z\text{rd}}\sqrt{L^2_{z\text{rd}} + r^2_{\rm g}} \ , \label{eq:polar_EoM}
\end{equation}
with $\Upsilon_{z \rm g} = \sqrt{a^2(1 - E^2_{\rm g}) + L^2_{z \rm g}}$ the frequency of harmonic oscillations near $z_{\rm g} =0$. Sec.~\ref{sec:polar_motion_spin_prec} presents the general solutions for Eq.~\eqref{eq:polar_EoM}.

\section{Equatorial motion} \label{sec:equatorial_orbits}
\subsection{Roots of the radial potential}\label{sec:roots_radial_potential}
Before solving the system~\eqref{eq:1st-order-EoM}, it is necessary to find the roots of the quintic polynomial $R_5(r)$. There are two separate cases to consider:  $\Lzrd \neq 0$ and $\Lzrd =0$. In this Section, we only consider orbits that satisfy the former constraint. The condition $\Lzrd =0$ leads to a drastic simplification of the radial potential $R_5(r)$, and it is only satisfied for special plunging orbits, which are discussed in Section~\ref{sec:special_plunge}.

When $\Lzrd \neq 0$, at least one of the $R_5(r)$ root is real, which we denote as $r_1$. The remaining  four roots can be all real, or form either one pair or two pairs of complex conjugate roots. Regardless, we expect that three of the roots of $R_5(r)$ are $\mathcal O (\epsilon \chi_\parallel)$ close to the geodesic roots, with
\begin{equation}
     r_1 = r_{1 \rm g} + \epsilon \chi_\parallel \delta r_1    
\end{equation}
while
\begin{align}
    r_2 &= r_{2 \rm g} + \epsilon \chi_\parallel \delta r_2 \ ,\\
    r_3 &= r_{3 \rm g} + \epsilon \chi_\parallel \delta r_3 \ ,
\end{align}
when $\{r_{2 \rm g}, r_{3 \rm g} \} \in \mathbb{R}$, and
\begin{align}
    r_2 &= \rho_{r\rm g} + i\rho_{i\rm g} + \epsilon \chi_\parallel (\delta \rho_r + i \delta \rho_i) \ ,\\
    r_3 & = \overline{r_2}
\end{align}
when $\{r_{2 \rm g},r_{3 \rm g}\} \in \mathbb{C}$, with overline denoting complex conjugation. 

It is always possible to impose a constraint between $\delta r_1$, and the shifts to the constants of motion as follows. First, we expand the right hand side of Eq.~\eqref{eq:R_of_r_velocity} using Eq.~\eqref{eq:trajectory_expansion}
\begin{equation}
\frac{R_5(r)}{r} = R_{\rm g}(r_{\rm g}) + \epsilon \chi_\parallel \big[ \delta R(r_{\rm g}) +  R'_{\rm g}(r_{\rm g}) \delta r(r_{\rm g}) \big]  \, ,   
\end{equation}
where the prime denotes derivation with respect to $r_{\rm g}$, while the functions $\delta R(r_{\rm g})$ is defined as
\begin{align}
&  \delta R(r_{\rm g}) = R_s(r_{\rm g}) + \displaystyle \sum^2_{i =1} \frac{\partial R_{\rm g}}{\partial C_{i \rm g}} \delta C_i  \ , \\
&   R_s(r_{\rm g}) = \frac{2 a \Lzrd^2}{r_{\rm g}} - 6 E_{\rm g} \Lzrd r_{\rm g} + 2 E_{\rm g} L_{z \rm g} r^2_{\rm g} \ .
\end{align}
where the subscript ``s" stands for spin. When convenient, we use a subscript or superscript ``s" to denote specific spin corrections. Then, the shift to the turning points $\delta r_1 = \delta r(r_{1\rm g})$ satisfies
\begin{equation}
    \delta R(r_{1\rm g}) = - \delta r_1 R'_{\rm g}(r_{1\rm g})  \ ,   \label{eq:constraint_delta r_1}     
\end{equation} 
which provides a constraint between $\delta r_1$, $\delta E$ and $\delta L_z$. When $(r_{2 \rm g},r_{3 \rm g}) \in \mathbb{R}$, an additional constraint can be imposed for the shift to the turning point $\delta r_2 = \delta r(r_{2\rm g})$, leading to the following system of equations
\begin{align} \label{eq:constraint_shifts_turning_points_constants_of_motion}
    \delta R(r_{1\rm g}) &= - \delta r_1 R'_{\rm g}(r_{1\rm g})  \ ,  \\
    \delta R(r_{2\rm g}) &= - \delta r_2 R'_{\rm g}(r_{2\rm g})  \ .
\end{align} 
The remaining roots  $r_4, r_5$ can be found using the following Vieta's formulas
\begin{subequations}
\label{eq:Vieta_formula}
    \begin{empheq}{align}
    -\frac{c_0}{c_5} &= r_1 r_2 r_3 r_4 r_5 \ ,\\
    -\frac{c_4}{c_5} &= r_1 + r_2 + r_3 + r_4 + r_5  \ ,\\
    0 &= (r_1 r_2 r_3 r_4) + (r_1 r_2 r_3 r_5)  \nonumber \\
    & + (r_1 r_3 r_4 r_5) + (r_1 r_2 r_4 r_5)  \nonumber  \\
    & + (r_2r_3 r_4 r_5) \ , \\
    -\frac{c_2}{c_5} &= (r_1 r_2 r_3) + (r_1 r_2 r_4) + (r_1 r_2 r_5) \nonumber \\
    & + (r_1 r_3 r_4) + (r_1 r_3 r_5) + (r_1 r_4 r_5)  \nonumber \\
    & + (r_2 r_3 r_4) + (r_2 r_3 r_5) + (r_2 r_4 r_5)  \nonumber \\
    & + (r_3 r_4 r_5) \ . 
    \end{empheq}
\end{subequations}
After expanding both sides of Eqs.~\eqref{eq:Vieta_formula} in $\epsilon \chi_\parallel$, and equating term by term, the roots $r_4$ and $r_5$ can be written as
\begin{align}
    \delta r_4 &= -\frac{a}{4 \Lzrd^2}\big(a^2(1 - E^2_{\rm g})+ L^2_{z \rm g} \big) \ , \\
    r_4 &= \sqrt{-\epsilon \chi_\parallel} \sqrt{a} + \epsilon \chi_\parallel \delta r_4 + \mathcal O(\epsilon^{3/2})  \, \\
    r_5 &= -\sqrt{-\epsilon \chi_\parallel} \sqrt{a} + \epsilon \chi_\parallel \delta r_4 + \mathcal O(\epsilon^{3/2}) \ .
\end{align}
$r_4$ and $r_5$ form a pair of complex conjugate roots when $\chi_\parallel >0$. Remarkably, the structure of $r_4$ and $r_5$ is the same regardless of the type of nearly equatorial orbits with $\Lzrd \neq 0$. Moreover, $r_4$ and $r_5$ are independent of the shifts to the constants of motion $\delta E$ and $\delta L_z$.

Additionally, when $(r_{2 \rm g},r_{3 \rm g}) \in \mathbb{R}$, Eqs.~\eqref{eq:Vieta_formula} provide the following constraint between $\delta r_3$ and the remaining roots
\begin{equation}
     \delta r_3 =  \frac{4 E_{\rm g} \delta E}{(1 - E^2_{\rm g})^2} - \delta r_1 - \delta r_2 - 2\delta r_4 \ ,  
\end{equation}

The shifts to the roots $\delta r_1, \delta r_2, \delta r_3$ and constants of motion $\delta E, \delta L_z$ are not unique. Instead, they depend on the fiducial geodesics chosen as reference to calculate the spin-corrections. As a result, the solutions of the linearized MPD equations admit a family of equivalent solutions, which are related to each other by differentiable maps. This intrinsic freedom in the phase space is called spin-gauge~\cite{Piovano:2024yks,Mathews:2025nyb}. We stress that the spin-gauges are different from spin supplementary conditions (SSCs). The latter encodes the freedom in choosing referential frames in which to calculate the stress-energy tensor multipoles~\cite{Semerak:1999qc,Kyrian:2007zz}. SSCs are needed to close both the full MPD equations~\eqref{eq:MPDeq}, and their truncated counterparts. By contrast, the concept of spin gauge only emerge when solving the MPD equations expanded at order $\mathcal O(\epsilon \chi)$~\eqref{eq:MPDeq_lin} or higher. For instance, all spin-gauges discussed in Refs.~\cite{Piovano:2024yks,Mathews:2025nyb} referred to the same Tulczyjew-Dixon SSC\footnote{It is also worth mentioning that several SSCs are identical at linear order in order in the spin, such as the Mathisson-Pirani, Tulczyjew-Dixon, and Ohashi-Kyrian-Semer\'{a}k conditions~\cite{Timogiannis:2022bks}.}.

In this work, we adopt the fixed eccentricity spin gauge~\cite{Piovano:2025aro}, which we refer to as FE spin gauge. This parametrization ensures that the shifts to periodic orbits remain finite at the geodesic separatrix, and continuously reduce to homoclinic orbits in the limit $(p_{\rm g}, e_{\rm g}) \to (p^*_{\rm g}, e^*_{\rm g})$. All other spins gauge introduce divergences of order $1/(r_{2 \rm g} - r_{3 \rm g})$ for $r_{3 \rm g} \to r_{2 \rm g}$ in the corrections to periodic orbits and frequencies~\cite{Piovano:2025aro}. 

The definition of the FE spin-gauge depends on the nature of the roots of the geodesic radial potential. In the following, we distinguish between two cases: all roots of $R_{\rm g}(r_{\rm g})$ are real, and $R_{\rm g}(r_{\rm g})$ admits complex roots.

\subsubsection{$R_{\rm g}(r_{\rm g})$ has only real roots}
The FE spin gauge is defined by the following constraints on the spin corrections to the roots
\begin{align}
    \delta r_1 = \frac{\delta p}{1 - e_{\rm g}}  \ , \qquad  \delta r_2 = \frac{\delta p}{1 + e_{\rm g}}  \ ,  \qquad \delta r_3 = \delta r_2 \ ,\label{eq:shift_turning_points_FE} 
\end{align}
while the shifts to the constants of motion are defined by
\begin{align} 
    \delta E &= \delta E^\text{s} + \frac{\partial E_{\rm g}}{\partial p} \delta p  \ , \label{eq:shift_energy_FE} \\
    \delta L_z &= \delta L^\text{s}_z + \frac{\partial L_{z\rm g}}{\partial p} \delta p  \ .  \label{eq:shift_angular_momentum_FE}
\end{align}
The quantities $\delta E^\text{s}$ and $\delta L^\text{s}_z$ are the solutions of the system
\begin{equation}
     \delta R(r_{1\rm g}) =0 \ , \qquad      \delta R(r_{2\rm g}) =0  \ . 
\end{equation} 
Explicit formulas for $\delta E^\text{s}$ and $\delta L^\text{s}_z$ can be found in Appendix~\ref{app:spin_constants_of_motion_corr}. The constraint~\eqref{eq:shift_turning_points_FE} and Eqs.~\eqref{eq:Vieta_formula} lead to the following expression for $\delta p$
\begin{equation}
    \delta p = \frac{2(1 - e^2_{\rm g})\big(2 E_{\rm g} \delta E^\text{s} - \big(1 - E^2_{\rm g}\big)^2 \delta r_4\big)}{4 E_{\rm g} \frac{\partial E_{\rm g}}{\partial p} \big(e^2_{\rm g} -1 \big) + (3 - e_{\rm g}) \big(1 - E^2_{\rm g}\big)^2}  \ . \label{eq:semi_latus_rectum_shift}
\end{equation}
Eq.~\eqref{eq:semi_latus_rectum_shift} can be considered as a function of the three orbital elements $(a,p_{\rm g},e_{\rm g})$. As a consequence of Eqs.~\eqref{eq:shift_turning_points_FE}, the eccentricity $e_{\rm g}$ is the same for both spinning and non-spinning particles. We stress that all previous expressions apply to periodic and homoclinic motion as well. 

\begin{figure}[!bth]
    {\centering
        \includegraphics[width=.47\textwidth]{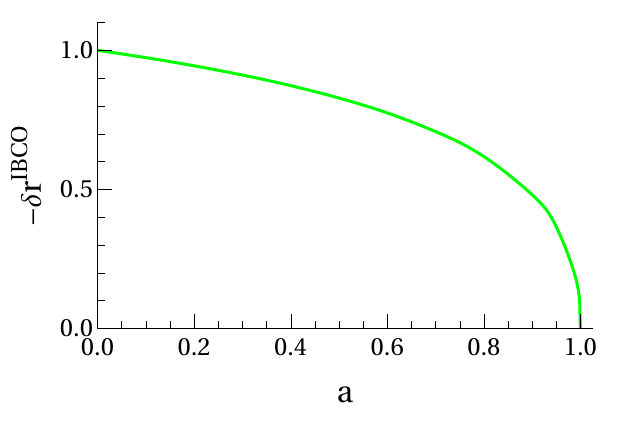}}
        \caption{Spin correction to the IBCO radius $ \delta r_\text{ibco}$ for different primary spins. 
}       \label{fig:correction_IBCO}
\end{figure}

As shown in Ref.~\cite{Piovano:2025aro}, Eq.~\eqref{eq:semi_latus_rectum_shift} gives the spin-corrections to the location of the separatrix (to the ISCO radius) for $e_{\rm g} = e^*_{\rm g}$ ($e^*_{\rm g} = 0$). We denote the shift to the separatrix location and ISCO radius as $\delta p^* \equiv \delta p(a,p^*_{\rm g},e^*_{\rm g}) $ and $\delta r_\text{isco} \equiv \delta p(a,p^*_{\rm g},0)$, respectively. Unlike other classes of orbits, the spin-corrections to ISCOs are spin-gauge independent. In fact, the condition that the radial potential has a inflection point (or equivalently, a triple root) up to  $\mathcal O(\epsilon \chi)$ order uniquely fixes the shifts to the constants of motion, which were obtained in Ref.~\cite{Jefremov:2015gza}. Likewise, the corrections to the IBCO do not depend on the spin-gauge. The shift to the IBCO radius can be obtained from Eq.~\eqref{eq:semi_latus_rectum_shift} as
\begin{equation}
    \delta r_\text{ibco} = \frac{1}{2} \delta p(a,p^*_{\rm g},1) 
\end{equation}
with $e^*_{\rm g} = 1$, or equivalently, $E_{\rm g} = 1$. Notice that $\delta E = 0$ for $e^*_{\rm g} = 1$. The shift $\delta r_\text{ibco}$ represents the spin correction to the location of the separatrix dividing bound and unbound motion. The closed-form expression for $\delta r_\text{ibco}$ is given in Appendix~\ref{app:spin_delta_p_corr}, and it is plotted in Fig.~\ref{fig:correction_IBCO} as a function of the primary spin. 

Finally, $\delta p^*$ simplifies considerably in Schwarzschild spacetime, and reduces to
\begin{equation}
    \delta p^*(0, e^*_{\rm g}) = - 2 \sqrt{\frac{2(1 + e^*_{\rm g})}{3 + e^*_{\rm g}}} \ ,
\end{equation}
which agrees with the expression found in Ref.~\cite{Witzany:2024ttz}.

\subsubsection{Generic plunges}
The FE spin gauge~\eqref{eq:shift_turning_points_FE} for generic plunges is defined by the following equations
\begin{align}
    \delta r_1 = \frac{\delta p^\gen}{1 - e_{\rm g}}  \ , \qquad  \delta \rho_r = \frac{\delta p^\gen}{1 + e_{\rm g}}  \ ,  \qquad \delta \rho_i = 0 \ , \label{eq:shift_turning_points_FE_cc} 
\end{align}
where $\delta p^\gen =\delta p^\gen(a,p_{\rm g},e_{\rm g})$ is obtained from the Eqs.~\eqref{eq:Vieta_formula} and the constraints~\eqref{eq:shift_turning_points_FE_cc}. The explicit expression for $\delta p^\gen$ is given in Appendix~\ref{app:spin_delta_p_corr}. The shifts to the constants of motion are defined as
\begin{align} 
    \delta E^\gen &= \delta E^{\text{s},\gen} + \frac{\partial E_{\rm g}}{\partial p} \delta p^\gen  \ , \label{eq:shift_energy_FE_cc} \\
    \delta L^\gen_z &= \delta L^{\text{s},\gen}_z + \frac{\partial L_{z\rm g}}{\partial p} \delta p^\gen  \ ,  \label{eq:shift_angular_momentum_FE_cc}
\end{align}
where $\gen$ denotes a generic plunge. $\delta L^\text{s,gen}_z$ is given as a solution of the following equation
\begin{equation}
     \delta R(r_{1\rm g}) =0 \ ,  
\end{equation} 
while $\delta E^\text{s,gen}$ can be obtained from Vieta's formula~\eqref{eq:Vieta_formula}. Expressions for $\delta E^\text{s,gen}$ and $\delta L^\text{s,gen}_z$ are presented in Appendix~\ref{app:spin_constants_of_motion_corr}.
The shifts to the roots and constants of motion for generic plunges are defined for $p_{\rm g} \leq p^*_{\rm g}$ and $e_{\rm g} \leq e^*_{\rm g}$. In particular,
\begin{align}
    &\delta p^\gen(a,p^*_{\rm g},e^*_{\rm g}) = \delta p^* \  ,  
\end{align}
while $(\delta E^\gen, \delta L^\gen_z) = (\delta E, \delta L_z)$ for $(p_{\rm g},e_{\rm g}) = (p^*_{\rm g},e^*_{\rm g})$. In other words, when $\rho_{i \rm g} = 0$, Eqs.~\eqref{eq:shift_energy_FE_cc}-\eqref{eq:shift_angular_momentum_FE_cc} reduce to the corrections of the constants of motion~\eqref{eq:shift_energy_FE}-\eqref{eq:shift_angular_momentum_FE} for homoclinic motion and critical plunges.
As a result, the spin-corrections for generic plunges are continuous in the limit $\rho_{i \rm g} \to 0$ (as occurs at geodesic order), and reduce the spin-corrections to homoclinic motion.

\subsection{Solutions by quadrature}\label{sec:quad_sol}
As shown in Ref.~\cite{Piovano:2025aro}, the linear-in-spin corrections to the 4-velocities can be written as
\begin{subequations}\label{eq:1st-order-EoM-lambda-fully-linearize}
    \begin{empheq}[]{align}
        \totder{\delta r}{\lambda} & = \pm \sqrt{R_{\rm g}(r_{\rm g})}\delta V_r(r_{\rm g}) \pm  \frac{\delta r (r_{\rm g})}{2\sqrt{R_{\rm g}(r_{\rm g})}} \totder{R_{\rm g}}{r_{\rm g}}  \, , \\
        \totder{\delta t}{\lambda} &= \delta T(r_{\rm g}) + \totder{T_{\rm g}(r_{\rm g})}{r_{\rm g}} \delta r(r_{\rm g}) \, , \\
        \totder{\delta \phi}{\lambda} &= \delta \Phi(r_{\rm g}) + \totder{\Phi_{\rm g}(r_{\rm g})}{r_{\rm g}} \delta r(r_{\rm g})  \, ,
  \end{empheq}
\end{subequations} 
where the upper (lower) sign corresponds to positive (negative) geodesic radial velocity. In general, the function $\delta V_r(r_{\rm g})$ is defined as
\begin{align}
  \delta V_r(r_{\rm g}) &= -\frac{1}{2}\bigg( \displaystyle \sum^3_{i =1} \frac{\delta r_i}{r_{\rm g} - r_{i \rm g}} + \frac{2\delta r_4}{r_{\rm g}} - \frac{a}{r^2_{\rm g}} + \frac{2 E_{\rm g} \delta E}{1 - E^2_{\rm g}} \bigg)  \, .  \label{eq:diff_radial_potential_general} 
\end{align}
except for generic and special plunges. In these cases, the function $\delta V_r(r_{\rm g})$ will be specified later on.
By changing variable from $\lambda$ to $r_{\rm g}$, the system~\eqref{eq:1st-order-EoM-lambda-fully-linearize} becomes
\begin{subequations}\label{eq:1st-order-EoM-r-fully-linearize}
   \begin{empheq}[]{align}
    &\totder{\delta r}{r_{\rm g}} = \frac{1}{2 R_{\rm g}(r_{\rm g})} \totder{R_{\rm g}}{r_{\rm g}}\delta r (r_{\rm g}) +\delta V_r(r_{\rm g}) \, ,  \label{eq:lin-corr-radial-motion}  \\  
	&\totder{\delta t}{r_{\rm g}} = \pm \frac{\delta T(r_{\rm g})}{\sqrt{R_{\rm g}(r_{\rm g})}} \pm \totder{T_{\rm g}(r_{\rm g})}{r_{\rm g}} \frac{\delta r(r_{\rm g})}{\sqrt{R_{\rm g}(r_{\rm g})}}  \, , \\ 
	&\totder{\delta \phi}{r_{\rm g}} = \pm \frac{\delta \Phi(r_{\rm g})}{\sqrt{R_{\rm g}(r_{\rm g})}} \pm \totder{\Phi_{\rm g}(r_{\rm g})}{r_{\rm g}}\frac{\delta r(r_{\rm g})}{\sqrt{R_{\rm g}(r_{\rm g})}}  \, .
   \end{empheq}
\end{subequations} 
while the solutions of the previous system are the following equations~\cite{Piovano:2025aro}
\begin{equation}
    \delta r(r_{\rm g}) = \sqrt{R_{\rm g}(r_{\rm g})} \int \! \frac{\big( \delta V_r(r') + \delta \hat \Upsilon_r \big) \dd r'}{\sqrt{R_{\rm g}(r')}}   \, \ , \label{eq:generic_solution_radial_shift}     
\end{equation}

\begin{align}
    \delta t(r_{\rm g}) &= \mp T_{\rm g}(r_{\rm g})\frac{\delta r(r_{\rm g})}{\sqrt{R_{\rm g}(r_{\rm g})}} \pm \int \frac{\delta T(r')\dd r'}{\sqrt{R_{\rm g}(r')}}  \nonumber \\ 
    &\mp \int \frac{T_{\rm g}(r') \delta V_r(r')}{\sqrt{R_{\rm g}(r')}} \dd r'  \,  \label{eq:generic_solution_time_shift}  \\
    \delta \phi(r_{\rm g}) &= \mp \Phi_{\rm g}(r_{\rm g})\frac{\delta r(r_{\rm g})}{\sqrt{R_{\rm g}(r_{\rm g})}} \pm \int \frac{\delta \Phi(r')\dd r'}{\sqrt{R_{\rm g}(r')}}  \nonumber \\ 
    & \mp \int \frac{\Phi_{\rm g}(r') \delta V_r(r')}{\sqrt{R_{\rm g}(r')}} \dd r'  \, . \label{eq:generic_solution_azimuthal_shift} 
\end{align} 
which are valid for any nearly-equatorial orbits. The constant $\delta\hat\Upsilon_r$ depends on the class of orbits and spin-gauge: $\delta\hat\Upsilon_r = \delta \Upsilon_r/\Upsilon_{r\rm g}$ for periodic orbits, with $\Upsilon_{r\rm g}$ the Mino-time radial frequency and $\delta \Upsilon_r$ its corresponding shift , which depend on the chosen spin-gauge. The term $\delta \Upsilon_r/\Upsilon_{r\rm g}$ is introduced to remove secular growing corrections from periodic orbits~\cite{Drummond:2022neareq,Drummond:2022efc}. It was shown in~\cite{Piovano:2025aro} that spin-corrections to periodic orbits and frequencies are divergent in the limit $r_{3 \rm g} \to r_{2 \rm g}$ in all spin-gauges, with the exception of the fixed eccentricity spin-gauge. Such singularities are nonphysical, and represent a by-product of the complete linearization in the secondary spin of the solutions to Eq.~\eqref{eq:R_of_r_velocity}. We stress that the analytic solutions for periodic geodesic are not divergent at the separatrix, but instead continuously reduce to the homoclinic geodesic. Thus, we expect that the shift to periodic orbits remain finite as well at the separatrix. 

The shifts to the frequencies are finite at the geodesic separatrix only in the FE spin gauge. Furthermore, this parametrization ensures that the shift to the separatrix location is well defined and agrees with previous results in Schwarzschild spacetime~\cite{Witzany:2023bmq} and for the correction to the ISCO~\cite{Jefremov:2015gza}. Thus, the solutions to the spin-corrections to bound plunges can be consistently defined only in the FE spin-gauge. From now on, we add a label to $\delta\hat\Upsilon^{\rm y}_r$ to identify this constant for each class of plunging orbits. Specifically,
\begin{align}
    \delta\hat\Upsilon^\per_r &= \frac{\delta \Upsilon^\FE_r}{\Upsilon_{r\rm g}} \ ,  \\
    \delta\hat\Upsilon^\homo_r &= \frac{a}{2r^2_{2\rm g}} - \frac{E_{\rm g} \delta E}{1 - E^2_{\rm g}} - \frac{\delta r_4}{r_{2 \rm g}} + \frac{\delta r_1}{2(r_{1 \rm g} - r_{2 \rm g})}  \nonumber \\
    &+ \frac{\big(r_{1 \rm g} - 2 r_{2 \rm g}\big)\delta r_2}{2(r_{1 \rm g} - r_{2 \rm g})r_{2 \rm g}} \ ,  \label{eq:shift_radial_frequency}\\
    \delta\hat\Upsilon^\gen_r &= \frac{A+B}{2 A r_{1 \rm g}} \delta r_1 - \frac{(A+B)}{B r_{1 \rm g}} \delta r_4 - \frac{E_{\rm g} \delta E^\gen}{1 - E^2_{\rm g}}  \nonumber \\
    &+ \frac{(A - B) \delta \rho_r}{B(r_{1 \rm g} - r_{2 \rm g}) + A r_{2 \rm g}} - \frac{a}{6 B^2} + \frac{a(A+B)}{3 B r^2_{1 \rm g}} \nonumber \\
    &+ a \frac{(2 A + B)\rho_{r \rm g}}{3 B^3 r_{1 \rm g}}  \ , \\
    \delta\hat\Upsilon^\spec_r &= 0 \ ,
\end{align}
where the constants $A$ and $B$ are given in Appendix~\ref{app:integrals_generic_plunge}, while the frequency shift $\delta \Upsilon^\FE_r$ is presented in Appendix E of~\cite{Piovano:2025aro}. The constant $\delta\hat\Upsilon^\per_r$ applies to both periodic orbits and PBOs, $\delta\hat\Upsilon^\homo_r$ for homoclinic and critical plunges, $\delta\hat\Upsilon^\gen_r$ is valid for generic plunges, and $\delta\hat\Upsilon^\spec_r$ for special plunges. The terms $\delta\hat\Upsilon^{\rm y}_r$ satisfies the following limits
\begin{align}
    & \lim_{r_{3 \rm g} \to r_{2 \rm g}} \delta\hat\Upsilon^\per_r = \delta\hat\Upsilon^\homo_r \ , \\
    & \lim_{\rho_{i \rm g} \to 0} \delta\hat\Upsilon^\gen_r = \delta\hat\Upsilon^\homo_r \ , 
\end{align}

Eqs.~\eqref{eq:1st-order-EoM-r-fully-linearize} and their solutions~\eqref{eq:generic_solution_radial_shift}-\eqref{eq:generic_solution_azimuthal_shift} are not define for certain radii. The geodesic trajectories  $t(r_{\rm g})$ and $\phi(r_{\rm g})$ and their spin corrections  $\delta t(r_{\rm g})$ and $\delta \phi(r_{\rm g})$ are divergent at $r_{\rm g} = r_\pm$. Such a singularity is simply due to the ill-behavior of the Boyer-Lindquist coordinates at the black hole horizons. Thus, $r_{\rm g} = r_\pm$ represent purely coordinate singularities. However,  $\delta t(r_{\rm g})$ and $\delta \phi(r_{\rm g})$ diverge as poles at $r_{\rm g} = r_\pm$, while $t(r_{\rm g})$ and $\phi(r_{\rm g})$ have only logarithmic divergences. The stronger divergence at $r_{\rm g} = r_\pm$ for $\delta t(r_{\rm g})$ and $\delta \phi(r_{\rm g})$ is an unfortunate consequence of any perturbative solutions of the MPD equations. Indeed, the spin corrections described in Eqs.~S12-S14 of Ref.~\cite{Skoupy:analytic-solutions} are also divergent as poles at $r_{\rm g} = r_\pm$.These poles cannot be removed by a smooth change of coordinates, therefore posing a problem in the modeling of Inspiral-Merger-Ringdown (IMR) waveforms (see Refs.~\cite{Kuchler:2025hwx,DellaRocca:2025zbe}). We discuss this issue in more detail in Sec.~\ref{sec:partial_expansion_plunge_solutions} and propose a convenient workaround.

Notice that the right hand side of Eqs.~\eqref{eq:1st-order-EoM-r-fully-linearize} is formally not defined also at the geodesic turning points $(r_{1\rm g}, r_{2\rm g}, r_{3\rm g})$. These singularities can be removed if the turning points are simple roots for the radial potential. In such a case, the limits $r_{\rm g} \to r_{i \rm g}$ for $i = 1, 2, 3$ exist and are finite for the system~\eqref{eq:1st-order-EoM-r-fully-linearize}. Thus, the shifts to the velocities~\eqref{eq:1st-order-EoM-lambda-fully-linearize} and trajectories~\eqref{eq:generic_solution_radial_shift}-~\eqref{eq:generic_solution_azimuthal_shift} become differentiable at the geodesic simple roots if they are suitably defined as piecewise functions~\footnote{As an example, the function $\text{sinc}(x) =\sin(x)/x$ is formally not defined for $x = 0$. However, the function becomes analytic if it is defined piecewise as $\text{sinc}(x) =\sin(x)/x$ for $x \neq 0$ and $\text{sinc}(0)=1$ for $ x = 0$.}. 

The situation is different when one of the turning points $(r_{1\rm g}, r_{2\rm g}, r_{3\rm g})$ is a repeated root of the radial potential, as in the case of critical and ISCO plunges. $R_{\rm g}(r_{\rm g})$ has a double root $r_{2 \rm g} = r_{3 \rm g}$ for critical plunges, and a triple root $r_{1 \rm g} = r_{2 \rm g} = r_{3 \rm g}$  for ISCO plunges. In both cases, the shifts to the radial velocities and radial trajectories can still be made differentiable at the repeated geodesic roots (see the left panel in Figs.~\ref{fig:shift_critical_plunge}-\ref{fig:shift_isco_plunge}). By contrast, the geodesic trajectories  $t(r_{\rm g})$ and $\phi(r_{\rm g})$ and their spin corrections  $\delta t(r_{\rm g})$ and $\delta \phi(r_{\rm g})$ diverge logarithmically at $r_{2 \rm g} = r_{3 \rm g}$ for a critical plunge, and as a pole $1/(r_{\rm g} - r_{\text{ISCO,g}})$ for an ISCO plunge.

Lastly, the solutions~\eqref{eq:generic_solution_radial_shift}-\eqref{eq:generic_solution_azimuthal_shift} are divergent for $r_{\rm g} =0$, i.e. at the ring singularity. Such a behavior is expected~\footnote{Notice that the the spin-corrections in Eqs.~S12-S14 of Ref.~\cite{Skoupy:analytic-solutions} diverge as well at $r_{\rm g} =0$.}, as the radial velocity~\eqref{eq:R_of_r_velocity} is also divergent for $r =0$, with the exception of the special plunges. Moreover, the solutions~\eqref{eq:generic_solution_time_shift}-\eqref{eq:generic_solution_azimuthal_shift} are finite at $r_{\rm g} =0$ in Schwarzschild spacetime.

Interestingly, the spinning particle can reach the ring singularity only when $\chi_\parallel >0$, since in this case $r_4$ and $r_5$ are complex conjugate roots. For $\chi<0$, the plunge stops instead at the turning point $r_4$ inside the horizon.

Before discussing the analytic solutions for plunging orbits, we present the general expression for the shift to the angular frequency $\delta \Omega_\phi$, which is given by
\begin{align}
    \delta \Omega_\phi(r_{\rm g}) &= \frac{\delta \Phi(r_{\rm g}) +  \Phi'_{\rm g}(r_{\rm g}) \delta r(r_{\rm g})}{T_{\rm g}(r_{\rm g})} \nonumber \\
    &-  \frac{\delta T(r_{\rm g}) +  T'_{\rm g}(r_{\rm g}) \delta r(r_{\rm g})}{T_{\rm g}(r_{\rm g})} \Omega_{\phi \rm g}(r_{\rm g}) \ , \\
    \Omega_{\phi \rm g}(r_{\rm g}) &= \frac{\Phi_{\rm g}(r_{\rm g})}{T_{\rm g}(r_{\rm g})} = \frac{L_{z \rm g} r_{\rm g} - 2 \Lzrd }{E_{\rm g} r^3_{\rm g} + a^2 E_{\rm g} r_{\rm g} - 2 a \Lzrd}  \ ,
\end{align}
After some algebraic manipulation, the frequency shift  $\delta \Omega_\phi$ simplifies to 
\begin{widetext}
\begin{align}
    \delta \Omega_\phi(r_{\rm g}) &= \frac{(\Lzrd^2 - E^2_{\rm g} r^3_{\rm g}) \Delta -  (L_{z \rm g} \delta E - E_{\rm g} \delta L_z) r^3_{\rm g} \Delta - 2 ( E_{\rm g} L_{z \rm g} r^3_{\rm g} - 3 E_{\rm g} \Lzrd r^2_{\rm g} +  a \Lzrd^2) r_{\rm g} \delta r(r_{\rm g})}{r_{\rm g}(E_{\rm g} r^3_{\rm g} + a^2 E_{\rm g} r_{\rm g} - 2 a \Lzrd)^2} \ . \label{eq:azimuthal_frequency_shift}
\end{align}    
\end{widetext}
Notice that Eq.~\eqref{eq:azimuthal_frequency_shift} is valid for any plunging orbits.

In the following, we present the solutions of Eqs.~\eqref{eq:1st-order-EoM-r-fully-linearize} for the different bound plunges. Numerical checks for all plunging orbits are presented in the GitHub repo~\cite{repoAnalyticHJproject}.

\subsection{Critical plunges}\label{sec:critical_plunge}
For critical plunges, we do not specify the initial radius $r_{\rm g}(0)$ at $\lambda =0$.  A common choice for geodesic plunging orbits is $r_{\rm g}(0)=0$, but the shifts to the orbits are divergent at this location. As such, there is not another particularly convenient radius to choose as initial condition. For this reason, all spin-corrections for the critical plunges include an integration constant to fix.

\begin{figure*}[!bth]
    {\centering
        \includegraphics[width=\textwidth]{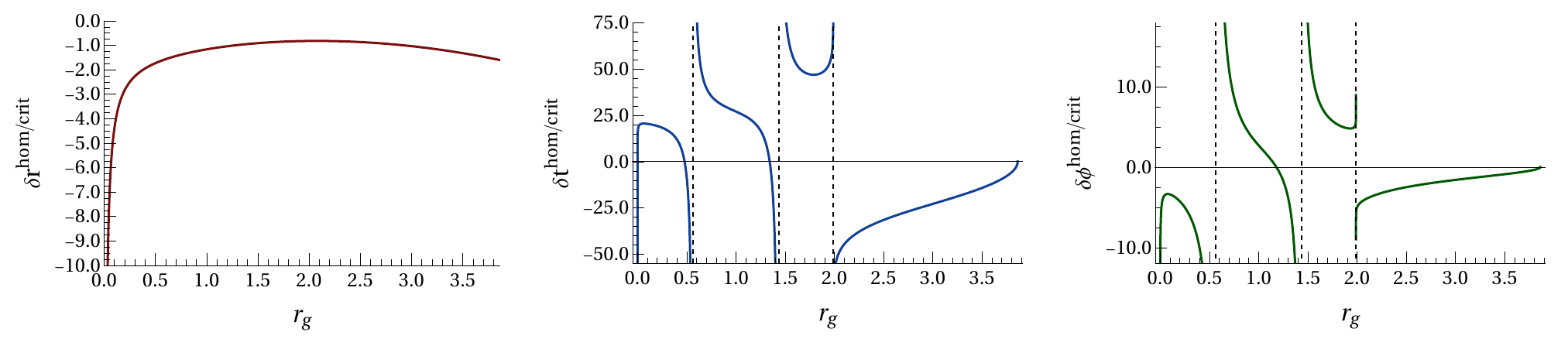}}
        \caption{Spin corrections to radial (left panel), coordinate time (middle panel) and azimuthal trajectories (right panel) for a prograde homoclinic orbit and prograde critical plunge for a spinning binary with $a =0.9$, $\chi_\parallel = 1/2$, $\chi_\perp =\sqrt{3}/2$. The underlying fiducial geodesic has eccentricity $e_{\rm g} =0.32$. In the middle and right panel, the dashed lines correspond, from left to right, to the event horizons $r_-$ and $r_+$, and the geodesic double root $r_{2\rm g}$. In all three panels, the shift to homoclinic orbits  starts at $r_{1 \rm g}$ and ends at $r_{2 \rm g}$. 
}       \label{fig:shift_critical_plunge}
\end{figure*}

\begin{figure}[!bth]
    {\centering
        \includegraphics[width=0.43\textwidth]{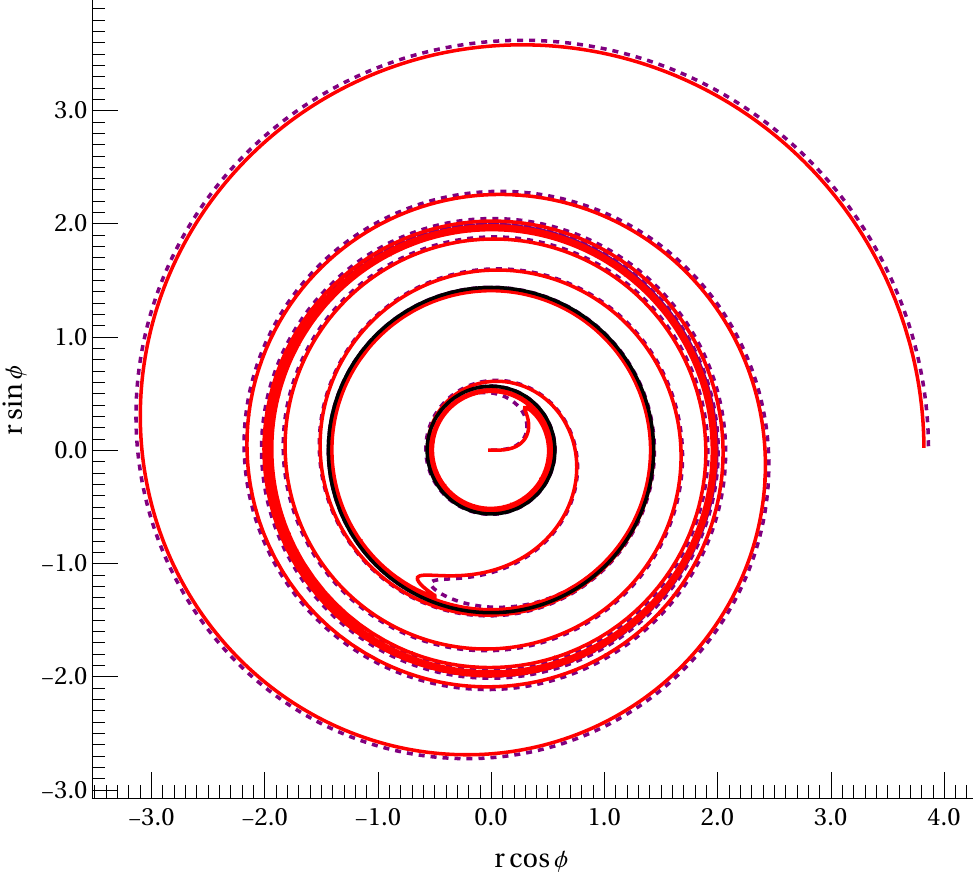} \\[0.4em]
        \includegraphics[width=0.43\textwidth]{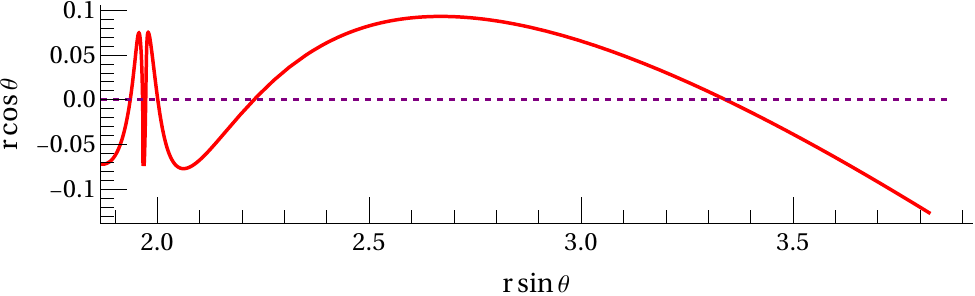}
    }
        \caption{Homoclinic orbits and critical plunges for a spinning particle (red-line) and its underlying fiducial geodesic (purple, dashed line).  The orbital parameters and the linear-in-spin corrections are the same of Fig.~\eqref{fig:shift_critical_plunge}, while the mass-ratio is fixed to $\epsilon = 1/20$. The two black rings correspond to the primary outer horizon $r_+ = 1+\sqrt{1 -a^2}$ and inner horizon $r_- = 1 - \sqrt{1 -a^2}$. Top panel: projection of the orbits onto the equatorial plane. Bottom panel: orthogonal projection of the orbits onto a co-rotating polar plane.
}       \label{fig:critical_plunge}
\end{figure}

The solution for the orbital radius shift $\delta r$ is given by
\begin{align} 
        \delta r(r_{\rm g}) &= - \delta r^\homo(r_{\rm g}) \nonumber \\
        &= - \sqrt{R_g(r_{\rm g})}\bigg[ \bigg(\frac{E_{\rm g}\delta E}{1 - E^2_{\rm g}} + \delta\hat\Upsilon^\homo \bigg) I^\crit(r_{\rm g})  \nonumber \\
        & + I^{\crit}_{1/(r - r_1)}(r_{\rm g})\frac{\delta r_1}{2} + 2I^{\crit}_{1/(r - r_2)}(r_{\rm g})\delta r_2   \nonumber \\
        & - \frac{a}{2} I^\crit_{1/r^2}(r_{\rm g}) + \delta r_{4} I^\crit_{1/r}(r_{\rm g})  \bigg]\ , \label{eq:rad_shift_trajectory_critical_plunge_old}
\end{align}
where $\delta r^\homo$ is the solution for homoclinic motion given in Section III of Ref.~\cite{Piovano:2025aro} (modulo an integration constant). Using the integrals of the Appendix~\eqref{app:integrals_critical_plunge} and Eq.~\eqref{eq:shift_radial_frequency}, the spin correction to the radial motion can be simplified to
\begin{align} 
        \delta r(r_{\rm g}) &= \frac{(r_{2\rm g} - r_{\rm g})(r_{1 \rm g} - r_{\rm g})}{r_{1 \rm g}r_{2 \rm g}}\bigg[2\delta r_4 - \frac{a}{3} \bigg( \frac{2}{r_{1 \rm g}} + \frac{3}{r_{2 \rm g}} \bigg)  \nonumber \\
        &- \frac{a}{3r_{\rm g}}\bigg]  + \frac{r_{\rm g}(r_{\rm g} - r_{2 \rm g})\delta r_1}{(r_{1 \rm g} - r_{2 \rm g})r_{1 \rm g}} + \frac{r_{\rm g}(r_{1\rm g} - r_{\rm g})\delta r_2}{(r_{1 \rm g} - r_{2 \rm g})r_{2 \rm g}}  \ , \label{eq:rad_shift_trajectory_critical_plunge}
\end{align}
while the solutions to the shifts to the coordinate time and azimutal trajectories read
    \begin{align}
        \delta t(r_{\rm g}) &= \frac{T_{\rm g}(r_{\rm g})\delta r(r_{\rm g})}{\sqrt{R_{\rm g}(r_{\rm g})}} - \frac{\delta r_1}{2} T_{\rm g}(r_{q \rm g})  I^\crit_{1/(r - r_1)}(r_{\rm g}) \nonumber \\ 
       &  - \delta r_2 T_{\rm g}(r_{2 \rm g}) I^\crit_{1/(r - r_2)}(r_{\rm g}) - 2E_{\rm g}\delta r_4 I^\crit(r_{\rm g}) \nonumber \\ 
       & - \frac{\delta E}{1 - E^2_{\rm g}} \Big(I^\crit_{r^2}(r_{\rm g}) + 2 I^\crit_{r}(r_{\rm g}) + 4 I^\crit(r_{\rm g}) \Big) \nonumber \\
       & - \frac{E_{\rm g}}{2} \big( \delta r_1 + 2 \delta r_2 + 2 \delta r_4 \big) I^\crit_r(r_{\rm g} \big)   \nonumber \\
       & - \frac{E_{\rm g}}{2} \big((2 + r_{1 \rm g})\delta r_1 + 2(2 + r_{2 \rm g})\delta r_2 - a \big) I^\crit(r_{\rm g}) \ , \label{eq:time_shift_trajectory_critical_plunge}
    \end{align}
    \begin{align}
        \delta \phi(r_{\rm g}) &= \frac{\Phi_{\rm g}(r_{\rm g})\delta r(r_{\rm g})}{\sqrt{R_{\rm g}(r_{\rm g})}} - \frac{\delta r_1}{2}  \Phi_{\rm g}(r_{1 \rm g}) I^\crit_{1/(r - r_i)}(r_{\rm g})  \nonumber \\
        & - \delta r_2 \Phi_{\rm g}(r_{1 \rm g})  I^\crit_{1/(r - r_i)}(r_{\rm g}) +E_{\rm g} I^\crit(r_{\rm g}) \nonumber \\
        & - \bigg(L_{z \rm g}\frac{E_{\rm g} \delta E}{1 - E^2_{\rm g}} + \delta L_z \bigg) I^\crit(r_{\rm g}) \ . \label{eq:azimuthal_shift_trajectory_critical_plunge}
    \end{align}    
Appendix~\ref{app:integrals_critical_plunge} lists the integrals $I^\crit(r_{\rm g})$, $I^\crit_{1/r}(r_{\rm g})$, $I^\crit_{1/r^2}(r_{\rm g})$, $I^\crit_{r}(r_{\rm g})$, $I^\crit_{r^2}(r_{\rm g})$, $I^\crit_{1/(r - r_1}(r_{\rm g})$, $I^\crit_{1/(r - r_2)}(r_{\rm g})$ , which are combinations of elementary functions.  Fig.~\ref{fig:shift_critical_plunge} presents the shifts to the homoclinic trajectories and critical plunges, with the former defined in the interval $(r_{2 \rm g}, r_{1 \rm g}]$, and the latter in $(0,r_{2 \rm g})$. Fig.~\ref{fig:critical_plunge} presents the projection onto the equatorial (co-rotating polar plane) in the top panel (bottom panel) for homoclinic and critical plunging motion.  
Notice that $\delta t(r_{\rm g})$ and $\delta \phi(r_{\rm g})$ diverge logarithmically at $r_{\rm g} = r_{2 \rm g}$ as their geodesic counterparts for both critical plunges and homoclinic motion. As a result, the coordinate time and azimuthal homoclinic trajectories are disconnected from the critical plunging trajectories. By contrast, Eq.~\eqref{eq:rad_shift_trajectory_critical_plunge} is valid for both homoclinic and critical plunges. 

\subsection{Plunges from the ISCO}\label{sec:ISCO_plunge}
\begin{figure*}[!bth]
    {\centering
        \includegraphics[width=\textwidth]{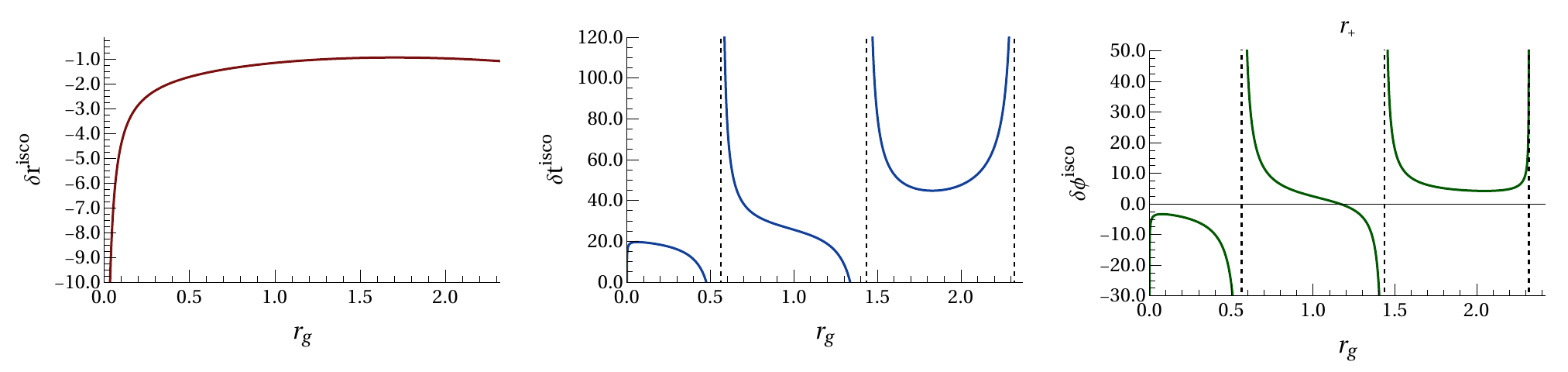}}
        \caption{Spin corrections to radial (left panel), coordinate time (middle panel) and azimuthal trajectories (right panel) for a prograde ISCO plunge for a spinning binary with $a =0.9$, $\chi_\parallel = 1/2$, $\chi_\perp = \sqrt{3}/2$. In the middle and right panel, the dashed lines correspond, from left to right, to the event horizons $r_-$ and $r_+$, and the geodesic ISCO $r_\text{g,isco}$. }   \label{fig:shift_isco_plunge}
\end{figure*}

\begin{figure}[!bth]
    {\centering
        \includegraphics[width=0.43\textwidth]{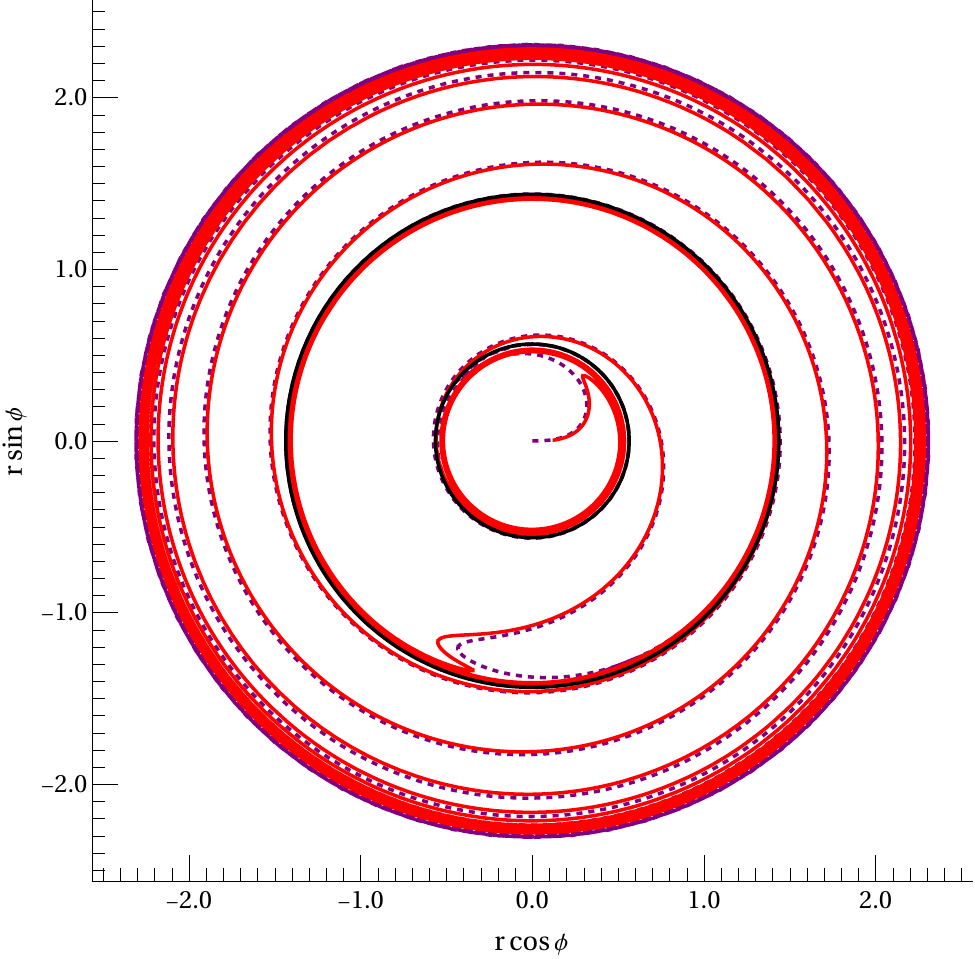}\\[0.4em]
        \includegraphics[width=0.43\textwidth]{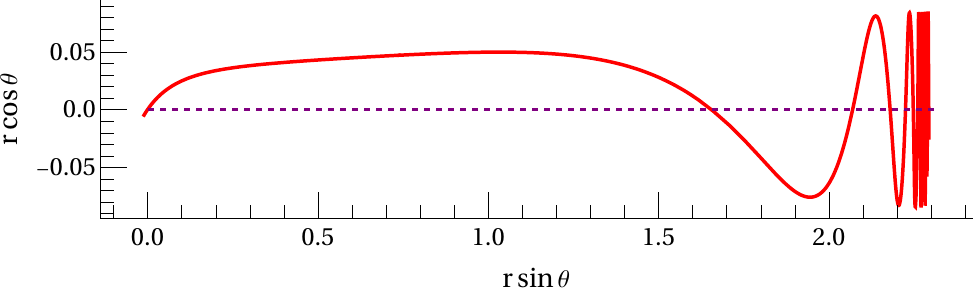}
    }
        \caption{Plots of ISCO plunges for a spinning particle (red-line) and its underlying fiducial geodesic (purple, dashed line). The orbital parameters and the linear-in-spin corrections are the same of Fig.~\eqref{fig:shift_isco_plunge}, while the mass-ratio is fixed to $\epsilon = 1/20$. The two black rings correspond to the primary outer horizon $r_+ = 1+\sqrt{1 -a^2}$ and inner horizon $r_- = 1 - \sqrt{1 -a^2}$.
        Top panel: projection of the orbits onto the equatorial plane. Bottom panel: orthogonal projection of the orbits onto a co-rotating polar plane.
}       \label{fig:isco}
\end{figure}

The trajectories for plunges from the ISCO can be obtained by simply taking the limit $r_{2 \rm g} \to r_{1 \rm g}$ of Eqs.~\eqref{eq:rad_shift_trajectory_critical_plunge}-\eqref{eq:azimuthal_shift_trajectory_critical_plunge}, which simplify to
\begin{align} 
        \delta r(r_{\rm g}) &= \frac{r_{\rm g}\delta r_\isco}{r_\text{g,isco}} - a \frac{(r_\text{g,isco} - r_{\rm g})^2 (r_\text{g,isco} + 5r_{\rm g})}{3 r^3_\text{g,isco} r_{\rm g}}   \nonumber \\
        &+ \frac{2}{r^2_\text{g,isco}} (r_\text{g,isco} - r_{\rm g})^2 \delta r_4 \ , \label{eq:rad_shift_trajectory_ISCO_plunge}
\end{align}

    \begin{align}
        \delta t(r_{\rm g}) &= \frac{T_{\rm g}(r_{\rm g})\delta r(r_{\rm g})}{\sqrt{R_{\rm g}(r_{\rm g})}} - \frac{3\delta r_\isco}{2} T_{\rm g}(r_\text{g,isco}) I^\isco_{1/(r_1 - r)}(r_{\rm g}) \nonumber \\
        & - \frac{\delta E}{1 - E^2_{\rm g}} \Big(I^\isco_{r^2}(r_{\rm g}) + 2 I^\isco_{r}(r_{\rm g}) + 4 I^\isco(r_{\rm g}) \Big) \nonumber \\
        & - \frac{E_{\rm g}}{2} \Big[ \big(3 \delta r_\isco + 2 \delta r_4 \big) I^\isco_r(r_{\rm g}) \Big] \nonumber \\
        & - \frac{E_{\rm g}}{2}\Big[ \big(3(2 +r_\text{g,isco})\delta r_\isco - a + 4\delta r_4 \big) I^\isco(r_{\rm g}) \Big]   \ , \label{eq:time_shift_trajectory_ISCO_plunge}
    \end{align}
    \begin{align}
        \delta \phi(r_{\rm g}) &= \frac{\Phi_{\rm g}(r_{\rm g})\delta r(r_{\rm g})}{\sqrt{R_{\rm g}(r_{\rm g})}} - \frac{3\delta r_\isco}{2}\Phi_{\rm g}(r_\text{g,isco})I^\isco_{1/(r_1 - r)}(r_{\rm g}) \nonumber \\
        & + \bigg(E_{\rm g} - L_{z \rm g}\frac{E_{\rm g} \delta E}{1 - E^2_{\rm g}} - \delta L_z \bigg) I^\isco(r_{\rm g})  \ , \label{eq:azimuthal_shift_trajectory_ISCO_plunge}
    \end{align}    
Appendix~\ref{app:integrals_ISCO_plunge} lists the integrals $I^\isco(r_{\rm g})$, $I^\isco_{1/r}(r_{\rm g})$, $I^\isco_{1/r^2}(r_{\rm g})$, $I^\isco_{r}(r_{\rm g})$, $I^\isco_{r^2}(r_{\rm g})$, $I^\isco_{1/(r_1 - r)}(r_{\rm g})$, which are combinations of elementary functions. Fig.~\ref{fig:shift_isco_plunge} presents the shifts to the ISCO plunge trajectories due to the secondary spin, while Fig.~\ref{fig:isco} presents the projection onto the equatorial (co-rotating polar plane) in the top panel (bottom panel) for ISCO plunging motion.
Notice that $\delta t (r_{\rm g})$ and $\delta \phi(r_{\rm g})$ have a pole at $r_{\rm g} = r_\text{g,isco}$ as their geodesic counterparts.

\subsection{Plunges related to bound orbits}\label{sec:pbo}
We set $r_{\rm g}(0) = r_{3 \rm g}$ at $\lambda =0$ as initial condition for plunging orbits related to bound motion. The solution for the radial shift is then
\begin{align} 
    \delta r(r_{\rm g}) &= -\sqrt{R_g(r_{\rm g})}\bigg[\bigg(\frac{E_{\rm g}\delta E}{1 - E^2_{\rm g}} + \delta\hat\Upsilon^\gen \bigg) I^\pbo(r_{\rm g})  \nonumber \\
        & + \displaystyle \sum^3_{i=1} I^\pbo_{1/(r - r_i)}(r_{\rm g})\frac{\delta r_i}{2} - \frac{a}{2} I^\pbo_{1/r^2}(r_{\rm g})  \nonumber \\
        & + \delta r_{4} I^\pbo_{1/r}(r_{\rm g})  \bigg]\ , \label{eq:radial_shift_pbo}
\end{align}

    \begin{align}
       \delta t(r_{\rm g}) &= \frac{T_{\rm g}(r_{\rm g})\delta r(r_{\rm g})}{\sqrt{R_{\rm g}(r_{\rm g})}} - \displaystyle \sum^3_{i=1} T_{\rm g}(r_{i \rm g}) \frac{\delta r_i}{2} I^\pbo_{1/(r - r_i)}(r_{\rm g}) \nonumber \\
       &- \frac{\delta E}{1 - E^2_{\rm g}} \Big(I^\pbo_{r^2}(r_{\rm g}) + 2 I^\pbo_{r}(r_{\rm g}) + 4 I^\pbo(r_{\rm g}) \Big) \nonumber \\
       &  - \frac{E_{\rm g}}{2}\Big( \displaystyle \sum^3_{i=1} \delta r_i + 2 \delta r_4 \Big) I^\pbo_r(r_{\rm g}) \nonumber \\
       & - \frac{E_{\rm g}}{2} \Big(\displaystyle \sum^3_{i=1} (2 + r_{i \rm g})\delta r_i - a + 4\delta r_4 \Big) I^\pbo(r_{\rm g})   \ , \label{eq:time_shift_pbo} 
    \end{align} 
\begin{align}
    \delta \phi(r_{\rm g}) &= \frac{\Phi_{\rm g}(r_{\rm g})\delta r(r_{\rm g})}{\sqrt{R_{\rm g}(r_{\rm g})}} - \displaystyle \sum^3_{i=1} \Phi_{\rm g}(r_{i \rm g}) \frac{\delta r_i}{2} I^\pbo_{1/(r - r_i)}(r_{\rm g})  \nonumber \\
    & + \bigg(E_{\rm g} - L_{z \rm g}\frac{E_{\rm g} \delta E}{1 - E^2_{\rm g}} - \delta L_z \bigg) I^\pbo(r_{\rm g})   \ , \label{eq:azimuthal_shift_pbo}
\end{align}
where the integrals $I^\pbo(r_{\rm g})$, $I^\pbo_{1/r}(r_{\rm g})$, $I^\pbo_{1/r^2}(r_{\rm g})$, $I^\pbo_{r}(r_{\rm g})$, $I^\pbo_{r^2}(r_{\rm g})$ and $I^\pbo_{1/(r - r_i)}(r_{\rm g})$ for $i= 1, \dots 3$ are given in Appendix~\ref{app:integrals_pbo}. 

Solutions for plunges related to stable circular orbits can be obtained by taking the limit $r_{2 \rm g} \to r_{1 \rm g}$ in Eqs.~\eqref{eq:radial_shift_pbo}-\eqref{eq:azimuthal_shift_pbo}. Moreover, the previous expressions reduce to the corrections for critical plunges when $r_{3 \rm g} \to r_{2 \rm g}$, as discussed in Appendix~\ref{app:integrals_pbo}. Fig.~\ref{fig:shift_pbo} presents the spin corrections to the PBO trajectories, while Fig.~\ref{fig:pbo} presents the projection onto the equatorial (co-rotating polar plane) in the top panel (bottom panel) for PBO motion.

\begin{figure*}[!bth]
    {\centering
        \includegraphics[width=\textwidth]{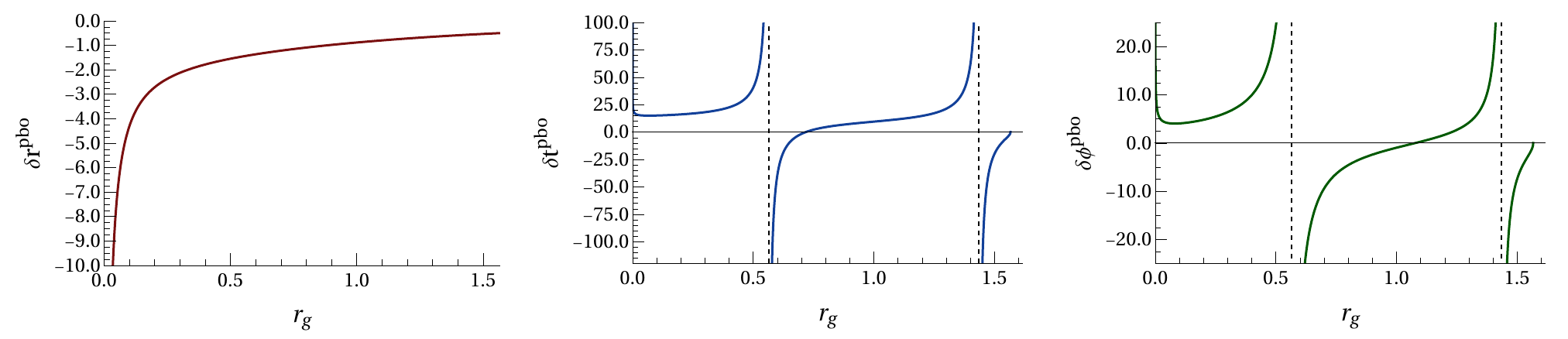}}
        \caption{Spin corrections to radial (left panel), coordinate time (middle panel) and azimuthal trajectories (right panel) for prograde PBOs for a spinning binary with $a =0.9$, $\chi_\parallel = 1/2$, $\chi_\perp = \sqrt{3}/2$. Orbital parameters of the fiducial geodesic: $e_{\rm g} =0.32$, $p_{\rm g} = p^*_{\rm g}+1$, with $p^*_{\rm g}$ the location of the geodesic separatrix at $e_{\rm g} =0.32$. In the middle and right panel, the dashed lines correspond, from left to right, to the event horizons $r_-$ and $r_+$.
}       \label{fig:shift_pbo}
\end{figure*}

\begin{figure}[!bth]
    {\centering
        \includegraphics[width=0.43\textwidth]{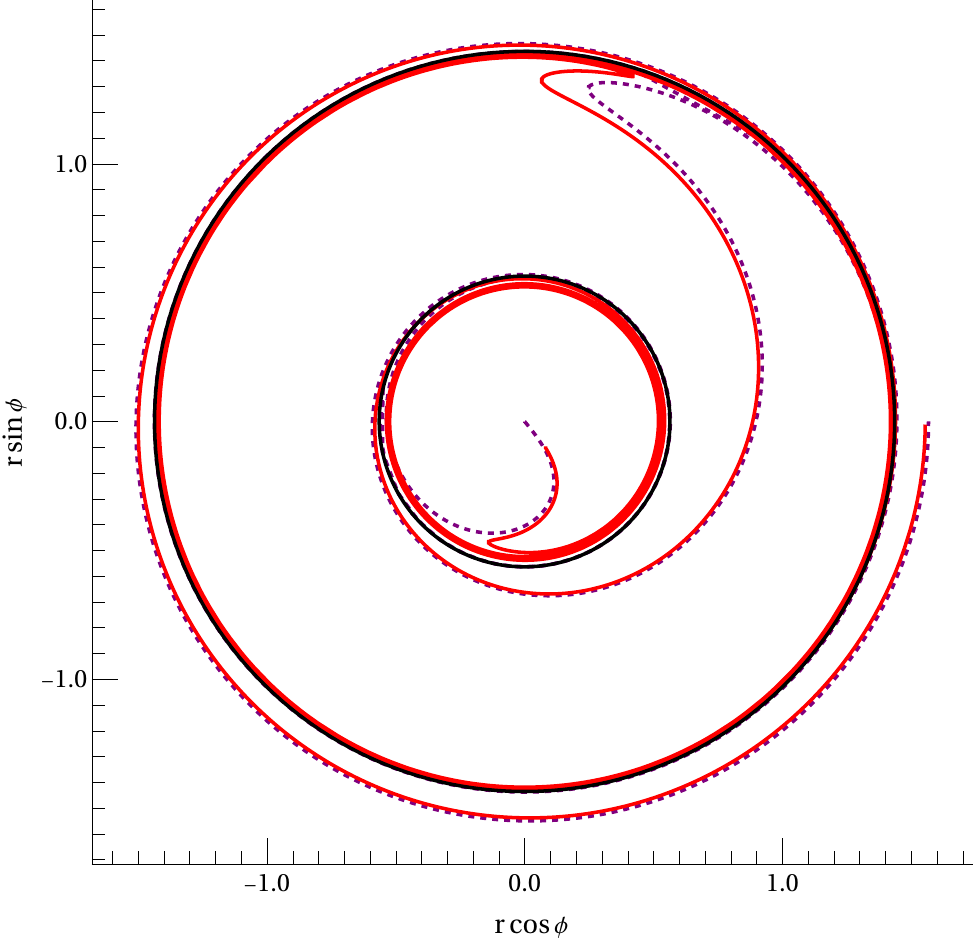}\\[0.4em]
        \includegraphics[width=0.43\textwidth]{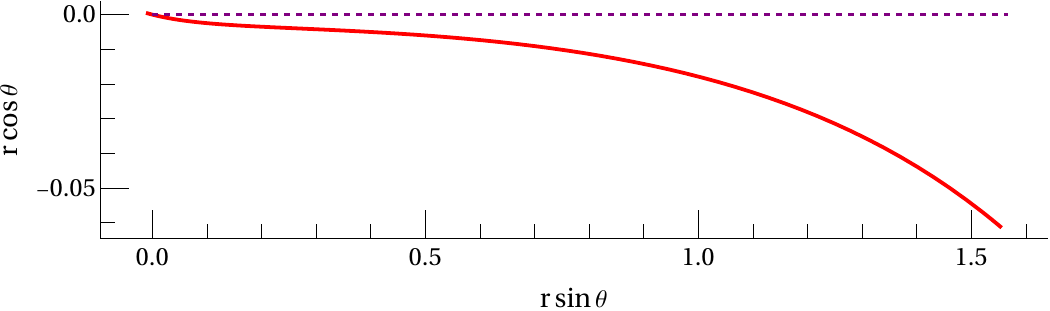}
    }
        \caption{Plots of a PBO trajectory for a spinning particle (red line) and its underlying fiducial geodesic (purple, dashed line). The orbital parameters and the linear-in-spin corrections are the same of Fig.~\eqref{fig:shift_pbo}, while the mass-ratio is fixed to $\epsilon = 1/20$. The two black rings correspond to the primary outer horizon $r_+ = 1+\sqrt{1 -a^2}$ and inner horizon $r_- = 1 - \sqrt{1 -a^2}$.
        Top panel:  projection of the orbits onto the equatorial plane. Bottom panel: orthogonal projection of the orbits onto a co-rotating polar plane.
}       \label{fig:pbo}
\end{figure}

\subsection{Generic plunges}\label{sec:generic_plunge}
Generic plunging orbits have as initial condition $r_{\rm g}(0) = r_{1 \rm g}$ at $\lambda =0$, while the function $\delta V_r(r_{\rm g})$ is defined as
\begin{align}
  \delta V_r(r_{\rm g}) &= -\frac{1}{2}\bigg(\frac{\delta r_1}{r_{\rm g} - r_{1 \rm g}} + \frac{2\delta r_4}{r_{\rm g}} - \frac{a}{r^2_{\rm g}} + \frac{2 E_{\rm g} \delta E^\gen}{1 - E^2_{\rm g}} \nonumber \\
  & + \frac{(r_{\rm g} -\rho_{\rm g})\delta \rho_r}{(r_{\rm g} - \rho_{r\rm g} - i \rho_{i\rm g})(r_{\rm g} - \rho_{r\rm g} + i \rho_{i\rm g})}    \bigg)  \, \ .
\end{align}
The shifts to the trajectories $\delta r(r_{\rm g})$, $\delta t(r_{\rm g})$ and $\delta \phi(r_{\rm g})$ can be written as
\begin{align} 
    \delta r(r_{\rm g}) &= \sqrt{R_{\rm g}(r_{\rm g})}\bigg[ \bigg(\frac{E_{\rm g}\delta E^\gen}{1 - E^2_{\rm g}} + \delta \hat \Upsilon^\gen_r \bigg) I^\gen(r_{\rm g})  \nonumber \\
    &  + \delta r_4 I^\gen_{1/r}(r_{\rm g}) - \frac{a}{2} I^\gen_{1/r^2}(r_{\rm g}) + \frac{\delta r_1}{2} I^\gen_{1/(r - r_1)}(r_{\rm g}) \nonumber\\
    & + \frac{\delta \rho_r}{2} \Big(I^\gen_{1/(r - r_2)}(r_{\rm g}) + I^\gen_{1/(r - \overline{r_2})}(r_{\rm g}) \Big)  \bigg]\ , \label{eq:radial_shift_plunge}
\end{align}

    \begin{align}
       \delta t(r_{\rm g}) &= - \frac{T_{\rm g}(r_{\rm g})\delta r(r_{\rm g})}{\sqrt{R_{\rm g}(r_{\rm g})}} + \frac{\delta r_1}{2} T_{\rm g}(r_{1 \rm g}) I^\gen_{1/(r - r_1)}(r_{\rm g})\nonumber \\
        & + \frac{\delta \rho_r}{2}  T_{\rm g}(r_{2 \rm g})I^\gen_{1/(r - r_2)}(r_{\rm g}) + 2 E_{\rm g} \delta r_4 I^\gen(r_{\rm g}) \nonumber \\
        & + \frac{\delta \rho_r}{2} T_{\rm g}(\overline{r_{2 \rm g}}) I^\gen_{1/(r - \overline{r_2})}(r_{\rm g})  - a \frac{E_{\rm g}}{2} I^\gen(r_{\rm g}) \nonumber \\ 
        & + \frac{\delta E^\gen}{1 - E^2_{\rm g}} \Big(I^\gen_{r^2}(r_{\rm g}) + 2 I^\gen_{r}(r_{\rm g}) + 4 I^\gen(r_{\rm g}) \Big)  \nonumber \\
        & + \frac{E_{\rm g}}{2} \Big( (2 + r_{1 \rm g})\delta r_1 + 2(2 + \rho_{r \rm g})\delta \rho_r \Big) I^\gen(r_{\rm g})  \nonumber \\
        & + \frac{E_{\rm g}}{2}\Big( \delta r_1 + 2 \rho_r + 2 \delta r_4 \Big) I^\gen_r(r_{\rm g})   \ , \label{eq:time_shif_plunge} 
    \end{align}    
    \begin{align}
        \delta \phi(r_{\rm g}) &= - \frac{\Phi_{\rm g}(r_{\rm g})\delta r(r_{\rm g})}{\sqrt{R_{\rm g}(r_{\rm g})}} + \Phi_{\rm g}(r_{1 \rm g}) \frac{\delta r_1}{2} I^{\gen}_{1/(r - r_1)}(r_{\rm g}) \nonumber \\
        & + \Phi_{\rm g}(r_{2 \rm g}) \frac{\delta \rho_r}{2} I^\gen_{1/(r - r_2)}(r_{\rm g}) + L_{z \rm g}\frac{E_{\rm g} \delta E^\gen}{1 - E^2_{\rm g}} I^\gen(r_{\rm g}) \nonumber \\
        & + \Phi_{\rm g}(\overline{r_{2 \rm g}}) \frac{\delta \rho_r}{2} I^\gen_{1/(r - \overline{r_2})}(r_{\rm g}) -\big(E_{\rm g} - \delta L^\gen_z \big) I^\gen(r_{\rm g})   \ , \label{eq:azimuthal_shift_plunge}        
    \end{align} 
where the integrals $I^\gen(r_{\rm g})$, $I^\gen_{1/r}(r_{\rm g})$, $I^\gen_{1/r^2}(r_{\rm g})$, $I^\gen_{r}(r_{\rm g})$, $I^\gen_{r^2}(r_{\rm g})$, $I^\gen_{1/(r - r_1)}(r_{\rm g})$, $I^\gen_{1/(r - r_2)}(r_{\rm g})$, and $I^\gen_{1/(r - \overline{r_2})}(r_{\rm g})$ are given in Appendix~\ref{app:integrals_generic_plunge}. 
Moreover, the previous expressions reduce to the corrections for homoclinic motion when $\rho_{i \rm g} \to 0$, as discussed in Appendix~\ref{app:integrals_generic_plunge}. Fig.~\ref{fig:shift_plunge} presents the spin corrections to generic plunges, while Fig.~\ref{fig:plunge} presents the projection onto the equatorial (co-rotating polar plane) in the top panel (bottom panel) for generic plunging motion.

\begin{figure*}[!bth]
    {\centering
        \includegraphics[width=\textwidth]{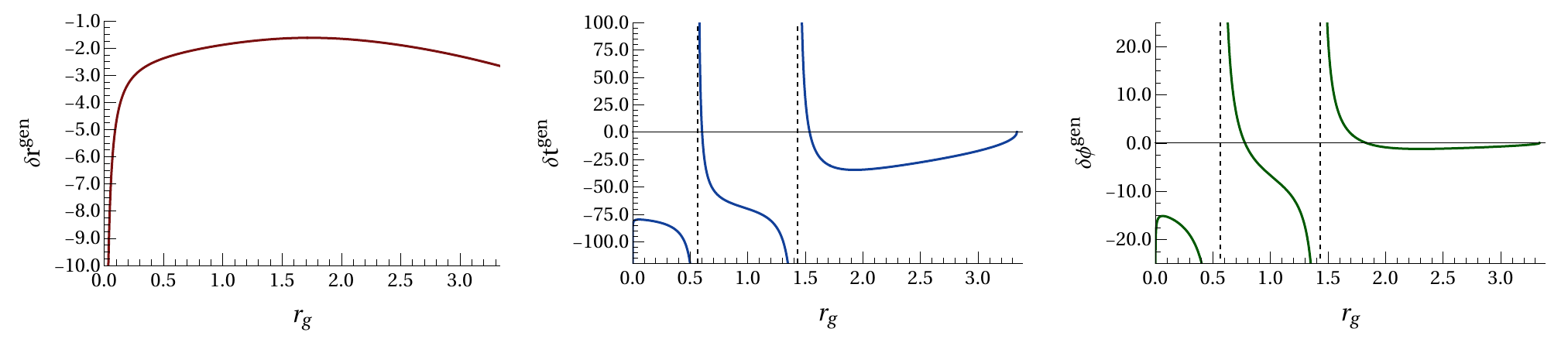}}
        \caption{Spin corrections to radial (left panel), coordinate time (middle panel) and azimuthal trajectories (right panel) for prograde generic plunging orbits for a spinning binary with $a =0.9$, $\chi_\parallel = 1/2$, $\chi_\perp = \sqrt{3}/2$. Orbital parameters of the fiducial geodesic: $e_{\rm g} =0.32$, $p_{\rm g} = 2.27$. In the middle and right panel, the dashed corresponds lines, from left to right, to the event horizons $r_-$ and $r_+$.}   \label{fig:shift_plunge}
\end{figure*}

\begin{figure}[!bth]
    {\centering
        \includegraphics[width=0.45\textwidth]{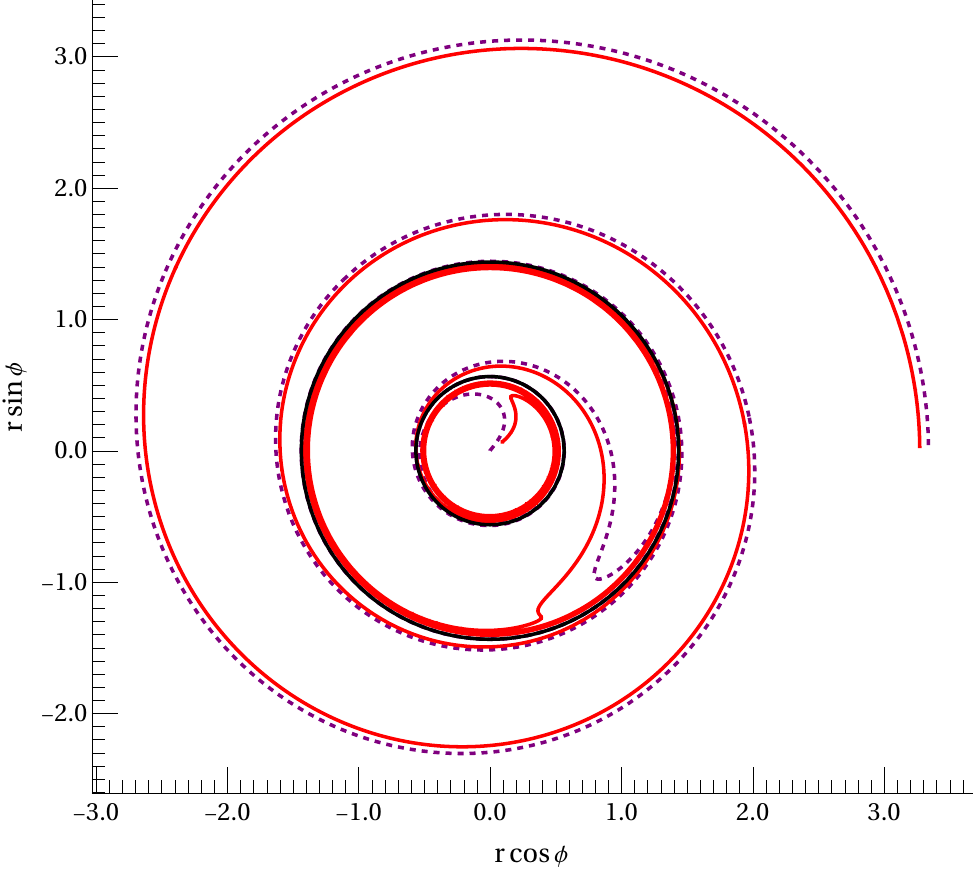}\\[0.4em]
        \includegraphics[width=0.45\textwidth]{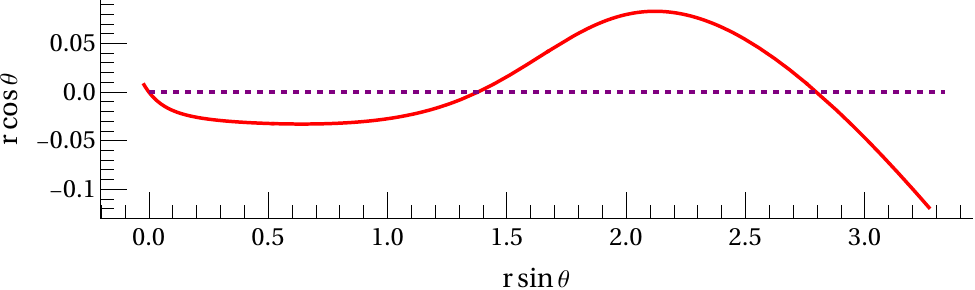}
    }
        \caption{Plots of generic plunges for a spinning particle (red-line) and its underlying fiducial geodesic (purple, dashed line). The orbital parameters and the linear-in-spin corrections are the same of Fig.~\eqref{fig:shift_plunge}, while the mass-ratio is fixed to $\epsilon = 1/20$. The two black rings correspond to the primary outer horizon $r_+ = 1+\sqrt{1 -a^2}$ and inner horizon $r_- = 1 - \sqrt{1 -a^2}$.
        Top panel:  projection of the orbits onto the equatorial plane. Bottom panel: orthogonal projection of the orbits onto a co-rotating polar plane.
}       \label{fig:plunge}
\end{figure}


\subsection{Special plunges}\label{sec:special_plunge}
Special plunges satisfy the constraint $L_{z \rm g} =  a E_{\rm g}$, which implies that
\begin{equation}
    \delta L_z = a \delta E \ .
\end{equation}
Thus, the radial potential $R_5(r)$ reduce to 
\begin{equation}
    R_5(r) = \big[(E^2_{\rm g} -1 + \epsilon \chi_\parallel 2 E_{\rm g} \delta E) r^2 + 2 r - a(a - \epsilon \chi_\parallel 2 E^2_{\rm g} ) \big] r^2 \ ,
\end{equation}
The corrections to the radial turning points $\delta r_1$ and $\delta r_2$ are
\begin{align}
    \delta r_1 = \frac{E_{\rm g} \big( a E_{\rm g}(1 - E^2_{\rm g}) - (a^2 -2 r_{1 \rm g})\delta E \big)}{(1 - E^2_{\rm g})\sqrt{1 - a^2(1 - E^2_{\rm g})}} \ , \\
    \delta r_2 = -\frac{E_{\rm g} \big( a E_{\rm g}(1 - E^2_{\rm g}) - (a^2 -2 r_{1 \rm g})\delta E \big)}{(1 - E^2_{\rm g})\sqrt{1 - a^2(1 - E^2_{\rm g})}}
\end{align}
The FE gauge for special plunging orbits corresponds to the condition $\delta r_1 = \delta r_2$, which fix the shift $\delta E$ as
\begin{equation}
    \delta E = - \frac{a E_{\rm g}(1 - E^2_{\rm g})^2}{2 - a^2(1 - E^2_{\rm g})}
\end{equation}
The correction $\delta r_1$ in the FE gauge then simplifies to
\begin{equation}
   \delta r_1 = - \frac{2 a E^2_{\rm g}}{2 - a^2(1 - E^2_{\rm g})} \ , 
\end{equation}
while the function $ \delta V_r(r_{\rm g})$ is defined as
\begin{align}
  \delta V_r(r_{\rm g}) &= -\frac{1}{2}\bigg( \displaystyle \delta r_1\sum^2_{i =1} \frac{1}{r_{\rm g} - r_{i \rm g}}  + \frac{2 E_{\rm g} \delta E}{1 - E^2_{\rm g}} \bigg)  \, .  \label{eq:diff_radial_potential_special_plunge} 
\end{align}

The solutions of Eqs.~\eqref{eq:1st-order-EoM-r-fully-linearize} for a special plunge reads
\begin{align} 
        \delta r(r_{\rm g}) &= \sqrt{R_g(r_{\rm g})}\bigg(\frac{E_{\rm g}\delta E}{1 - E^2_{\rm g}} I^\spec(r_{\rm g})  \nonumber \\
        &+ \frac{\delta r_1}{2} \displaystyle \sum^2_{i=1} I^\spec_{1/(r - r_i)}(r_{\rm g}) \bigg) \ , \label{eq:rad_shift_trajectory_special_plunge}
\end{align}
    \begin{align}
        \delta t(r_{\rm g}) &= \frac{T_{\rm g}(r_{\rm g})\delta r(r_{\rm g})}{\sqrt{R_{\rm g}(r_{\rm g})}} - \frac{\delta r_1}{2}\displaystyle \sum^2_{i=1} T_{\rm g}(r_{i \rm g}) I^\spec_{1/(r - r_i)}(r_{\rm g}) \nonumber \\
       & - \frac{\delta E}{1 - E^2_{\rm g}} \Big(I^\spec_{r^2}(r_{\rm g}) + 2 I^\spec_{r}(r_{\rm g}) + 4 I^\spec(r_{\rm g}) \Big) \nonumber \\
       &  - \frac{E_{\rm g}}{2} \delta r_1 \Big[2 I^\spec_r(r_{\rm g}) + \big(4+r_{1 \rm g} + r_{2 \rm g} \big) I^\spec(r_{\rm g}) \Big]   \ , \label{eq:time_shift_trajectory_special_plunge}
    \end{align}
    \begin{align}
        \delta \phi(r_{\rm g}) &= \frac{\Phi_{\rm g}(r_{\rm g})\delta r(r_{\rm g})}{\sqrt{R_{\rm g}(r_{\rm g})}}  + \frac{\delta r_1}{2}\displaystyle \sum^2_{i=1} \Phi_{\rm g}(r_{i \rm g}) I^\spec_{1/(r - r_i)}(r_{\rm g})  \nonumber \\
        & + \bigg(E_{\rm g} - \frac{a \delta E}{1 - E^2_{\rm g}} \bigg) I^\spec(r_{\rm g})   \ , \label{eq:azimuthal_shift_trajectory_special_plunge}
    \end{align}    
where the integrals $I^\spec(r_{\rm g})$, $I^\spec_{r}(r_{\rm g})$, $I^\spec_{r^2}(r_{\rm g})$, $I^\spec_{1/(r - r_1)}(r_{\rm g})$ and $I^\spec_{1/(r - r_2)}(r_{\rm g})$ are given in Appendix~\ref{app:integrals_special_plunge}. As a initial condition, we set $r_{\rm g}(0) = r_{1 \rm g}$ for $\lambda=0$. Unlike the other kind of plunging orbits, the spin corrections to special plunges are finite at $r_{\rm g} = 0$. 
\subsection{Partially expanded solutions of the equations of motion}\label{sec:partial_expansion_plunge_solutions}
As shown in the previous subsections, the spin corrections to the trajectories diverge as poles at $r_{\rm g} = r_\pm$, which is an issue for applications in waveform modeling. Near the horizon, the perturbative solutions fail, since the spin-corrections diverge faster than the leading order geodesic term. As mentioned previously, a suitable change of coordinates cannot simultaneously remove the logarithmic divergences at geodesic order, and the pole singularities at order $\mathcal O(\epsilon \chi)$ without introducing discontinuities for $\epsilon \to 0$.
Moreover, it is not possible to remove the poles at $r_{\rm g} = r_\pm$ in $\delta t(r_{\rm g})$ and $\delta \phi(r_{\rm g})$ by adding quadratic or higher order spin corrections, since these terms would introduce higher order poles.
Adopting a spin-gauge different than the FE-gauge may remove the pole divergences at $r = r_\pm$ for some class of plunges, but not in the case of the ISCO plunge, since the spin corrections to the ISCO are unique. Moreover, a different spin-gauge may introduce discontinuities at the parameter space boundaries between different type of plunges.
A workaround to the pole singularity issue consists in not fully expanding the trajectories into a fiducial geodesic and linear corrections to the spin in the system~\eqref{eq:1st-order-EoM}. Instead, we solve the equations of motion for $t(r)$ and $\phi(r)$ using the unexpanded orbital radius $r$ as a parameter. Fortunately, an explicit solution for the plunging radial trajectory $r(\lambda)$ is unnecessary in several applications, especially for building IMR waveforms~\cite{Kuchler:2025hwx,DellaRocca:2025zbe}~\footnote{Non-perturbative solutions for $r(\lambda)$ generally require the inversion of a hyper-elliptic integral~\cite{Hackmann:2008zz,Hackmann:2014tga}.}.
Likewise, the constants of motion $(E, L_z)$, and the radial potential roots $r_1, r_2, r_3$ are kept unexpanded in the equations of motion and the corresponding solutions. The same choice was adopted in Ref.~\cite{Witzany:2024ttz} to derive the spin corrections to the actions and frequencies, and in Ref.~\cite{Witzany:2023bmq} to directly solve by quadrature the equations of motion.
The roots $r_4$ and $r_5$ of the radial potential $R_5(r)$ are instead expanded in Eqs.~\eqref{eq:1st-order-EoM}. Such a practical choice ensure that the equations of motion can be solved by quadrature in terms of elementary functions and elliptic integrals. Moreover, expanding the roots $r_4$ and $r_5$ avoid large numerical cancellations when evaluating analytic solutions in the limit $\epsilon\to 0$ or $a \to 0$.~\footnote{For instance, if the roots $r_4$ and $r_5$ are kept unexpanded, the critical and ISCO plunges are described by elliptic integrals instead of elementary functions as in the geodesic case. The same argument applies to the Schwarzschild limit, since $r_4 = r_5 =0$ for $a=0$.}
The partially expanded equations of motion~\eqref{eq:1st-order-EoM} can be written as
\begin{align}
    & \totder{t}{r} = \frac{T_{\rm g}(r)}{\sqrt{\tilde R(r)}}\bigg[1 + \epsilon \chi_\parallel \bigg( \frac{\delta r_4}{r} - \frac{a}{2r^2} \bigg) \bigg] + \epsilon \chi_\parallel \frac{\delta \tilde T(r)}{\sqrt{\tilde R(r)}}  \label{eq:T_of_r_velocity_partial_expansion} \, ,\\
    & \totder{\phi}{r} = \frac{\Phi_{\rm g}(r)}{\sqrt{\tilde R(r)}}\bigg[1 + \epsilon \chi_\parallel \bigg( \frac{\delta r_4}{r} - \frac{a}{2r^2} \bigg) \bigg] + \epsilon \chi_\parallel \frac{\delta \tilde \Phi(r)}{\sqrt{\tilde R(r)}} \label{eq:Phi_of_r_velocity_partial_expansion} \, , 
\end{align}
where $\tilde R_{\rm g}(r)$, $\tilde T_{\rm g}(r)$ and $\tilde \Phi_{\rm g}(r)$ correspond to Eqs.~\eqref{eq:1st-order-EoMgeo}-\eqref{eq:1st-order-EoMgeoPhi} with $E_{\rm g}$ and $L_{z\rm g}$  replaced with $E$ and $L_z$, while
\begin{align}
    & \delta \tilde T(r) = \frac{2}{\Delta}( a E r - (L_z - a E)))  -\frac{(L_z - a E)}{r}  \, ,\\
    & \delta \tilde \Phi(r) = \frac{a}{r \Delta}( a E r  - (L_z - a E)) -  E  \, , 
\end{align}
After partial fraction decomposition, the solutions for Eq.~\eqref{eq:T_of_r_velocity_partial_expansion} and Eq.~\eqref{eq:Phi_of_r_velocity_partial_expansion} can be written as follows:
\begin{equation}
    t^\mathsf{y}(r) = - t^{\mathsf{y},\text{sgn}}_{\rm g}(r) \ , \qquad \phi^\mathsf{y}(r) = - \phi^{\mathsf{y},\text{sgn}}(r)
\end{equation}
with $\text{y} = (\crit, \isco, \pbo, \spec)$ and
\begin{widetext}
    \begin{align}
        t^{\mathsf{y},\text{sgn}}(r) &= t^{\mathsf{y},\text{sgn}}_{\rm g}(r) + \epsilon \chi_\parallel\frac{E}{2} \big[(4 \delta r_4 - a) I^\mathsf{y}(r) + \delta r_4 I^\mathsf{y}_{r}(r) \big] \nonumber \\
        &- \epsilon \chi_\parallel \frac{I^\mathsf{y}_{1/(r - r_+)}(r)}{2(r_+ - r_-)} \bigg[a^3 E + 2 L_z r_+ + a E (r_+ - 6) r_+ + 2\big(E r^3_+ + a^2 E(2 + r_+) -2 a L_z \big) \bigg]  + (+\leftrightarrow-)   \ ,  \label{eq:coordinate_time_partial_exp_trajectory} \\ 
        \phi^{\mathsf{y},\text{sgn}}(r) &= \phi^{\mathsf{y},\text{sgn}}_{\rm g}(r) - \epsilon \chi_\parallel E I^\mathsf{y}(r) - \epsilon \chi_\parallel \frac{I^\mathsf{y}_{1/(r - r_+)}(r)}{2(r_+ - r_-)} \bigg[a (L_z - 2 a E) - 2(2 a E + L_z(r_+ -2)) \delta r_4 \bigg] + (+\leftrightarrow-)  \ . \label{eq:azimuthal_partial_exp_trajectory}
    \end{align}    
\end{widetext}
The integrals $I^\mathsf{y}(r), I^\mathsf{y}_{r}(r), I^\mathsf{y}_{1/(r - r_\pm)}(r)$ are given in Appendix~\ref{app:integrals_critical_plunge}-\ref{app:integrals_pbo} for critical plunges ($\mathsf{y} = \crit$), ISCO plunges ($\text{y} = \isco$) and PBOs ($\text{y} = \pbo$), respectively, and in Appendix~\ref{app:integrals_special_plunge} for special plunges ($\text{y} = \spec$). For generic plunges
\begin{equation}
    t^\gen_{\rm g}(r_{\rm g}) = t^{\mathsf{gen},\text{sgn}}_{\rm g}(r_{\rm g}) \ , \quad \phi^\gen_{\rm g}(r_{\rm g}) = \phi^{\mathsf{gen},\text{sgn}}_{\rm g}(r_{\rm g}) \ ,
\end{equation}
and the integrals $I^\mathsf{y}(r_{\rm g}), I^\mathsf{y}_{r}(r_{\rm g}), I^\mathsf{y}_{1/(r - r_\pm)}(r_{\rm g})$ can be found in Appendix~\ref{app:integrals_ISCO_plunge}. The sign choice ensures that generic plunges match homoclinic orbits in the limit $\rho_{i \rm g} \to 0$. We checked the validity of Eqs.~\eqref{eq:coordinate_time_partial_exp_trajectory}-\eqref{eq:azimuthal_partial_exp_trajectory} for each kind of plunges in a Mathematica notebook available in the GitHub repo~\cite{repoAnalyticHJproject}. 
Moreover, it is easy to check that Eqs.~\eqref{eq:coordinate_time_partial_exp_trajectory}-\eqref{eq:azimuthal_partial_exp_trajectory} for $a =0$ match Eq.29-30 of~\cite{Witzany:2023bmq} for plunging orbits in Schwarzschild spacetime. It is sufficient to plug their Eqs.(35)-(37) into Eqs.(29)-(30) of~\cite{Witzany:2023bmq} and perform a simple partial fraction decomposition; additionally, $\delta r_4 =0$ for $a=0$.

The partially expanded solutions~\eqref{eq:coordinate_time_partial_exp_trajectory} and \eqref{eq:azimuthal_partial_exp_trajectory} do not have any pole singularities at $r = r_\pm$, thus they present a convenient alternative to the fully linearized solutions shown in Sec.~\ref{sec:critical_plunge}-\ref{sec:special_plunge}. In practical applications, one fixes a referential plunging geodesics by choosing the values $(a, p_{\rm g}, e_{\rm g})$, which in turn determines the spin corrections in the FE-gauge to the constants of motion, radial potential roots and orbital elements, as described in Sec.~\ref{sec:roots_radial_potential}.  The roots $r_1, r_2 , r_3$ and constants of motion $E, L_z$ in~\eqref{eq:coordinate_time_partial_exp_trajectory} and \eqref{eq:azimuthal_partial_exp_trajectory} are then fixed by choosing $(\epsilon, \chi)$. If needed, the linear and higher-order corrections to the waveform amplitudes can then be extracted via interpolation or numerical derivatives.

\section{Polar motion and spin-precession} \label{sec:polar_motion_spin_prec}
The polar motion is governed by Eq.~\eqref{eq:polar_EoM}, which is the equation of motion of a forced harmonic oscillator driven by the spin-curvature force for $\Lzrd \neq 0$. When $\Lzrd =0$, Eq.~\eqref{eq:polar_EoM} reduces to the equation of motion of a free harmonic oscillator. We impose that the reference geodesics is constrained on the equatorial plane through the initial conditions $(\delta z_\text{h}(0),\dd \delta z_\text{h}(0)/ \dd \lambda) = (0,0)$ for the homogeneous solutions $\delta z_\text{h}(\lambda)$ of Eq.~\eqref{eq:polar_EoM}. As a result, the general solution of Eq.~\eqref{eq:polar_EoM} is given by~\cite{Witzany:2023bmq,Piovano:2024yks,Piovano:2025aro} 
\begin{equation}
    \delta z(\lambda) = - \chi_\perp \frac{\cos\psi_{\rm p} \sqrt{r^2_{\rm g} + \Lzrd^2}}{r_{\rm g} \Lzrd}  \ , \quad \text{ for } \Lzrd \neq 0 \label{eq:polar_solution}   
\end{equation}
When $\Lzrd = 0$ (i.e. for special plunges), the motion is constrained on the equatorial plane, which implies that the secondary spin is not precessing. Notice that the shift $\delta z(\lambda)$ is independent of the corrections to the constants of motion, therefore it is the same in any spin-gauge. Moreover, the polar motion orbits diverge at $r_{\rm g} = 0$. Such behavior is expected, since the radial motion is also divergent at the ring singularity. Eq.~\eqref{eq:polar_solution} is only valid when $\Lzrd$ is not too small. For instance, if $\Lzrd \sim \epsilon$, then $\epsilon \delta z(\lambda) \sim \mathcal O(1)$, which is clearly inconsistent with our initial assumptions. In SF applications, values of $\Lzrd$  can only be made arbitrarily small by fine tuning the initial conditions. In fact, at the end of the inspiral and transition-to-plunge, the particle will typically move along an ISCO or a critical plunge, at leading order in the mass ratio (see Ref.~\cite{Faggioli:2025hff}).

With the solution~\eqref{eq:polar_solution} at hand, we only need to find an analytic expression for the spin-precession phase $\psi_p(\lambda)$, whose evolution satisfies Eq.~\eqref{eq:spin-precession-angle}. The latter can be written as
\begin{equation}
    \totder{\psi_p}{r_{\rm g}} = - \frac{\Psi_r(r_{\rm g})}{\sqrt{R_{\rm g}(r_{\rm g})}}  \, , 
\end{equation}
Using partial fraction decomposition, the previous equation becomes
\begin{align}
    \totder{\psi_p}{r_{\rm g}} &= \frac{(a + E_{\rm g} \Lzrd)\Lzrd}{2i\sqrt{R_{\rm g}(r_{\rm g})}}\bigg( \frac{1}{r_{\rm g} - i |\Lzrd|} - \frac{1}{r_{\rm g} + i |\Lzrd|}  \bigg) \nonumber \\
    &  - \frac{a \sgn(\Lzrd) + E_{\rm g}|\Lzrd|}{\sqrt{R_{\rm g}(r_{\rm g})}}  \ ,
\end{align}
which can be immediately solved in terms of elementary functions or elliptic integrals as
\begin{equation}
    \psi^\mathsf{y}_p(r_{\rm g}) = \psi^{\mathsf{y},\text{sgn}}_p(r_{\rm g}) \ ,   \qquad \mathsf{y} = \crit, \isco, \pbo
\end{equation}
with
\begin{align}
    \psi^{\mathsf{y},\text{sgn}}_p(r_{\rm g}) &= - \big(a\sgn(\Lzrd) + E_{\rm g}|\Lzrd|\big) I^\mathsf{y}(r_{\rm g})  \nonumber \\
    & + \frac{\Lzrd}{2} (a + E_{\rm g}\Lzrd)\Im \!\Big( I^\mathsf{y}_{1/(r - i |\Lzrd|)}(r_{\rm g})  \nonumber \\
    & - I^\mathsf{y}_{1/(r + i |\Lzrd|)}(r_{\rm g}) \Big)  \ , \label{eq:spin_precession_phase}
\end{align}
The integrals $I^\mathsf{y}(r_{\rm g}), I^\mathsf{y}_{1/(r \pm i |\Lzrd|)}(r_{\rm g})$ with $\mathsf{y} = \crit, \isco, \pbo$ are given in Appendices~\ref{app:integrals_critical_plunge}-\ref{app:integrals_pbo} for critical plunges, ISCO plunges, and PBOs, respectively. For generic plunges, the sign convention is different, and
\begin{equation}
    \psi^\gen_p(r_{\rm g}) = -\psi^{\mathsf{y},\text{sgn}}_p(r_{\rm g}) \ ,   \qquad \mathsf{y} = \gen
\end{equation}
with $I^\gen(r_{\rm g}), I^\gen_{1/(r \pm i |\Lzrd|)}(r_{\rm g})$ given in Appendix~\ref{app:integrals_generic_plunge}. The sign choice for  $\psi^\gen_p$ ensures that, in the limit $\rho_{i \rm g} \to 0$, generic plunging orbits continuously reduce to homoclinic orbits.

For all types of bound plunging motion, the  Mino-time spin precession frequency is simply given by $\Upsilon_p = \Psi_r(r_{\rm g})$. Finally, the bottom panels of Figs.~\ref{fig:critical_plunge},~\ref{fig:isco},~\ref{fig:pbo} and~\ref{fig:plunge} present the orthogonal projection onto a co-rotating polar plane for critical plunges, ISCO plunges, PBOs and generic plunge, respectively, with $\delta z = \delta z(r_{\rm g})$ parametrized using the geodesic orbital radius $r_{\rm g}$.

\section{Discussion}\label{sec:discussion}
In this work, we present the first, complete study of bound plunging motion of a spinning test-particle in Kerr spacetime. We solve the equations of motion in the near equatorial case, including the precession of the particle's spin and orbital plane. The solutions are given in closed-form as elliptic integrals and Jacobi elliptic functions or elementary functions. Additionally, we derive a closed-form expression for the shift to the IBCO, which is useful to the characterization of both plunging and parabolic scattering. We also introduce a new, convenient parametrization for generic plunging orbits using Keplerian-like orbital elements. Finally, we discuss a major issue affecting perturbative solutions of the MPD equations for plunging motion: the presence of pole singularities near the horizon in the linear-in-spin corrections to coordinate time and azimuthal trajectories. This issue arises in both our direct solutions for the motion and the ``virtual geodesics'' formalism of Ref.~\cite{Skoupy:analytic-solutions}. We circumvent it by deriving analytic trajectories that are only partially expanded in the secondary spin. These alternative solutions are particularly convenient for waveform modeling applications.

Our analytic solutions can be employed to add conservative spin corrections to IMR waveform models constructed via self-force and black hole perturbation theory~\cite{Kuchler:2025hwx,Faggioli:2025hff,Compere:2021zfj,Compere:2021iwh,Becker:2024xdi,Becker:2025zzw,DellaRocca:2025zbe,Kuchler:2024esj,Lhost:2024jmw,Compere:2021zfj,Compere:2021iwh}. 
Moreover, they represent a useful starting point to calculate quadratic corrections in the particle's spin~\cite{Compere:2021bkk,Ramond:2026fpi,Ramond:2024ozy}, which are required in the modeling of the transition-to-plunge phase~\cite{Honet:2025lmk,Honet:2025dho,Kuchler:2025hwx}. The radial potential can still be factorized in terms of the roots of a polynomial, and the roots can be derived by applying Vieta's formulas. The FE spin-gauge Eqs.~\eqref{eq:shift_turning_points_FE} and \eqref{eq:shift_turning_points_FE_cc} can then be imposed on the quadratic corrections of the geodesic roots $r_{2\rm g}$ and $r_{3\rm g}$. Moreover, the partially expanded solutions shown in Sec.~\ref{sec:partial_expansion_plunge_solutions} can also be extended to quadratic order. It is sufficient to expand, in the equations of motion, all radial potential roots that do not have a geodesic counterpart, i.e. all the roots that vanish for $\epsilon \to 0$.

In a follow-up work~\cite{PiovanoInPrep}, we will extend the methods presented here to complete the analysis of the nearly equatorial motion for a spinning particle. In particular, we will derive analytic solutions for plunging orbits related to scattering orbits ~\cite{Liu:2023tcy}, unbound plunging orbits from infinity~\cite{chandrasekhar1983mathematical,Mummery:2023hlo}, and scattering orbits~\cite{Bini:2017pee,Mummery:2023hlo,Liu:2023tcy}. Such solutions would find applications, for instance, in the calculation of gravitational radiation from hyperbolic and parabolic orbits~\cite{Saijo:1998mn,Warburton:2025ymy,Barack:2026izc}.

\begin{acknowledgments}
This work makes use of the following \textit{Mathematica} packages of the Black Hole Perturbation Toolkit~\cite{BHPToolkit}: ``KerrGeodesics"\cite{BHPToolkitKerrGeodesics}.
The code used for the calculation of the orbits, and all numerical checks is available in the GitHub repository~\cite{repoAnalyticHJproject}.
G.A.P. also acknowledges the support of the Win4Project grant ETLOG of the Walloon Region for the Einstein Telescope and of the Institut Complexys of the University of Mons.

\end{acknowledgments}

\newpage


\appendix

\section{Convention for elliptic integrals} \label{app:elliptic_integrals_convention}
Table~\ref{tab:elliptic_integrals} lists the conventions for the Legendre elliptic integrals and Jacobi elliptic functions adopted in this work, which follow the Mathematica conventions.
\begin{table}[h!]
  \begin{center}
    \caption{Elliptic Integral Conventions}
    \label{tab:elliptic_integrals}
    \begin{tabular}{l|c}
      \toprule 
      \textbf{Function} & \textbf{Definition} \\
      \midrule 
       $\mathsf{F}(x|m)$& $\int_{0}^{x}\frac{1}{\sqrt{1-m\sin^{2}(\theta)}}\mathrm{d}\theta$\\
       $\mathsf{K}(m)$& $\mathsf{F}(\pi/2|m)$\\
       $\mathsf{E}(x|m)$& $\int_{0}^{x}\sqrt{1-m\sin^{2}(\theta)}\mathrm{d}\theta$\\
       $\mathsf{E}(m)$& $\mathsf{E}(\pi/2|m)$\\
       $\mathsf{\Pi}(n,x|m)$&$\int_{0}^{x}\frac{1}{(1-n\sin^{2}(\theta))\sqrt{1-m\sin^{2}(\theta)}}\mathrm{d}\theta$\\
        $\mathsf{\Pi}(n|m)$&$\mathsf{\Pi}(n,\pi/2|m)$\\
       $\mathsf{am}(u|m)$& $u=\mathsf{F}(\mathsf{am}(u|m)|m)$\\
       $\mathsf{sn}(u|m)$& $\sin(\mathsf{am}(u|m))$\\
      \bottomrule 
    \end{tabular}
  \end{center}
\end{table}

When the argument $m=1$, the elliptic integrals reduce to elementary functions
\begin{align} \label{eq:limit_elliptic_integrals}
    \mathsf{F}(x|1) &= \log\big(\sec(x) + \tan(x)\big) \ , \\
    \mathsf{E}(x|1) &= \sin(x) \ , \\
    \mathsf{\Pi}(n,x|1) &= \frac{\sqrt{n}}{n-1}  \arctanh(\sqrt{n}\sin{x}) \nonumber \\
    &- \frac{1}{n-1}\log\big(\sec(x) + \tan(x)\big) \Big) \ .
\end{align}
The above expressions are useful to check the edge cases of certain orbits, such as generic plunges in the limit $\rho_{i \rm g} \to 0$, or PBOs for $r_{3 \rm g} \to r_{2 \rm g}$.

\section{Integrals for critical plunges} \label{app:integrals_critical_plunge}
This Section lists the integrals for the analytic solutions of a critical plunge. The following expressions are equal to the integrals for homoclinc motion given in Appendix B of Ref.~\cite{Piovano:2025aro}, apart for an integration constant.

\begin{widetext}    
\begin{align}
    I^\crit(r_{\rm g}) &= \int \frac{\dd r'}{\sqrt{R_{\rm g}(r')}} = \frac{1}{2\sqrt{(1 - E^2_{\rm g})(r_{1 \rm g} - r_{2 \rm g})r_{2 \rm g}}} \log \!\Bigg[\Bigg( \frac{\sqrt{r_{\rm g}(r_{1 \rm g} - r_{2 \rm g})} -\sqrt{r_{2\rm g}(r_{1 \rm g} - r_{\rm g})}}{\sqrt{r_{\rm g}(r_{1 \rm g} - r_{2 \rm g})} + \sqrt{r_{2 \rm g}(r_{1 \rm g} - r_{\rm g)}} } \Bigg)^{\!\!2} \Bigg] = \lambda  \ , \\
    I^\crit_{1/r}(r_{\rm g}) & = \int \frac{\dd r'}{r'\sqrt{R_{\rm g}(r')}} =  -\frac{2\sqrt{r_{1 \rm g} - r_{\rm g}}}{r_{1 \rm g} r_{2 \rm g}\sqrt{(1 - E^2_{\rm g})r_{\rm g}}} + \frac{I^\crit(r_{\rm g})}{r_{2 \rm g}} \ , \\
    I^\crit_{1/r^2}(r_{\rm g}) &= \int \frac{\dd r'}{r'^2\sqrt{R_{\rm g}(r')}} = -\frac{2}{3} \frac{\sqrt{(r_{1 \rm g} - r_{\rm g})}}{r_{1 \rm g} r_{2 \rm g} \sqrt{(1 - E^2_{\rm g})r_{\rm g}}} \bigg[ \frac{2}{r_{1 \rm g}} + \frac{1}{r_{2 \rm g}} \bigg(3 + \frac{r_{2 \rm g}}{r_{\rm g}} \bigg) \bigg]  + \frac{I^\crit(r_{\rm g})}{r^2_{2\rm g}}  \ , \\
    I^\crit_{r}(r_{\rm g}) &= \int \frac{r'\dd r'}{\sqrt{R_{\rm g}(r')}} = \frac{2}{\sqrt{1 - E^2_{\rm g}}} \arccos\bigg(\sqrt{\frac{r_{\rm g}}{r_{1\rm g}}} \bigg) + r_{2 \rm g} I^\crit(r_{\rm g}) \ , \\
    I^\crit_{r^2}(r_{\rm g}) &= \int \frac{r'^2\dd r'}{\sqrt{R_{\rm g}(r')}} = \frac{\sqrt{(r_{1\rm g} - r_{\rm g})r_{\rm g}}}{\sqrt{1 - E^2_{\rm g}}} + \frac{r_{1\rm g} + 2 r_{2 \rm }}{\sqrt{1 - E^2_{\rm g}}}  \arccos\bigg(\sqrt{\frac{r_{\rm g}}{r_{1\rm g}}} \bigg) + r^2_{2 \rm g} I^\crit(r_{\rm g})   \ , \\
    I^\crit_{1/(r - \alpha)}(r_{\rm g}) &= \int\frac{\dd r'}{(r' - \alpha)\sqrt{R_{\rm g}(r')}} = \frac{I^\crit(r_{\rm g})}{r_{2 \rm g} - \alpha} + \frac{\log \!\Bigg[\Bigg( \frac{\sqrt{r_{\rm g}(r_{1 \rm g} -\alpha)} \;\, - \;\, \sqrt{\alpha(r_{1 \rm g} - r_{\rm g})}}{\sqrt{r_{\rm g}(r_{1 \rm g} - \alpha)} \;\, + \;\, \sqrt{\alpha(r_{1 \rm g} - r_{\rm g)}} } \Bigg)^{\!\!2} \Bigg]}{2(r_{2 \rm g} - \alpha)\sqrt{(1 - E^2_{\rm g})(r_{1 \rm g} - \alpha)\alpha}} \ , \\
    I^\crit_{1/(r - r_1)}(r_{\rm g}) &= \int \frac{\dd r'}{(r' - r_{1 \rm g})\sqrt{R_{\rm g}(r')}} =  \frac{2}{r_{1 \rm g}(r_{1 \rm g} - r_{2 \rm g})}\frac{1}{\sqrt{1 - E^2_{\rm g}}}\frac{\sqrt{r_{\rm g}}}{\sqrt{r_{1 \rm g} - r_{\rm g}}}  - \frac{I^\crit(r_{\rm g})}{(r_{1 \rm g} - r_{2 \rm g})} \ , \\
    I^\crit_{1/(r - r_2)}(r_{\rm g}) &= \int \frac{\dd r'}{(r' - r_{2 \rm g})\sqrt{R_{\rm g}(r')}} = \frac{1}{r_{2 \rm g}(r_{1 \rm g} - r_{2 \rm g})}\frac{1}{\sqrt{1 - E^2_{\rm g}}}\frac{\sqrt{(r_{1 \rm g} - r_{\rm g})r_{\rm g}}}{r_{\rm g} - r_{2\rm g}} - \frac{(r_{1 \rm g} - 2 r_{2 \rm g})I^\crit(r_{\rm g})}{2(r_{1 \rm g} - r_{2 \rm g})r_{2 \rm g}}  \ .    
\end{align}
\end{widetext}

\section{Integrals for plunges from the ISCO} \label{app:integrals_ISCO_plunge}
We list here the integrals that appear in the closed-form solutions for the ISCO plunges. These integrals can be obtained by simply taking the limit $r_{1 \rm g} \to r_{2 \rm g}$ in the expressions of Appendix~\ref{app:integrals_critical_plunge}. 
\begin{widetext}    
\begin{align}
    I^\isco(r_{\rm g}) &= \int\!\frac{\dd r'}{\sqrt{R_{\rm g}(r')}} = \frac{2}{r_{1 \rm g}} \frac{1}{\sqrt{(1 - E^2_{\rm g})}} \sqrt{\frac{r_{\rm g}}{r_{1 \rm g} - r_{\rm g}}} = \lambda  \ , \\
    I^\isco_{1/r}(r_{\rm g}) &= \int\! \frac{\dd r'}{r'\sqrt{R_{\rm g}(r')}} = \frac{1}{\sqrt{(1 - E^2_{\rm g})}}\frac{2(2r_{\rm g} - r_{1 \rm g})}{\sqrt{r_{\rm g}(r_{1 \rm g} - r_{\rm g})}r^2_{1 \rm g}}  \ , \\
    I^\isco_{1/r^2}(r_{\rm g}) &= \int\!\frac{\dd r'}{r'^2\sqrt{R_{\rm g}(r')}} = \frac{2}{3r^2_{1 \rm g}\sqrt{(1 - E^2_{\rm g})}}\frac{1}{\sqrt{r_{\rm g}(r_{1 \rm g} -r_{\rm g})}}\Big(-4 + \frac{8 r_{\rm g}}{r_{1 \rm g}} - \frac{r_{1 \rm g}}{r_{\rm g}}  \Big)  \ , \\
    I^\isco_{r}(r_{\rm g}) &= \int\! \frac{r'\dd r'}{\sqrt{R_{\rm g}(r')}} = \frac{2}{\sqrt{(1 - E^2_{\rm g})}} \arccos \bigg( { \sqrt{\frac{r_{\rm g}}{r_{1 \rm g}}} } \bigg) + r_{1 \rm g}  I^\isco(r_{\rm g}) \ , \\
    I^\isco_{r^2}(r_{\rm g}) &= \int\! \frac{r'^2\dd r'}{\sqrt{R_{\rm g}(r')}} = \frac{1}{\sqrt{(1 - E^2_{\rm g})}}\Bigg( r_{1 \rm g} \arccos \bigg( \sqrt{\frac{r_{\rm g}}{r_{1 \rm g}}}  \bigg) + \sqrt{r_{\rm g}(r_{1 \rm g} - r_{\rm g})} \Bigg) + r_{1 \rm g}  I^\isco_{r}(r_{\rm g})   \ , \\
    I^\isco_{1/(r - \alpha)}(r_{\rm g}) &= \int\! \frac{\dd r'}{(r' - \alpha)\sqrt{R_{\rm g}(r')}} = \frac{I^\isco(r_{\rm g})}{r_{1 \rm g} - \alpha} + \frac{\log \!\Bigg[\Bigg( \frac{\sqrt{r_{\rm g}(r_{1 \rm g} -\alpha)} \;\, - \;\, \sqrt{\alpha(r_{1 \rm g} - r_{\rm g})}}{\sqrt{r_{\rm g}(r_{1 \rm g} - \alpha)} \;\, + \;\, \sqrt{\alpha(r_{1 \rm g} - r_{\rm g)}} } \Bigg)^{\!\!2} \Bigg]}{2\sqrt{(1 - E^2_{\rm g})\alpha}(r_{1 \rm g} - \alpha)^{3/2}} \ , \\
    I^\isco_{1/(r_1 - r)}(r_{\rm g}) &= \int \! \frac{\dd r'}{(r_{1 \rm g} - r')\sqrt{R_{\rm g}(r')}} = -\frac{2}{3}\frac{\sqrt{r_{\rm g}}(2 r_{\rm g} - 3r_{1 \rm g})}{r^2_{1 \rm g}(r_{1 \rm g} - r_{\rm g})^{3/2}} \ ,  
\end{align}
\end{widetext}

\section{Integrals for plunges related to bound orbits} \label{app:integrals_pbo}

All roots of the geodesic radial potential are real in the case of plunges related to bound orbits. When all roots are distinct, PBO motion can be described in terms of elliptic integrals in a similar fashion to bound orbits.  For PBOs, the $m$ argument in the elliptic integrals of Table~\ref{tab:elliptic_integrals} is $k_{r\rm g}$, which is defined as
\begin{align}
    k_{r \rm g} &= \frac{(r_{1 \rm g} - r_{2 \rm g})r_{3 \rm g}}{(r_{1 \rm g} - r_{3 \rm g})r_{2 \rm g}} \, , 
\end{align}
while the elliptic characteristic $n$ in $\mathsf{\Pi}(n|m)$ is one of the following terms
\begin{align}
    \gamma_r &= \frac{r_{3\rm g}}{r_{2\rm g}}  \,,\\
    h_\alpha &= \frac{r_{3\rm g}(r_{2\rm g} - \alpha)}{r_{2\rm g}(r_{3\rm g} - \alpha)} \, ,
\end{align}
with $\alpha \in \mathbb{C}$. Moreover, the elliptic amplitude $x$ is defined as
\begin{equation}
  \zeta(r_{\rm g}) = \arcsin \Bigg(\sqrt{\frac{r_{2\rm g}(r_{3\rm g} - r_{\rm g})}{r_{3\rm g}(r_{2\rm g} - r_{\rm g})}}   \Bigg) \ , \label{eq:elliptic_amplitude_arcsin}
\end{equation}
which can be also defined using the Jacobi amplitude $\mathsf{am(u|m)}$ as
\begin{align}
    \zeta(\lambda) &= \mathsf{am}\Big(\frac{1}{2}Y_r \lambda \Big \rvert k_{r \rm g}\Big) \ ,  \label{eq:elliptic_amplitude_am_pbo} \\
    Y_r &= \sqrt{(1 - E^2_{\rm g})(r_{1\rm g} - r_{3\rm g})r_{2\rm g}} \ .
\end{align}

\subsection{Plunges related to bound orbits - all distinct roots} 
We list here the integrals that appear in the analytic solutions for plunges related to bound orbits, i.e. when all roots of the geodesic radial potential are distinct.

\begin{widetext}
\begin{align}
    I^\pbo(r_{\rm g}) &= \int^{r_{3\rm g}}_{r_{\rm g}} \frac{\dd r'}{\sqrt{R_{\rm g}(r')}} = \frac{\mathsf{F}(\zeta| k_{r \rm g})}{Y_r} = \lambda  \ , \\
    I^\pbo_{1/r}(r_{\rm g}) &= \int^{r_{3\rm g}}_{r_{\rm g}} \frac{\dd r'}{r'\sqrt{R_{\rm g}(r')}} = \frac{2}{Y_r r_{3 \rm g}}\Big(\frac{r_{3\rm g}}{r_{1\rm g}} -1\Big) \bigg( \mathsf{E}(\zeta|k_{r\rm g}) - \sqrt{1 - k_{r\rm g}\sin(\zeta)^2}\tan(\zeta) \bigg) + \frac{I^\pbo(r_{\rm g})}{r_{3\rm g}} \ , \\
    I^\pbo_{1/r^2}(r_{\rm g}) &= \int^{r_{3\rm g}}_{r_{\rm g}} \frac{\dd r'}{r'^2\sqrt{R_{\rm g}(r')}} = \frac{2(r_{1\rm g} - r_{3\rm g})\big(r_{2\rm g}r_{3\rm g} + r_{1\rm g}(r_{2\rm g} + r_{3\rm g})\big)}{3r^2_{1\rm g}r_{2\rm g}r^2_{3\rm g} Y_r} \Big(\sqrt{1 - k_{r\rm g}\sin^2(\zeta)}\tan(\zeta) - 2\mathsf{E}(\zeta|k_{r\rm g})\Big) \nonumber \\
    &\phantom{=\int^{r_{1\rm g}}_{r_{\rm g}} \frac{\dd r'}{r'^2\sqrt{R_{\rm g}(r')}}} + \frac{(r_{2\rm g} - r_{3\rm g})r_{3\rm g} + r_{1\rm g}(2r_{2\rm g} + r_{3\rm g})}{3r_{1\rm g}r_{2\rm g} r^2_{3\rm g}} I^\pbo(r_{\rm g})  + \frac{2 r_{3\rm g}\sin(\zeta) \cos(\zeta)}{3r^2_{1\rm g}(r_{1\rm g} - r_{3\rm g})Y_r} \sqrt{\frac{r_{2\rm g} - r_{\rm g}}{r_{1\rm g} - r_{\rm g}}}     \nonumber \\
    &\phantom{=\int^{r_{1\rm g}}_{r_{\rm g}} \frac{\dd r'}{r'^2\sqrt{R_{\rm g}(r')}}} + \frac{2\big(r^2_{3\rm g} + r_{3\rm g} r_{\rm g} + r^2_{\rm g} - r_{1\rm g}(r_{3\rm g} + r_{\rm g}) \big)(r_{3\rm g} - r_{\rm g})}{3(r_{1\rm g} - r_{3\rm g})r^2_{3\rm g}r_{\rm g} R_{\rm g}(r_{\rm g})}   \ , \\
    I^\pbo_{r}(r_{\rm g}) &= \int^{r_{3\rm g}}_{r_{\rm g}} \frac{r'\dd r'}{\sqrt{R_{\rm g}(r')}} = r_{2\rm g}I^\pbo(r_{\rm g}) - \frac{2(r_{2\rm g} - r_{3\rm g})}{Y_r}\mathsf{\Pi}(\gamma_r,\zeta|k_{r\rm g})  \ , \\
    I^\pbo_{r^2}(r_{\rm g}) &= \int^{r_{3\rm g}}_{r_{\rm g}} \frac{r'^2\dd r'}{\sqrt{R_{\rm g}(r')}} = \frac{r_{2\rm g}(r_{2\rm g} + r_{3\rm g})}{Y_r}I^\pbo(r_{\rm g}) - \frac{r_{3\rm g}\sin(\zeta)\cos(\zeta)}{\sqrt{1 - E^2_{\rm g}}}\sqrt{\frac{(r_{1\rm g} - r_{\rm g})(r_{2\rm g} - r_{\rm g})}{r_{2\rm g}(r_{2\rm g} - r_{3\rm g})}}  \nonumber \\
    &\phantom{=\int^{r_{1\rm g}}_{r_{\rm g}} \frac{r'^2\dd r'}{\sqrt{R_{\rm g}(r')}}} + \frac{\sqrt{r_{2\rm g}(r_{1\rm g} - r_{3\rm g})}}{\sqrt{1 - E^2_{\rm g}}}\mathsf{E}(\zeta|k_{r\rm g}) - \frac{(r_{2\rm g} - r_{3\rm g})(r_{1\rm g} + r_{2\rm g} + r_{3\rm g})}{Y_r}\mathsf{\Pi}(\gamma_r,\zeta|k_{r\rm g}) \ , \\
    I^\pbo_{1/(r - r_1)}(r_{\rm g}) &= \int^{r_{3\rm g}}_{r_{\rm g}} \frac{\dd r'}{(r' - r_{1\rm g})\sqrt{R_{\rm g}(r')}} = \frac{2r_{2\rm g}\mathsf{E}(\zeta|k_{r\rm g})}{r_{1\rm g}(r_{1\rm g} - r_{2\rm g})Y_r} - \frac{I^\pbo(r_{\rm g})}{r_{1\rm g} - r_{2\rm g}}  - \frac{2r_{3\rm g}}{r_{1\rm g}(r_{1\rm g} - r_{3\rm g})Y_r} \frac{\cos(\zeta)\sin(\zeta)}{\sqrt{1 - k_{r\rm g}\sin(\zeta)^2}} \ , \\
    I^\pbo_{1/(r - r_2)}(r_{\rm g}) &= \int^{r_{3\rm g}}_{r_{\rm g}} \frac{\dd r'}{(r' - r_{2\rm g})\sqrt{R_{\rm g}(r')}} = \frac{I^\pbo(r_{\rm g})}{r_{1\rm g} - r_{2\rm g}} - \frac{2(r_{1\rm g} - r_{3\rm g})\mathsf{E}(\zeta|k_{r\rm g})}{(r_{1\rm g} - r_{2\rm g})(r_{2\rm g} - r_{3\rm g})Y_r} \ , \\
    I^\pbo_{1/(r - \alpha)}(r_{\rm g}) &= \int^{r_{3\rm g}}_{r_{\rm g}} \frac{\dd r'}{(r' - \alpha)\sqrt{R_{\rm g}(r')}} = \frac{I^\pbo(r_{\rm g})}{r_{2\rm g} -\alpha}  + \frac{2(r_{2\rm g} - r_{3\rm g}) \mathsf{\Pi}(h_\alpha,\zeta | k_{r\rm g})}{(r_{2\rm g} - \alpha)(r_{3\rm g} - \alpha)Y_r} \ .
\end{align} 
\end{widetext}
Additionally, we used the following diverging integral
\begin{align}
    I^\pbo_{1/(r - r_3)}(r_{\rm g}) &= \int^{r_{3\rm g}}_{r_{\rm g}} \frac{\dd r'}{(r' - r_{3\rm g})\sqrt{R_{\rm g}(r')}} \nonumber \\
    & = I^{\pbo,\mathsf{sing}}_{1/(r - r_3)}(r_{\rm g}) + I^{\pbo,\mathsf{reg}}_{1/(r - r_3)}(r_{\rm g}) \ ,  
\end{align}
which is composed of a diverging piece
\begin{align}
    I^{\pbo,\mathsf{sing}}_{1/(r - r_3)}(r_{\rm g}) &= \frac{2(r_{1\rm g} - r_{\rm g})r_{\rm g}}{(r_{1\rm g} - r_{3\rm g})r_{3\rm g}\sqrt{R_{\rm g}(r_{\rm g})}}   \ , 
\end{align}
and a finite part
\begin{align}
    I^{\pbo,\mathsf{reg}}_{1/(r - r_3)}(r_{\rm g}) &= - \frac{I^\pbo(r_{\rm g})}{r_{3\rm g}} + \frac{2r_{2\rm g}\mathsf{E}(\zeta|k_{r\rm g})}{r_{3\rm g}(r_{2\rm g} - r_{3\rm g})Y_r} \ .
\end{align}    

In the limit $r_{3 \rm g} \to r_{2 \rm g}$, the integrals for PBOs become the integrals for critical plunges given in Appendix~\ref{app:integrals_critical_plunge}. In fact, $k_{r \rm g} \to 1$ for $r_{3 \rm g} \to r_{2 \rm g}$, therefore the elliptic integrals reduce to elementary functions according to Eqs.~\eqref{eq:limit_elliptic_integrals}.

\subsection{Plunges related to bound orbits - one double root}
In the case $r_{1 \rm g} = r_{2 \rm g}$, the previous integrals reduce to
\begin{widetext}
\begin{align}
    I^\pbo(r_{\rm g}) &= \int^{r_{3\rm g}}_{r_{\rm g}} \frac{\dd r'}{\sqrt{R_{\rm g}(r')}} = \frac{2}{Y_r}\arcsin\!\Bigg(\sqrt{\frac{r_{1\rm g}(r_{3\rm g} - r_{\rm g})}{r_{3\rm g}(r_{1\rm g} - r_{\rm g})}} \Bigg) = \lambda  \ , \\
    I^\pbo_{1/r}(r_{\rm g}) &= \int^{r_{3\rm g}}_{r_{\rm g}} \frac{\dd r'}{r'\sqrt{R_{\rm g}(r')}} = \frac{2}{r_{1 \rm g}r_{3 \rm g}\sqrt{1 - E^2_{\rm g}}}\sqrt{\frac{r_{3\rm g}}{r_{\rm g}} - 1} + \frac{I^\pbo(r_{\rm g})}{r_{1\rm g}}  \ , \\
    I^\pbo_{1/r^2}(r_{\rm g}) &= \int^{r_{3\rm g}}_{r_{\rm g}} \frac{\dd r'}{r'^2\sqrt{R_{\rm g}(r')}} = -\frac{2\big(r^2_{3\rm g} + r^2_{\rm g} + r_{3\rm g}r_{\rm g} - r_{1\rm g}(r_{\rm g} +r_{3\rm g})\big)}{3\sqrt{1 - E^2_{\rm g}}(r_{1\rm g} - r_{3\rm g}) r^2_{3\rm g}  (r_{1 \rm g} - r_{\rm g})r_{\rm g}}\sqrt{\frac{r_{3\rm g}}{r_{\rm g}} -1} + \frac{ I^\pbo(r_{\rm g})}{r^2_{1\rm g}}  \nonumber \\
    &\phantom{=\int^{r_{3\rm g}}_{r_{\rm g}} \frac{\dd r'}{r'^2\sqrt{R_{\rm g}(r')}}}  +  \frac{2r_{3\rm g}\sin(\zeta)\cos(\zeta)}{3r^2_{1\rm g}(r_{1\rm g} - r_{3\rm g})Y_r} + \frac{2(r_{1\rm g} - r_{3\rm g})(r_{1\rm g} +2r_{3\rm g})\tan(\zeta)}{3r^2_{1\rm g} r^2_{3\rm g}Y_r}  \ , \\
    I^\pbo_{r}(r_{\rm g}) &= \int^{r_{3\rm g}}_{r_{\rm g}} \frac{r'\dd r'}{\sqrt{R_{\rm g}(r')}} = r_{1\rm g}  I^\pbo(r_{\rm g}) - i \log \bigg(\frac{\sqrt{r_{\rm g}} - i \sqrt{r_{3\rm g} - r_{\rm g}}}{\sqrt{r_{\rm g}} + i \sqrt{r_{3\rm g} - r_{\rm g}}} \bigg)  \ , \\
    I^\pbo_{r^2}(r_{\rm g}) &= \int^{r_{3\rm g}}_{r_{\rm g}} \frac{r'^2\dd r'}{\sqrt{R_{\rm g}(r')}} =  r^2_{1 \rm g} I^\pbo(r_{\rm g}) - \frac{(2r_{1\rm g} + r_{3\rm g})}{\sqrt{1 - E^2_{\rm g}}} \frac{i}{2} \log \bigg(\frac{\sqrt{r_{\rm g}} - i \sqrt{r_{3\rm g} - r_{\rm g}}}{\sqrt{r_{\rm g}} + i \sqrt{r_{3\rm g} - r_{\rm g}}}\bigg)  - \frac{\sqrt{(r_{3\rm g} - r_{\rm g})r_{\rm g}}}{\sqrt{1 - E^2_{\rm g}}}  \ , \\
    I^\pbo_{1/(r - r_1)}(r_{\rm g}) &= \int^{r_{3\rm g}}_{r_{\rm g}} \frac{\dd r'}{(r' - r_{1\rm g})\sqrt{R_{\rm g}(r')}} = - \frac{(2r_{1\rm g} - r_{3\rm g})I^\pbo(r_{\rm g})}{r_{1\rm g}(r_{1\rm g} - r_{3\rm g})Y_r} - \frac{2}{\sqrt{1 - E^2_{\rm g}}r_{1\rm g}(r_{1\rm g} - r_{3\rm g})}\frac{\sqrt{(r_{3\rm g} - r_{\rm g})r_{\rm g}}}{r_{1\rm g} - r_{\rm g}} \ , \\
    I^\pbo_{1/(r - \alpha)}(r_{\rm g}) &= \int^{r_{3\rm g}}_{r_{\rm g}} \frac{\dd r'}{(r' - \alpha)\sqrt{R_{\rm g}(r')}} =  \frac{I^\pbo(r_{\rm g})}{r_{1\rm g} - \alpha} - \frac{\log \!\Bigg[\Bigg( \frac{\sqrt{r_{1\rm g}(r_{3 \rm g} -\alpha)} \;\, - \;\, \tan(\zeta)\sqrt{\alpha(r_{1 \rm g} - r_{3\rm g})}}{\sqrt{r_{1\rm g}(r_{3 \rm g} - \alpha)} \;\, + \;\,\tan(\zeta)\sqrt{\alpha(r_{1\rm g} - r_{3\rm g)}} } \Bigg)^{\!\!2} \Bigg]}{2(r_{1 \rm g} - \alpha)\sqrt{(1 - E^2_{\rm g})(r_{3 \rm g} - \alpha)\alpha}} \ .
\end{align} 
\end{widetext}
while the following diverging integral becomes
\begin{align}
    I^\pbo_{1/(r - r_3)}(r_{\rm g}) &= \int^{r_{3\rm g}}_{r_{\rm g}} \frac{\dd r'}{(r' - r_{1 \rm g})\sqrt{R_{\rm g}(r')}} \nonumber \\
    & = I^{\pbo,\mathsf{sing}}_{1/(r - r_1)}(r_{\rm g}) + I^{\pbo,\mathsf{reg}}_{1/(r - r_1)}(r_{\rm g}) \ ,  
\end{align}
which is separated into a diverging piece
\begin{align}
    I^{\pbo,\mathsf{sing}}_{1/(r - r_3)}(r_{\rm g}) &= \frac{2\sqrt{r_{1\rm g}}r_{\rm g}}{r_{3\rm g}Y_r\sqrt{(r_{1\rm g} - r_{3\rm g})(r_{3\rm g} - r_{\rm g})r_{\rm g}}}   \ , 
\end{align}
and a regular part
\begin{align}
    I^{\pbo,\mathsf{reg}}_{1/(r - r_3)}(r_{\rm g}) &= \frac{I^\pbo(r_{\rm g})}{r_{1\rm g} - r_{3\rm g}}  \ .
\end{align}

\section{Integrals for generic plunges} \label{app:integrals_generic_plunge}
For generic plunges, the $m$ argument in Table~\ref{tab:elliptic_integrals} is the quantity $k_{r\rm g}$ defined as
\begin{align}
    k_{r \rm g} &= \frac{r^2_{1 \rm g} - (A-B)^2}{4AB} \, , 
\end{align}
with $A$ and $B$ given by
\begin{equation}
    A = \sqrt{\rho^2_{i \rm g} + (r_{1 \rm g} - \rho_{r \rm g})^2} \ , \qquad B = \sqrt{\rho^2_{r \rm g} + \rho^2_{i \rm g}} \ .
\end{equation}
The parameter $k_{r\rm g} \in [0,1]$, and $k_{r \rm g} = 1$ when $\rho_{i \rm g} =0$. The elliptic characteristic $n$ in $\mathsf{\Pi}(n|m)$ is one of the following terms
\begin{align}
    \gamma_r &= - \frac{(A - B)^2}{4 A B}  \,,\\
    h_\alpha &= \bigg(\frac{B r_{1 \rm g} + (A - B)\alpha}{B r_{1 \rm g} - (A + B)\alpha} \bigg)^{\!2} \, ,
\end{align}
with $\alpha \in \mathbb{C}$. Moreover, the elliptic amplitude $x$ of Table~\ref{tab:elliptic_integrals} is given by
\begin{equation}
    \xi(r_{\rm g}) = \arccos{\!\bigg(\frac{(r_{1 \rm g} - r_{\rm g})B-r_{\rm g}A}{(r_{1 \rm g} -r_{\rm g})B + r_{\rm g}A}\bigg)} \ . \label{eq:elliptic_amplitude_arccos}
\end{equation}
which, in our case, is defined in the domain $0 \leq \xi \leq  \pi$.  
The standard domain for the elliptic amplitude $x$ is $[0,\pi/2]$, which can be extended by taking advantage of the periodicity of elliptic integrals and functions (cf.~\cite{handbook-ellpitic-integrals}). Moreover, the elliptic amplitude $\xi$ can be defined using the Jacobi amplitude $\mathsf{am(u|m)}$ as
\begin{equation}
    \xi(\lambda) = \mathsf{am}\Big(\frac{1}{2}Y_r \lambda \Big \rvert k_{r \rm g}\Big) \ ,  \label{eq:elliptic_amplitude_am_plunge}
\end{equation}
where $Y_r = \sqrt{(1 - E^2_{\rm g})AB}$. 

In the following, we lists the elliptic integrals used to construct the analytic solutions for generic plunges:
\begin{widetext}
\begin{align}
    I^\gen(r_{\rm g}) &= \int^{r_{1\rm g}}_{r_{\rm g}} \frac{\dd r'}{\sqrt{R_{\rm g}(r')}} = \frac{\mathsf{F}(\pi - \xi| k_{r \rm g})}{Y_r} = \lambda  \ , \\
    I^\gen_{1/r}(r_{\rm g}) &= \int^{r_{1\rm g}}_{r_{\rm g}} \frac{\dd r'}{r'\sqrt{R_{\rm g}(r')}} = \frac{A+B}{B r_{1 \rm g}} I^\gen(r_{\rm g}) - \frac{2 A \mathsf{E}(\pi -\xi| k_{r \rm g})}{B r_{1 \rm g} Y_r} + \frac{(A r_{\rm g} + B(r_{1 \rm g} -r_{\rm g}))\sin(\xi) \sqrt{1 - k_{r \rm g} \sin^2(\xi)}}{B r_{1 \rm g}Y_r r_{\rm g}}  \ , \\
    I^\gen_{1/r^2}(r_{\rm g}) &= \int^{r_{1\rm g}}_{r_{\rm g}} \frac{\dd r'}{r'^2\sqrt{R_{\rm g}(r')}} = \frac{2 (r_{1 \rm g} - r_{\rm g})}{3 r_{1 \rm g} r_{\rm g}R_{\rm g}(r_{\rm g})} - \frac{1}{3B^3 Y_r} \frac{\sin(\xi)}{\sqrt{1 - k_{r \rm g} \sin(\xi)^2} } \bigg( \frac{\rho^2_{r \rm g} - \rho^2_{i \rm g}}{B} + \frac{4A k_{r \rm g} \rho_{r \rm g}}{r_{1 \rm g}} \cos(\xi) \bigg) \nonumber \\
    &\phantom{=\int^{r_{1\rm g}}_{r_{\rm g}} \frac{\dd r'}{r'^2\sqrt{R_{\rm g}(r')}}} + \frac{2 B^2(A+B) - B r^2_{1 \rm g} + 2 (2A +B) r_{1 \rm g}\rho_{r \rm g}}{3 B^3 r^2_{1 \rm g}}I^\gen(r_{\rm g}) - \frac{4 A (B^2 + 2r_{1 \rm g} \rho_{r \rm g})}{3r^2_{1 \rm g}B^3Y_r}\mathsf{E}(\pi -\xi| k_{r \rm g})  \nonumber \\
    &\phantom{=\int^{r_{1\rm g}}_{r_{\rm g}} \frac{\dd r'}{r'^2\sqrt{R_{\rm g}(r')}}} + \frac{2(B^2 + r_{1 \rm g} \rho_{r \rm g})}{3r^2_{1 \rm g}B^3Y_r r_{\rm g}}\big(A r_{\rm g} + B(r_{1 \rm g} - r_{\rm g}) \big)\sin(\xi) \sqrt{1 - k_{r \rm g} \sin^2(\xi)} \ , \\
    I^\gen_{r}(r_{\rm g}) &= \int^{r_{1\rm g}}_{r_{\rm g}} \frac{r'\dd r'}{\sqrt{R_{\rm g}(r')}} = \frac{\sqrt{A B}}{Y_r} \arctan\!\Bigg( \frac{r_{1 \rm g} \sin(\xi)}{2 \sqrt{A B} \sqrt{1 - k_{r \rm g} \sin^2(\xi)}} \Bigg) - \frac{B r_{1 \rm g}}{A - B}I^\gen(r_{\rm g}) \nonumber \\
    &\phantom{= \int^{r_{1\rm g}}_{r_{\rm g}} \frac{r'\dd r'}{\sqrt{R_{\rm g}(r')}}} + \frac{(A + B)r_{1 \rm g}}{2(A - B)Y_r} \mathsf{\Pi}\big(\!\left.\gamma_r,\pi - \xi \right\rvert k_{r \rm g} \big) \ , \\
    I^\gen_{r^2}(r_{\rm g}) &= \int^{r_{1\rm g}}_{r_{\rm g}} \frac{r'^2\dd r'}{\sqrt{R_{\rm g}(r')}} =  \frac{AB}{Y_r}\mathsf{E}(\pi - \xi| k_{r \rm g}) + \frac{(A-B)R_{\rm g}(r_{\rm g})}{(1 - E^2_{\rm g})(A r_{\rm g} + B(r_{1\rm g} - r_{\rm g}))} + \frac{1}{2}(r_{1 \rm g} + 2\rho_{r\rm g})I^\gen_{r}(r_{\rm g}) \nonumber \\
    &\phantom{=\int^{r_{1\rm g}}_{r_{\rm g}} \frac{r'^2\dd r'}{\sqrt{R_{\rm g}(r')}}} - \frac{B r_{1 \rm g}(r_{1 \rm g} - 2 \rho_{\rm g})}{2(A-B)}I^\gen(r_{\rm g}) \ , \\
    I^\gen_{1/(r - r_2)}(r_{\rm g}) &= \int^{r_{1\rm g}}_{r_{\rm g}} \frac{\dd r'}{(r' - r_{2 \rm g})\sqrt{R_{\rm g}(r')}} = \frac{1}{Y_r} \frac{r_{1 \rm g}(r_{\rm g} - \overline{r_{2 \rm g}})}{(r_{\rm g} - \overline{r_{2 \rm g}})(r_{1 \rm g} - r_{2 \rm g})r_{2 \rm g} - A B(r_{\rm g} - r_{2 \rm g})} \frac{\sin(\xi)}{\sqrt{1 - k_{r \rm g}\sin^2(\xi)}} \nonumber \\
    &\phantom{=\int^{r_{1\rm g}}_{r_{\rm g}} \frac{\dd r'}{(r' - r_{2 \rm g})\sqrt{R_{\rm g}(r')}}} -\frac{(A-B)}{A r_{2 \rm g} + B(r_{1 \rm g} - r_{2 \rm g})} I^\gen(r_{\rm g}) - \frac{2 A B r_{1 \rm g}\mathsf{E}(\pi -\xi| k_{r \rm g})}{\big(A^2 r^2_{2\rm g} - B^2(r_{1 \rm g} - r_{2 \rm g})^2\big)Y_r} \ , \\
    I^\gen_{1/(r - \overline{r_2})}(r_{\rm g}) &= \int^{r_{1\rm g}}_{r_{\rm g}} \frac{\dd r'}{(r' - \overline{r_{2 \rm g}})\sqrt{R_{\rm g}(r')}} = \overline{I^\gen_{1/(r - r_{2 \rm g})}(r_{\rm g})} \ , \\
    I^\gen_{1/(r - \alpha)}(r_{\rm g}) &= \int^{r_{1\rm g}}_{r_{\rm g}} \frac{\dd r'}{(r' - \alpha)\sqrt{R_{\rm g}(r')}} = - \frac{(A-B)I^\gen(r_{\rm g})}{(A \alpha + B(r_{1 \rm g} -\alpha))} - \frac{\log \!\Bigg[\Bigg( \frac{\sin(\xi)\sqrt{(h_\alpha-1)k_{r \rm g} - h_\alpha} \;\, - \;\, \sqrt{1 - h_\alpha}\sqrt{1 - k_{r \rm g} \sin^2(\xi)}}{\sin(\xi)\sqrt{(h_\alpha-1)k_{r \rm g} - h_\alpha} \;\, + \;\, \sqrt{1 - h_\alpha}\sqrt{1 - k_{r \rm g}\sin^2(\xi)} } \Bigg)^{\!\!2} \Bigg]}{4\sqrt{(1- E^2_{\rm g})(r_{1 \rm g} - \alpha)\alpha(\rho^2_{i \rm g} + (\alpha - \rho_{r \rm g})^2)}} \nonumber \\
    &\phantom{=\int^{r_{1\rm g}}_{r_{\rm g}} \frac{\dd r'}{(r - \alpha)\sqrt{R_{\rm g}(r')}}} - \frac{4 A B r_{1 \rm g}  \Big[ \mathsf{\Pi}\Big(h_\alpha \Big\rvert \frac{k_{r \rm g}}{k_{r \rm g} -1} \Big) + \mathsf{\Pi}\Big(\!\left.h_\alpha,\frac{\pi}{2} - \xi \right\rvert \frac{k_{r \rm g}}{k_{r \rm g} -1}\Big) \Big]}{\sqrt{2(1 - E^2_{\rm g})(B^2 + A B - \rho_{r\rm g} r_{1 \rm g})}(A^2 \alpha^2 -B^2(r_{1 \rm g} - \alpha)^2)} \ .
\end{align} 
\end{widetext}
where the bar denotes complex conjugation, and $r_{2 \rm g} = \rho_{r \rm g} + i \rho_{i \rm g}$. Additionally, we used the following diverging integral 
\begin{align}
    I^\gen_{1/(r - r_1)}(r_{\rm g}) &= \int^{r_{1\rm g}}_{r_{\rm g}} \frac{\dd r'}{(r' - r_{1 \rm g})\sqrt{R_{\rm g}(r')}} \nonumber \\
    & = I^{\gen,\mathsf{sing}}_{1/(r - r_1)}(r_{\rm g}) + I^{\gen,\mathsf{reg}}_{1/(r - r_1)}(r_{\rm g}) \ . 
\end{align}
The diverging part of the previous integral is 
\begin{align}
    I^{\gen,\mathsf{sing}}_{1/(r - r_1)}(r_{\rm g}) &= \frac{2}{\sqrt{R_{\rm g}(r_{\rm g})}}   \ , 
\end{align}
while the regular part is given by
\begin{widetext}
\begin{align}
    I^{\gen,\mathsf{reg}}_{1/(r - r_1)}(r_{\rm g}) &= \frac{2B\mathsf{E}(\pi - \xi|k_{r\rm g})}{r_{1 \rm g}AY_r}  - \frac{\big(A r_{\rm g} + B (r_{1 \rm g} - r_{\rm g})\big)(A B + r_{\rm g}(r + r_{1 \rm g} - 2 \rho_{r\rm g}))}{ A r_{1 \rm g}Y_r(\rho^2_{i \rm g} + (r_{\rm g} - \rho_{r \rm g})^2)r_{\rm g}} \sin(\xi) \sqrt{1 - k_{r \rm g} \sin^2(\xi)} \nonumber \\
    & - \frac{(A -B)I^\gen(r_{\rm g})}{A r_{1 \rm g}}   \ ,
\end{align}    
\end{widetext}
The integrals for generic plunges reduce to the ones of Appendix~\ref{app:integrals_critical_plunge} in the limit $\rho_{i \rm g} \to 0$ since  $k_{r \rm g} \to 1 \text{ as } \rho_{i \rm g} \to 0$. In this case, all elliptic integrals reduce to elementary functions (see Eqs.~\eqref{eq:limit_elliptic_integrals}). 

\section{Integrals for special plunges} \label{app:integrals_special_plunge}
This Section presents the integrals that appear in the closed-form solutions for the special plunges. 
\begin{widetext}    
\begin{align}
    I^\spec(r_{\rm g}) &= \int^{r_{1 \rm g}}_{r_{\rm g}}\!\frac{\dd r'}{\sqrt{R_{\rm g}(r')}} = -\frac{2}{\sqrt{(1 - E^2_{\rm g})r_{1\rm g} r_{2\rm g}}} \arctan \!\Bigg( \sqrt{\frac{(r_{1\rm g} - r_{\rm g})r_{2\rm g}}{r_{1 \rm g}(r_{\rm g} - r_{2\rm g})}} \Bigg) = \lambda \ , \\
    I^\spec_{r}(r_{\rm g}) &= \int^{r_{1 \rm g}}_{r_{\rm g}}\! \frac{r'\dd r'}{\sqrt{R_{\rm g}(r')}} = - \frac{2}{\sqrt{(1 - E^2_{\rm g})}} \arctan \!\Bigg( \sqrt{\frac{r_{1\rm g} - r_{\rm g}}{r_{\rm g} - r_{2\rm g}}} \Bigg) \ , \\
    I^\spec_{r^2}(r_{\rm g}) &= \int^{r_{1 \rm g}}_{r_{\rm g}}\! \frac{r'^2\dd r'}{\sqrt{R_{\rm g}(r')}} = - \frac{ \sqrt{(r_{1\rm g} - r_{\rm g})(r - r_{2\rm g})}}{\sqrt{(1 - E^2_{\rm g})}} + \frac{1}{2} (r_{1\rm g} - r_{2\rm g})I^\spec_{r}(r_{\rm g})   \ , \\
    I^\spec_{1/(r - r_\pm)}(r_{\rm g}) &= \int^{r_{1 \rm g}}_{r_{\rm g}}\! \frac{\dd r'}{(r' - r_\pm)\sqrt{R_{\rm g}(r')}} = -\frac{I^\spec(r_{\rm g})}{r_\pm} + \frac{\log \!\Bigg[\Bigg( \frac{\sqrt{(r_{1 \rm g} - r_\pm)(r_{\rm g} -r_{2\rm g})} \;\, - \;\, \sqrt{(r_{1 \rm g} - r_{\rm g})(r_\pm - r_{2 \rm g})}}{\sqrt{(r_{1 \rm g} - r_\pm)(r_{\rm g} -r_{2\rm g})}  \;\, + \;\, \sqrt{(r_{1 \rm g} - r_{\rm g})(r_\pm - r_{2 \rm g})}} \Bigg)^{\!\!2} \Bigg]}{2\sqrt{(1 - E^2_{\rm g})}\sqrt{(r_{1 \rm g} - r_\pm)(r_\pm - r_{2 \rm g})}} \ , \\
    I^\spec_{1/(r - r_1)}(r_{\rm g}) &= \int^{r_{1 \rm g}}_{r_{\rm g}}\!\frac{\dd r'}{(r' - r_{1\rm g})\sqrt{R_{\rm g}(r')}} = - \frac{2}{r_{1\rm g}(r_{1\rm g} - r_{2\rm g})\sqrt{1 - E^2_{\rm g}}} \sqrt{\frac{r_{\rm g} - r_{2\rm g}}{r_{1\rm g} - r_{\rm g}}} - \frac{1}{r_{1\rm g}} I^\spec(r_{\rm g}) \ , \\ 
    I^\spec_{1/(r - r_2)}(r_{\rm g}) &= \int^{r_{1 \rm g}}_{r_{\rm g}}\!\frac{\dd r'}{(r' - r_{2\rm g})\sqrt{R_{\rm g}(r')}} = - \frac{2}{r_{2\rm g}(r_{1\rm g} - r_{\rm g})\sqrt{1 - E^2_{\rm g}}} \sqrt{\frac{r_{1\rm g} - r_{\rm g}}{r_{\rm g} - r_{2\rm g}}} - \frac{1}{r_{2\rm g}} I^\spec(r_{\rm g}) \ ,
\end{align}
\end{widetext}

\section{Analytic solutions for geodesic plunge motion on the equatorial plane} \label{app:geo_analytic}
This Section provides the analytic solutions for geodesic bound plunges on the equatorial plane. closed-form expressions for the critical plunges and homoclinic motion on the equatorial plane were first derived in Ref.~\cite{Levin:2008yp}, while explicit solutions for equatorial ISCO plunges and PBOs with a double root were presented in Ref.~\cite{Mummery:2023hlo}. Ref.~\cite{Dyson:2023fws} extended these solutions in the case of inclined, plunging orbits, while presenting, for the first time, generic plunges and PBOs with a single root in analytic form. Analytic solutions for bound and unbound plunging motion in Kerr-Newman spacetime were given in Ref.~\cite{Ko:2023igf}. 
Finally, plunging orbits with $\Lzrd =0$ and $E_{\rm g}>1$ were already derived in Ref.~\cite{chandrasekhar1983mathematical}, although a closed-form solution for the radial motion is missing~\footnote{The concept of Mino-time was introduced much later on in~\cite{Mino:2003yg}.}. Here we present the complete solution for plunging orbits with $\Lzrd =0$ and $E_{\rm g} < 1$ (or special plunges).

The geodesic equations of motion~\eqref{eq:1st-order-EoMgeo} can be parametrized in the radius $r_{\rm g}$ as follows
\begin{subequations}
   \begin{empheq}[]{align}
    &\lambda = \pm \int \frac{\dd r'}{\sqrt{R_{\rm g}(r')}} \, , \label{eq:sol_geo_radial_motion}  \\  
	&\totder{t}{r_{\rm g}} = \pm \frac{T_{\rm g}(r_{\rm g})}{\sqrt{R_{\rm g}(r_{\rm g})}} \, , \\ 
	&\totder{\phi}{r_{\rm g}} = \pm \frac{\Phi_{\rm g}(r_{\rm g})}{\sqrt{R_{\rm g}(r_{\rm g})}}  \, .
   \end{empheq}
\end{subequations} 
The solutions for geodesic coordinate-time and azimuthal motion can in general be written as
\begin{equation}
    t^\mathsf{y}_{\rm g}(r_{\rm g}) = - t^{\mathsf{y},\text{sgn}}_{\rm g}(r_{\rm g}) \ , \qquad \phi^\mathsf{y}_{\rm g}(r_{\rm g}) = - \phi^{\mathsf{y},\text{sgn}}_{\rm g}(r_{\rm g})
\end{equation}
with $\text{y} = (\crit, \isco, \pbo, \spec)$ and
\begin{widetext}
    \begin{align}
        t^{\mathsf{y},\text{sgn}}_{\rm g}(r_{\rm g}) &= E_{\rm g} \big(4 I^\mathsf{y}(r_{\rm g}) +  2 I^\mathsf{y}_{r}(r_{\rm g}) + I^\mathsf{y}_{r^2}(r_{\rm g}) \big) - \frac{2}{r_+ - r_-} \big(2 a^2 E_{\rm g} + (a L_{z \rm g} - 4 E_{\rm g}) r_+ \big)I^\mathsf{y}_{1/(r - r_+)}(r_{\rm g}) + (+\leftrightarrow-)   \ ,  \label{eq:coordinate_time_geo_trajectory} \\ 
        \phi^{\mathsf{y},\text{sgn}}_{\rm g}(r_{\rm g}) &= L_{z\rm g} I^\text{y}(r_{\rm g}) + \frac{a (2 r_+ E_{\rm g} - a L_{z \rm g})}{(r_+ - r_-)} I^\text{y}_{1/(r - r_+)}(r_{\rm g}) +  (+\leftrightarrow-)  \ . \label{eq:azimuthal_geo_trajectory}
    \end{align}    
\end{widetext}
The integrals $I^\mathsf{y}(r_{\rm g}), I^\mathsf{y}_{r}(r_{\rm g}), I^\text{y}_{r^2}(r_{\rm g}), I^\mathsf{y}_{1/(r - r_\pm)}(r_{\rm g})$ are given in Appendix~\ref{app:integrals_critical_plunge}-\ref{app:integrals_pbo} for critical plunges ($\mathsf{y} = \crit$), ISCO plunges ($\text{y} = \isco$) and PBOs ($\text{y} = \pbo$), respectively, and in Appendix~\ref{app:integrals_special_plunge} for the special plunges ($\text{y} = \spec$). For generic plunges
\begin{equation}
    t^\gen_{\rm g}(r_{\rm g}) = t^{\mathsf{gen},\text{sgn}}_{\rm g}(r_{\rm g}) \ , \quad \phi^\gen_{\rm g}(r_{\rm g}) = \phi^{\mathsf{gen},\text{sgn}}_{\rm g}(r_{\rm g}) \ ,
\end{equation}
and the integrals $I^\mathsf{y}(r_{\rm g}), I^\mathsf{y}_{r}(r_{\rm g}), I^\mathsf{y}_{r^2}(r_{\rm g}), I^\mathsf{y}_{1/(r - r_\pm)}(r_{\rm g})$ can be found in Appendix~\ref{app:integrals_ISCO_plunge}. The sign choice ensures that generic plunging orbits match homoclinic orbits in the limit $\rho_{i \rm g} \to 0$. 

\subsection{Radial motion} 
The geodesic radial motion in Mino-time can be readily solved by inverting Eq.~\eqref{eq:sol_geo_radial_motion}. We list here the solutions for $r_{\rm g}(\lambda)$ for the various classes of bound plunges.\\
Critical plunge:
\begin{align} 
    r_{\rm g}(\lambda) &= \frac{r_{1 \rm g} (r_{1 \rm g} - r_{2 \rm g}) \tanh^2\big(\frac{1}{2}Y_r \lambda\big)}{r_{2 \rm g} + (r_{1 \rm g} - r_{2 \rm g}) \tanh^2\big(\frac{1}{2}Y_r \lambda\big)} + r^{(0)} \ , \\
    Y_r &= \sqrt{(1 - E^2_{\rm g})r_{2 \rm g}(r_{1 \rm g} - r_{2 \rm g})} \ ,
\end{align}
with $r^{(0)}$ an integration constant. \\
ISCO plunges
\begin{equation} 
    r_{\rm g}(\lambda) = r_\text{g,isco} + r^{(0)} -\frac{4 r_\text{g,isco}}{4 + (1 - E^2_{\rm g}) r^2_\text{g,isco} \lambda^2}  \ .
\end{equation}
with $r^{(0)}$ an integration constant. \\
PBOs - single roots:
\begin{align} 
     r_{\rm g}(\lambda) &= \frac{r_{2 \rm g} r_{3 \rm g} \mathsf{cn}^2\big(\frac{1}{2} Y_r \lambda |k_{r \rm g}\big)}{r_{2 \rm g} - r_{3 \rm g}\mathsf{sn}^2\big(\frac{1}{2} Y_r \lambda |k_{r \rm g}\big)} \ ,   \\
    Y_r &= \sqrt{(1 - E^2_{\rm g})r_{2 \rm g}(r_{1 \rm g} - r_{3 \rm g})} \ . 
\end{align}
with $k_{r\rm g}$ defined in Appendix~\ref{app:integrals_pbo}. \\
PBOs - double root $r_{1 \rm g} = r_{2 \rm g}$:
\begin{align} 
    r_{\rm g}(\lambda) &= \frac{r_{1 \rm g} r_{3 \rm g} \cos^2\big(\frac{1}{2} Y_r \lambda\big)}{r_{1 \rm g} - r_{3 \rm g}\sin^2\big(\frac{1}{2} Y_r \lambda\big)} \ , \\
    Y_r &= \sqrt{(1 - E^2_{\rm g})r_{1 \rm g}(r_{1 \rm g} - r_{3 \rm g})} \ . 
\end{align}
Generic plunges:
\begin{align} 
     r_{\rm g}(\lambda) &= \frac{r_{1 \rm g} B \big[1 + \mathsf{cn}^2\big(Y_r \lambda |k_{r \rm g}\big) \big]}{A + B - (A - B)\mathsf{sn}^2\big( Y_r \lambda |k_{r \rm g}\big)} \ ,   \\
    Y_r &= \sqrt{(1 - E^2_{\rm g})AB} \ . 
\end{align}
with $A,B, k_{r\rm g}$ defined in Appendix~\ref{app:integrals_generic_plunge}. \\
Special plunges:
\begin{align} 
    r_{\rm g}(\lambda) &= \frac{r_{1 \rm g} r_{2 \rm g} \Big[1 + \tan\big(\frac{1}{2}Y_r \lambda\big) \Big]}{r_{2 \rm g} + r_{1 \rm g} \tan\big(\frac{1}{2}Y_r \lambda\big)} \ , \\
    Y_r &= \sqrt{(1 - E^2_{\rm g})r_{1 \rm g} r_{2 \rm g}} \ . 
\end{align}

\section{Corrections to the constants of motion} \label{app:spin_constants_of_motion_corr}
\subsection{Critical plunges, ISCO plunges and PBOs}
The expressions $\delta E^\text{s}$ and $\delta L^\text{s}_z$ appearing in the main text are the following
\begin{widetext}
    \begin{align}
        \delta E^\text{s} &=\frac{\Lzrd}{\mathcal D} \big[ -2 a \Lzrd^2(r_{1 \rm g} + r_{2 \rm g}) - L_{z\rm g} E_{\rm g} r^2_{1 \rm g} r^2_{2 \rm g} -a L_{z\rm g} \Lzrd r_{1 \rm g} r_{2 \rm g} + a \Lzrd L_{z\rm g} (r_{1 \rm g} + r_{2\rm g})^2 \big] \ , \label{eq:delta_E_FT} \\
        \delta L_z^\text{s} &=\frac{E_{\rm g}r_{1\rm g} r_{2\rm g}}{\mathcal D} \Lzrd \big[ a^3 \Lzrd - a (a E_{\rm g} - 3 \Lzrd) r_{1 \rm g} r_{2 \rm g} - 3 E_{\rm g} r^2_{1\rm g} r^2_{2\rm g} \big] \nonumber \\
        & + \frac{a \Lzrd^2}{\mathcal D}\big[ - 2 a\Lzrd (r_{1\rm g} + r_{2\rm g}) + a^2 E_{\rm g} (r^2_{1\rm g} + r^2_{2\rm g}) + E_{\rm g} (r^4_{1\rm g} + r^4_{2\rm g}) \big]  \nonumber\\            & + \frac{E_{\rm g}r_{1\rm g} r_{2\rm g}}{\mathcal D} \big[ E_{\rm g} L_{z \rm g}r^2_{1\rm g} r^2_{2\rm g}(r_{1\rm g} + r_{2\rm g}) - 3E_{\rm g}\Lzrd r_{1\rm g} r_{2\rm g}(r^2_{1\rm g} + r^2_{2\rm g}) + a \Lzrd^2 (r^2_{1\rm g} + r^2_{2\rm g}) \big]  \ ,  \label{eq:delta_Lz_FT}      
    \end{align}   
    with
    \begin{equation}
        \mathcal D = r^2_{1 \rm g} r^2_{2 \rm g} \Big[ E_{\rm g} L_{z\rm g} r_{1\rm g} r_{2\rm g} (r_{1\rm g} + r_{2\rm g}) + 2 a \Lzrd^2 - 2 \Lzrd  E_{\rm g}(r_{1\rm g} + r_{2\rm g})^2 + 2 \Lzrd  E_{\rm g}r_{1\rm g} r_{2\rm g}  \Big] \ .
    \end{equation}
\end{widetext}
For critical plunges, the turning points $r_{1 \rm g}$ and $r_{2 \rm g}$ in Eqs.~\eqref{eq:delta_E_FT}-\eqref{eq:delta_Lz_FT} are
\begin{align}
    r_{1 \rm g} = \frac{p^*_{\rm g}}{1- e^*_{\rm g}} \ , \quad  r_{2 \rm g} =  \frac{p^*_{\rm g}}{1 + e^*_{\rm g}} \ , \quad \text{ critical plunges } \ .
\end{align}

In the case of plunges from the ISCO, Eqs.~\eqref{eq:delta_E_FT}-\eqref{eq:delta_Lz_FT} simplify to
\begin{widetext}
    \begin{align}
        \delta E^\text{s} &= -\frac{\Lzrd (4 a \Lzrd^2 - 3a L_{z \rm g}\Lzrd r_\text{g,isco} + E_{\rm g} L_{z \rm g}r^3_\text{g,isco} )}{2r^3_\text{g,isco} \big(a\Lzrd^2 - 3 E_{\rm g}\Lzrd r^2_\text{g,isco} + E_{\rm g} L_{z \rm g}r^3_\text{g,isco} \big)} \ ,  \label{eq:shift_energy_s_ISCO}  \\
        \delta L_z^\text{s} &= \frac{3 a^3 \Lzrd^2 r_\text{g,isco} + 7a E_{\rm g} \Lzrd L_{z \rm g} r^3_\text{g,isco} + E^2_{\rm g}r^5_\text{g,isco}(2 L_{z \rm g}r_\text{g,isco} -9\Lzrd) - 4a^2 \Lzrd(\Lzrd^2 + 2 E^2_{\rm g} r^3_\text{g,isco})}{2r^3_\text{g,isco} \big(a\Lzrd^2 - 3 E_{\rm g}\Lzrd r^2_\text{g,isco} + E_{\rm g} L_{z \rm g}r^3_\text{g,isco} \big)} \ ,  \label{eq:shift_angular_momentum_s_ISCO}
    \end{align}   
\end{widetext}
 
\subsection{Generic plunges}
The shifts to the constants of motion $\delta E^{\text{s},\gen}$ and $\delta L^{\text{s},\gen}_z$ for generic plunges are the following
\begin{widetext}
\begin{align}
    \delta E^{\text{s},\gen} &= -\frac{(1 - e^2_{\rm g})^2a \big( a^2 (1 -E^2_{\rm g}) + L^2_{z\rm g}\big)}{2(3-e_{\rm g})^2 E_{\rm g} \Lzrd^2 p^2_{\rm g}}  \ , \\
    \delta L^{\text{s},\gen}_z &= \frac{1}{r_{1 \rm g}L_{z \rm g} - 2 \Lzrd} \bigg(\frac{1}{r^2_{1 \rm g}}\big(a \Lzrd^2 - 3 E_{\rm g} \Lzrd r^2_{1 \rm g} + E_{\rm g} L_{z\rm g} r^3_{1 \rm g} \big) + \big(a^2 E_{\rm g} r_{1 \rm g} -2a \Lzrd + E_{\rm g}r^3_{1 \rm g}  \big) \delta E^{\text{s},\gen} \bigg) \ , 
\end{align}
\end{widetext}

\strut\newpage

\section{Corrections to the semilatus-rectum, IBCO and ISCO} \label{app:spin_delta_p_corr}
\subsection{Shift to the IBCO}
The shift to the IBCO radius is given by
\begin{equation}
  \delta r_\text{ibco} = \frac{\delta r_\text{ibco,num}}{\delta r_\text{ibco,den}}\ ,
\end{equation}
with
\begin{widetext}
    \begin{align}
        \delta r_\text{ibco,num} &= L_{z \rm g}(\tilde r_{\rm g}-2)^3\sqrt{\Delta(\tilde r_{\rm g})}\big[2\tilde L_{z \rm g}(a\tilde L_{z \rm g}^3 - \tilde L_{z \rm g}^2\tilde r^2_{\rm g} - a L_{z \rm g}\tilde r^2_{\rm g}) + a L^2_{z \rm g} \tilde r^3_{\rm g} \big] \ , \\
        \delta r_\text{ibco,den} &= 2\tilde L_{z \rm g}^2(L_{z \rm g}\tilde r_{\rm g} - 2\tilde L_{z \rm g})\tilde r^2_{\rm g}\sqrt{\Delta(\tilde r_{\rm g})}\big[10a^2 -16 + 8 \tilde r_{\rm g} + (\tilde r_{\rm g} -4)\Delta(\tilde r_{\rm g})\big] \nonumber \\
        &  - \sgn(L_{z \rm g}) 2\sqrt{2}a\tilde L_{z \rm g}^2\tilde r^{3/2}_{\rm g}(L_{z \rm g}\tilde r_{\rm g} - 2\tilde L_{z \rm g})\big[2a^2(\tilde r_{\rm g} - 1) + (\tilde r_{\rm g}+4)\Delta(\tilde r_{\rm g})\big] \ ,
    \end{align}
\end{widetext}
where we use the shorthand notation $\tilde r_{\rm g} \equiv r_\text{g,ibco}$ and $\tilde L_{z \rm g} = L_{z \rm g} - a$, $\Delta(\tilde r_{\rm g}) = \tilde r^2_{\rm g} - 2 \tilde r_{\rm g} + a^2$.

\subsection{Shift to the ISCO}
The shift to the ISCO can be obtained from Eq.~\eqref{eq:semi_latus_rectum_shift} with $(p_{\rm g}, e_{\rm g}) = (r_\text{g,isco},0)$. After some algebraic manipulations, the final result is  
\begin{equation}
  \delta r_\text{ibco} = \frac{a \big(a^2(1 - E^2_{\rm g}) + L^2_{z \rm g} \big)}{6\Lzrd^2} - \frac{2}{3} \frac{E_{\rm g}2\delta E^\text{s}}{(1 - E^2_{\rm g})^2}  \ , \label{eq:shift_ISCO}
\end{equation}    
with $\delta E^\text{s}$ given by Eq.~\eqref{eq:shift_energy_s_ISCO}. Eq.~\eqref{eq:shift_ISCO} agrees with $r_1$ given in Eq. 64 of~\cite{Jefremov:2015gza}.

\subsection{Generic plunges}
The correction to the semilatus-rectum for generic plunging orbits is the following
\begin{equation}
    \delta p^\gen= \frac{\delta p^\gen_{\text{num}}}{\delta p^\gen_{\text{den}}}\ ,
\end{equation}
with
\begin{widetext}
    \begin{align}
        \delta p^\gen_{\text{num}} &= \frac{(1 - e^2_{\rm g})}{(e_{\rm g} -3)^2} \frac{8 \delta r_{4}}{\rho^2_{r \rm g}r_{1 \rm g}} \bigg( (1 - E^2_{\rm g})(2 \Lzrd^2 + r^2_{1 \rm g}) + \frac{2r_{1 \rm g}(1 - E^2_{\rm g})^2\Lzrd(E_{\rm g}r^2_{1 \rm g} - a \Lzrd )}{E_{\rm g}(\Lzrd r_{1 \rm g} -2 \Lzrd)} -2 r_{1 \rm g}\bigg) \nonumber \\
        & - 8 \frac{(1 - E^2_{\rm g})^2}{r^2_{1 \rm g}} \Lzrd^2 \delta r_4 - \frac{4 a}{r_{1 \rm g}}(1 - E^2_{\rm g})^2 +  \frac{8 a (1 -E^2_{\rm g})^2 \Lzrd^3}{(L_{z \rm g} r_{1 \rm g} - 2 \Lzrd)r^3_{1 \rm g}} - \frac{4(1 - E^2_{\rm g})^2\Lzrd E_{\rm g} L_{z \rm g}}{L_{z \rm g}r_{1 \rm g} - 2 \Lzrd} \nonumber \\
        & - (1 -E^2_{\rm g})^2 2\delta r_4 \big(2 - (1- E^2_{\rm g})r_{1 \rm g} \big) \ , 
    \end{align}
    \begin{align}
        \delta p^\gen_{\text{den}} &=  - \frac{\partial L_{z \rm g}}{\partial p} (1 - E^2_{\rm g})^2 \frac{\Lzrd}{r_{1 \rm g}} +\frac{\partial E_{\rm g}}{\partial p} \Big(8 E_{\rm g} + \frac{8}{r_{1 \rm g}}(1 - E^2_{\rm g}) \big(a(1 - E^2_{\rm g})\Lzrd - E_{\rm g}\Lzrd^2 \big) -4 E_{\rm g}(1 - E^2_{\rm g})r_{1 \rm g} \Big) \nonumber \\ 
        & + \frac{(1 - E^2_{\rm g})^2}{(1 -e_{\rm g})r^2_{1 \rm g}}\big( 4 \Lzrd^2 + r^2_{1 \rm g}(-2 + (1 - E^2_{\rm g})r_{1 \rm g}) \big) \ .
    \end{align}   
\end{widetext}

\bibliographystyle{utphys}
\bibliography{Ref}

\end{document}